%% file: correlations_final_final_arxiv.tex
\documentclass[aps,pra,twocolumn,superscriptaddress]{revtex4-1}
\usepackage{CJKutf8}
\usepackage{graphics}
\usepackage[normalem]{ulem}
\usepackage{amssymb}
\usepackage{color}
\usepackage{graphicx}
\usepackage{amsthm}
\usepackage{paralist}
\usepackage{verbatim}
\usepackage{chemarr}
\usepackage{braket}
\usepackage{enumitem}
\usepackage{float}
\usepackage{tikzit}
\input{figure_styles.tikzstyles}
\usetikzlibrary{circuits.ee.IEC}
\usetikzlibrary{patterns}
\pgfdeclarelayer{nodelayer} 
\pgfdeclarelayer{edgelayer}   
\pgfsetlayers{nodelayer,edgelayer} 

\usepackage[caption=false]{subfig}
\usepackage{commath}
\usepackage{hyperref}

\newcommand{\spc}[1]{\mathcal{#1}}

\def\>{\rangle}
\def\<{\langle}

\newcommand{\ketbra}[2]{\ket{#1} \!  \bra{#2}}

\newcommand{\bs}[1]{\boldsymbol{#1}}     

\newcommand{\map}[1]{\mathcal{#1}}
\newcommand{\Tr}{\operatorname{Tr}}

\newcommand{\Chan}{{\mathsf{Chan}}}

\newtheorem{theo}{Theorem}
\newtheorem{prop}[theo]{Proposition}
\newtheorem{cor}[theo]{Corollary}
\newtheorem{lem}[theo]{Lemma}

\newcommand{\Proof}{{\bf Proof. \,}}

\usepackage{xr}

\newcommand{\hk}[1]{{\color{magenta}#1}}

\begin{document}
\begin{CJK*}{UTF8}{gbsn}
	
\title{Witnessing latent time correlations with a single quantum particle}
\author{Hl\'er Kristj\'ansson}
\affiliation{Quantum Group, Department of Computer Science, University of Oxford, Wolfson Building, Parks Road, Oxford, OX1 3QD, United Kingdom}
\affiliation{HKU-Oxford Joint Laboratory for Quantum Information and Computation}
\author{Wenxu Mao (貌文徐)}
\affiliation{QICI Quantum Information and Computation Initiative, Department of Computer Science, The University of Hong Kong, Pokfulam Road, Hong Kong}
\affiliation{Cornell Tech, Cornell University, 2 West Loop Road, New York, NY 10044, United States of America}
\author{Giulio Chiribella}
\email{giulio.chiribella@cs.ox.ac.uk}
\affiliation{QICI Quantum Information and Computation Initiative, Department of Computer Science, The University of Hong Kong, Pokfulam Road, Hong Kong}
\affiliation{Quantum Group, Department of Computer Science, University of Oxford, Wolfson Building, Parks Road, Oxford, OX1 3QD, United Kingdom}
\affiliation{HKU-Oxford Joint Laboratory for Quantum Information and Computation}
\affiliation{Perimeter Institute for Theoretical Physics, 31 Caroline Street North, Waterloo, Ontario, N2L 2Y5, Canada}



\begin{abstract}
When a noisy communication channel is used multiple times, the errors occurring at different times generally exhibit correlations.  Classically, these correlations do not affect the evolution of individual particles:  a single classical particle can only traverse the channel at a definite moment of time, and its evolution is insensitive to  the correlations between subsequent uses of the channel.  In stark contrast, here we show that a single quantum particle can sense the correlations between multiple uses of a channel at different moments of time. 
Taking advantage of this phenomenon, it is possible to enhance the amount of information that the  particle can reliably carry through the channel.          
In an extreme example, we show that a   channel that outputs white noise whenever the particle is sent at  a definite  time  can exhibit correlations that enable a perfect transmission of classical bits when the  particle is sent at a superposition of two distinct times. In contrast, we show that,  in the lack of  correlations, a single particle sent at a superposition of two times undergoes  an effective   channel with classical capacity of at most 0.16 bits. When multiple transmission lines are available, time correlations can be used to simulate the application of quantum channels in a coherent superposition of alternative causal orders, and even to provide   communication advantages that are not accessible through the  superposition of causal orders. 
\end{abstract}

\maketitle
\end{CJK*}

\section{Introduction}

Quantum communication enables new possibilities that were unthinkable in the classical world, notably including  secure key distribution 
 \cite{BennettCh1984, ekert1991quantum}.   The main hurdle to the implementation of quantum communication, however, is  the fragility of quantum states to noise.  To tackle this problem,  quantum error correction schemes  encode information into  multiple quantum  particles,  using  redundancy  to mitigate the effects of
 noise \cite{shor95error,gottesman2010introduction,lidar2013quantum}.

When the same communication channel is used multiple times,  the noisy processes experienced by particles sent at different  times  are generally correlated \cite{macchiavello02correlations,kretschmann2005quantum,caruso2014quantum,pollock2018nonmarkov}. For example, photons transmitted through an optical fibre are subject to random  changes in their polarisation \cite{ball2005hybrid}, and since such changes  happen on a finite timescale, photons sent at  nearby times experience approximately the same noisy  processes.  A similar situation arises in satellite quantum communication,  where the satellite's motion induces dynamical mismatches of reference frame with respect to the ground station \cite{bonato2006influence}.

The presence of correlations is both a threat and an opportunity for communication. On the one hand,  it can undermine the effectiveness of standard error correcting schemes, which  assume independent errors on  the transmitted particles. On the other hand, 
tailored codes that exploit the correlations among different particles can enhance the transmission of information \cite{macchiavello02correlations,chiribella2011quantum,giovannetti2005information,macchiavello2004transition,ruggeri2005information,cerf2005quantum,giovannetti2005bosonic,ball2004exploiting,banaszek2004experimental,caruso2014quantum,bowen04correlations,plenio2007spin,bayat2008memory,karpov2006entanglement,memarzadeh2011recovering,xiao2016protecting,darrigo2012transmission}.  

Like most error correcting schemes, the existing  codes for correlated noise  use multiple physical particles to encode a single logical message. Classically,  the use of multiple particles is essential: 
since  a single classical particle can only traverse  a communication channel at a definite  moment of time,   correlations between different uses of the  channel do not affect the particle's evolution. The same conclusion holds even if the moment of transmission is chosen at random: in this case, the resulting evolution is simply the average of the evolutions associated to each individual moment of time, and the overall evolution is independent of the time correlations.   

\begin{figure}
	\includegraphics[width=0.4\textwidth]{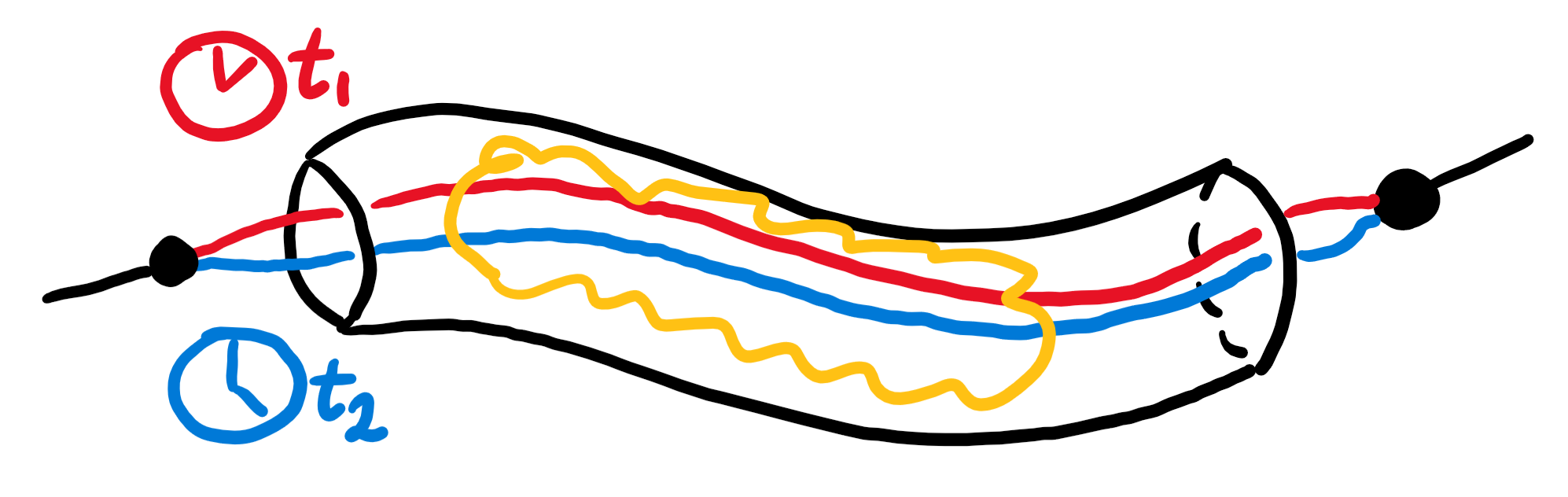}
	\caption{\label{fig:freestyle}  A single quantum particle can travel through a transmission line at a superposition of two different moment of time $t_{1}$ (red) and $t_{2}$ (blue). Along the way, the particle experiences errors (yellow region), and the errors occurring at time $t_1$ are generally correlated with the errors occurring  at time $t_2$.    By taking advantage of these correlations, the errors can be mitigated or even completely removed.}
\end{figure}

 In stark contrast, here we show that a single quantum particle can sense the correlations between multiple uses of the same quantum communication channel.  At the fundamental level, this effect is  made possible by  the ability of quantum particles to  experience a coherent superposition of multiple time-evolutions \cite{Aharonov1990,oi2003interference,aaberg2004subspace,gisin2005error,abbott2018communication,chiribella2019shannon2q,dong2019controlled,vanrietvelde2021universal}.   In particular, we will consider the situation in which the particle is in a superposition of travelling at different moments of time, as illustrated in Figure \ref{fig:freestyle}. 
  Taking advantage of the time correlations in the noise, we show that it is possible to enhance the amount of information that a single  particle can carry from a sender to a receiver, beating the ultimate limit achievable in the lack of correlations.     
  
    We demonstrate this effect with an  extreme example, in which a single quantum particle carries one  bit of classical information  through a transmission line  that completely erases information at every definite time step.    This phenomenon witnesses  the presence of correlations between different uses of the transmission line:   in the lack of correlations, we show that the number of bits that can be reliably transmitted by sending a single particle at a superposition of two different times does not exceed 0.16.


\begin{figure}
	\includegraphics[width=\linewidth]{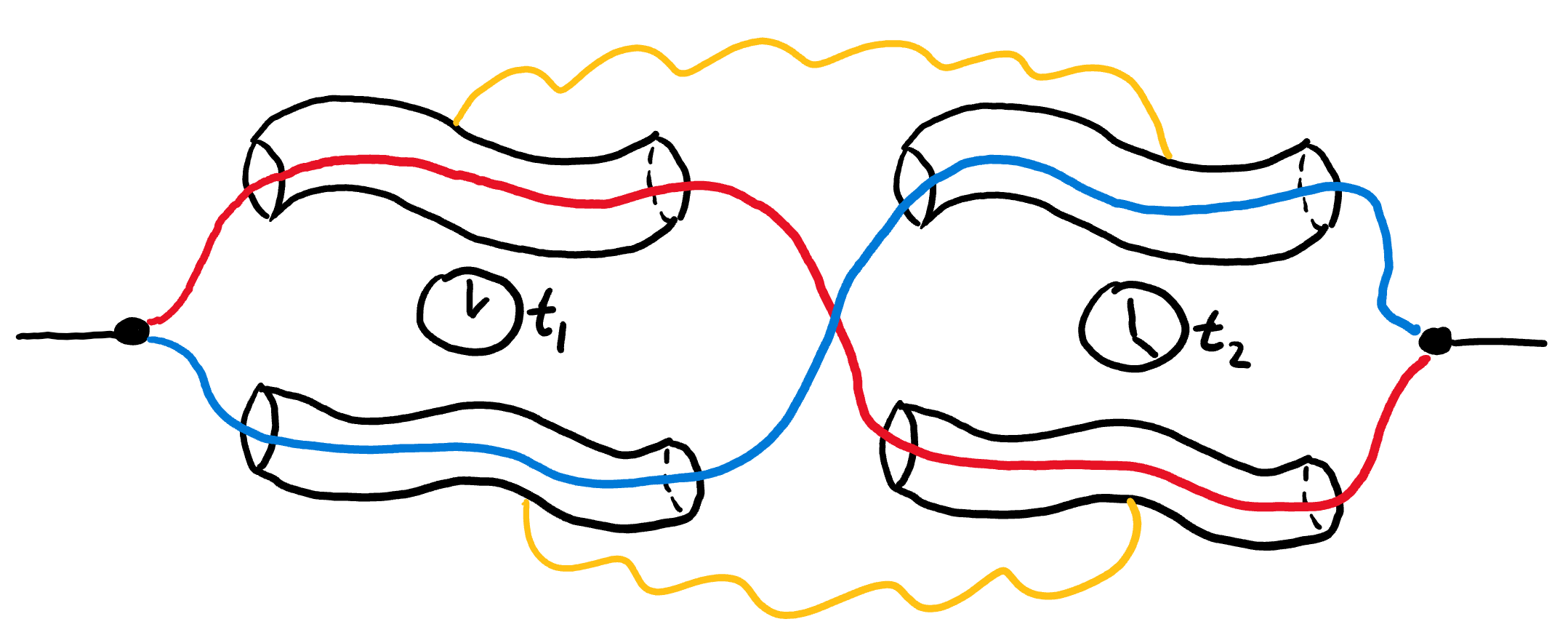}
	\caption{\label{fig:freestyle_switch}  A single particle can travel on a superposition of two different paths (red and blue), which traverse  two transmission lines  (top and bottom) at two moments of time $t_1$ and $t_2$.   The errors occurring on successive uses of the same transmission line are correlated (yellow lines), so the particle experiences  correlated errors across the two branches   (red and blue) of the superposition.   These time correlations are a resource that can be used to mimic the use of quantum channels in a superposition of orders, and even to achieve larger communication advantages.}
\end{figure}

It is worth stressing  that the above advantage is not specific to  time correlations, but applies more generally to spatial correlations, or to other types of correlations: as long as two different uses of a channel are correlated, one may take advantage of the correlations by sending a quantum particle in a superposition of going through one use or the other. 

Time-correlated channels  are also interesting for foundational reasons. Recently, they have been proposed as a way to reproduce the use of quantum channels in a superposition of different causal orders \cite{oreshkov2019time,chiribella2019shannon2q}. In particular, they have been used to reproduce the action of the   quantum {\tt SWITCH}  \cite{chiribella2009beyond,chiribella2013quantum}, a higher-order operation that combines two variable quantum channels in a superposition of two alternative  orders.  In practice, time-correlated channels  underlie all the existing experimental setups inspired by the quantum   {\tt SWITCH} \cite{procopio2015experimental,rubino2017experimental,goswami2018indefinite,guo2020experimental,goswami2018communicating,goswami2020experiments,rubino2021experimental}.

The quantum {\tt SWITCH} is known to offer a number of advantages in quantum communication \cite{ebler2018enhanced,salek2018quantum,chiribella2018indefinite,goswami2018communicating,procopio2019communication,procopio2020sending,chiribella2020quantum}.   Here we show that   {\em (1)}  time correlations are essential in order to reproduce the advantages of the quantum {\tt SWITCH}, and {\em (2)}  the access to time-correlated channels is an even more powerful resource than the ability to combine ordinary quantum channels in a superposition of alternative orders.   

To make the above points, we consider  the scenario illustrated in Figure \ref{fig:freestyle_switch}, where a single particle is sent on a superposition of two paths, traversing  two independent channels,  each with the property that different uses of the same channel at different moments of time are correlated, while the action of the channel at any given time  is completely depolarising.   When the noise is perfectly correlated, the network in Figure \ref{fig:freestyle_switch} reproduces the quantum {\tt SWITCH}  of two completely depolarising channels, which is known to   achieve  a communication capacity of     0.049 \cite{ebler2018enhanced,chiribella2020quantum}.   In contrast, we show that in the  lack of time correlations the maximum capacity achieved by sending a particle on a superposition of paths is at most 0.024 bits.  This result proves that, in this scenario,  the physical origin of the  communication advantage of the quantum {\tt SWITCH}  is not merely the superposition of  paths, but rather the interplay between the superposition of paths and the time correlations in the noise.

Remarkably, we also find that the time correlations that reproduce the action of the quantum  {\tt SWITCH} are not the most favourable for the transmission of classical information: while the quantum  {\tt SWITCH} of two completely depolarising channels can at most yield  0.049 bits of classical communication \cite{ebler2018enhanced,chiribella2020quantum}, a more sophisticated pattern of time correlations yields the communication of  at least  0.31 bits.   The gap between these two values further highlights the power of time correlations, which are  not only capable of reproducing the  benefits of the superposition of causal orders, but also of surpassing them.  

The remainder of the paper is structured as follows.  In Section \ref{sec:framework} we describe the formalism of time-correlated channels and derive  the effective evolution experienced by a  single particle upon entering a time-correlated channel at a superposition of times. In Section \ref{sec:comms}, we consider the transmission of a single particle at a superposition of times, as in Figure \ref{fig:freestyle}, and  we demonstrate that the correlations between different uses of the channels offer a communication advantage over all communication scenarios where the channels are uncorrelated.    In Section \ref{sec:network}, we consider the network scenario of Figure \ref{fig:freestyle_switch}, and we show that time correlations are necessary to reproduce  the advantages of the quantum {\tt SWITCH},  and that certain time correlations can even offer higher advantages. Finally, we discuss the effects of noise on the control degree of freedom in Section \ref{sec:noisypath} and conclude in Section \ref{sec:concl}.

\section{Transmission of a single particle at a superposition of different times}\label{sec:framework} 

\subsection{Time-correlated   channels}

A transmission line that can be accessed at $k$  different times  is described by a correlated quantum channel
\cite{macchiavello02correlations,kretschmann2005quantum,caruso2014quantum}. 
Mathematically, the correlated channel is  a 
 linear map  transforming density matrices of the composite system    $S_1 \otimes \dots \otimes S_k$, where $ S_j$ denotes the system sent at the $j$-th time.  Note that, in general, the $k$ systems sent  at $k$ different times can be initially prepared in an arbitrary entangled state. 
 
  Correlated quantum channels are also known as quantum memory channels \cite{bowen04correlations,kretschmann2005quantum,caruso2014quantum}, quantum combs \cite{chiribella2008quantum,chiribella2009theoretical},  or non-Markovian  quantum processes \cite{breuer2016nonmarkov,pollock2018nonmarkov}.   In the following we will focus on the  $k=2$ case, corresponding to a transmission line that can be accessed at two different time steps, hereafter denoted by $t_1$ and $t_2$.  We consider    random unitary channels of the form
\begin{align}\label{correlatedRU}
\map R  (  \rho_{12})    =  \sum_{m,n}   p(m,n)   \,    (U_m \otimes U_n)  \,\rho_{12}\,  (U_m\otimes U_n)^\dag    \, ,
\end{align} 
where  $U_m$ and $U_n$ are unitary gates in a given set,  and  $p(m,n)$ is a joint probability distribution.  Here, the system sent at time $t_1$ experiences the unitary gate $U_m$, while the system sent at time $t_2$ experiences the gate $U_n$.   The density matrix $\rho_{12}$ represents the joint state of the two systems sent at the two times $t_1$ and $t_2$, that is, $\rho_{12}$ is a density matrix  on the Hilbert space of the composite system $S_1 \otimes S_2$.    The probability distribution $p(m,n)$  specifies the correlations between the random unitary evolutions experienced by system $S_1$ and system $S_2$. 

Note that, while in this paper we will focus on time correlations, the correlations in  Eq.\ (\ref{correlatedRU}) are not specific to time.  The same expression can be used also to describe correlated  channels acting on two  spatially separated systems,  or on any other type of independently addressable systems.  

Physically, a time-correlated random unitary channel of the form (\ref{correlatedRU}) can arise  in a photonic setup where the systems $S_1$ and $S_2$ are modes of the electromagnetic field associated to two different time bins \cite{humphreys2013linear,donohue2013coherent,li2015hyperentanglement,donohue2014ultrafast}. The noisy channel can correspond e.g.\ to the action of an optical fibre, where the random unitary changes of the photon polarisation arise from  random fluctuations in the birefringence.  Correlations between the unitaries at different times can arise when the time difference $t_2-t_1$  between successive uses of the channels is smaller than the timescale on which the birefringence fluctuates.  

\subsection{Sending a single particle through a time-correlated channel}
Consider now the situation where the input of the  correlated channel (\ref{correlatedRU}) is a single particle, carrying information in its internal degrees of freedom.   Classically, the particle must be sent either at time $t_1$, or at time $t_2$, or at some random mixture of $t_1$ and $t_2$.   When the particle is sent at time $t_1$,  its evolution is given by the reduced channel $\map R_1   (\rho)  :=  \sum_m  \, p_1 (m)  \,  U_m \rho  U_m^\dag $, where $p_1(m)   :  =  \sum_n  \, p(m,n)$ is the marginal probability distribution of the unitaries at time $t_1$.  Similarly, if the particle is sent at time $t_2$, its evolution is given by the channel $\map R_2   (\rho)  :=  \sum_m  \, p_2 (n)  \,  U_n \rho  U_n^\dag $, with $p_2(n)   :  =  \sum_m  \, p(m,n)$.   A random choice of transmission times then results into a random mixture of the evolutions corresponding to channels $\map R_1$ and $\map R_2$.   Crucially, the evolution of the particle is independent of  any  correlation that may be present in the probability distributions $p(m,n)$, that is, of  any correlation between the first and the second use of the transmission line.


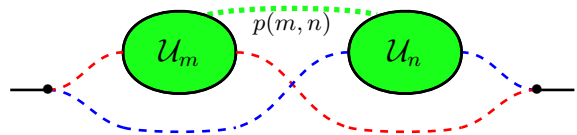
\begin{figure}
	\input{fig_freestyle_corsup.tikz}
	\caption{\label{fig:corsup}A single particle is sent at a superposition of two times (red and blue dashed lines), through the same transmission line (green ovals).  The green dotted line represents the correlations between random unitary processes $\map U_m$ and $\map U_n$ taking place with probability $p(m,n)$ at  the two subsequent uses of the transmission line, respectively. }
\end{figure}

In contrast, quantum mechanics allows one  to transmit a single particle in a way  that is sensitive to the correlations  between noisy processes at  different times.     The key idea is that the time when the particle is transmitted  can be   indefinite, as  the particle could be sent through the transmission line at  a coherent superposition of times $t_1$ and $t_2$ (see illustration in Figure \ref{fig:corsup}).  The superposition of transmission times could be achieved by adding an interferometric setup before the transmission line, letting  the particle travel on a coherent superposition of two paths,  one of which includes a delay \cite{ghafari2019interfering}. This results in a time-bin qubit, described by a superposition of amplitudes corresponding to localisation at two different points in time, separated by a time difference much greater than a photon's coherence time  \cite{marcikic2002time}.

Before developing the general theory of single particle transmission through time-correlated channels, it is instructive to look at a concrete example.   Consider the case of a single photon, and denote by $H_1$ and $V_1$  ($H_2$ and $V_2$) the horizontal and vertical polarisation modes in the first (second) time bin.  Here we take  the polarisation state to be  the same on both paths,  so that the only role of the interferometric setup is to coherently control the moment of transmission. 
The result is a linear combination  of states of the form  $  (\alpha  |1\>_{H1}  |0\>_{V1}  +  \beta  |0\>_{H1}  |1\>_{V1} )  \otimes |0\>_{H2}   |0\>_{V2}        $ and states of the form $  |0\>_{H1}   |0\>_{V1}            \otimes   (\alpha  |1\>_{H2}  |0\>_{V2}  +  \beta  |0\>_{H2}  |1\>_{V2} ) $.   The composite system of the two modes in the first  (second) time  bin can be regarded as system $S_1$  ($S_2$) in Eq.\ \eqref{correlatedRU}.      The states produced by the interferometric setup can then be written as  a linear combination of states of the form $|\psi\>_1 \otimes |{\rm vac}\>_2$ and states of the form 
$|{\rm vac}\>_1\otimes |\psi\>_2$,  where, for $i  \in \{1,2\}$,   $|{\rm vac}\>_i  : =   |0\>_{Hi}  |0\>_{Vi}$  is the vacuum state of the modes in system $S_i$, and $|\psi\>_i :  =  \alpha \, |1\> _{Hi}  |0\>_{Vi} + \beta\,  |0\>_{H_i} |1\>_{Vi}$ is a single-photon  polarisation state. The change in the particle's state upon the transmission is then computed by applying the channel   (\ref{correlatedRU}) to the appropriate state.

Generalising  the above  example, we  model the transmission of a single particle through channel (\ref{correlatedRU}) by interpreting systems $S_1$ and $S_2$ as {\em abstract modes}, each of which can  contain  a variable number of particles equipped with an internal degree of freedom, such as the photon's polarisation.  For $i\in  \{1,2\}$, the Hilbert space of system $S_i$ has  two orthogonal subspaces:  a one-particle subspace, denoted by $A^{(i)} $, and a vacuum subspace, denoted by  ${\rm Vac}^{(i)}$.  We assume that the dimension of the one-particle subspace is the same for both $S_1$ and $S_2$,  as in the example of the single-photon polarisation. Under this assumption, we have $A^{(1)} \simeq A^{(2)}  \simeq M$, where $M$ is the internal degree of freedom of the particle.   Also, we assume that each vacuum subspace is one-dimensional, and is spanned by a vacuum state $|{\rm vac}\>_i$, $i\in \{1,2\}$, as in our motivating example.

A single particle sent at a superposition of two moments of times will then be described by states of the form $\alpha\,  |\psi\>_1 \otimes |{\rm vac}\>_2  +  \beta\,  |{\rm vac}\>_1 \otimes |\psi\>_2$, where $|\psi\>  \in  M$  is the state of the particle's internal degree of freedom.   For the transmission of the particle, we will consider channels that conserve the number of particles, i.e.\ that map states of a given sector into states of the same sector.  This is the case, for example, for linear optical elements, which preserve the photon number.    For the channel  \eqref{correlatedRU},   preservation of the particle number means that the operators $U_m$ have the form   
\begin{align}\label{UVphi} 
U_m  =  V_m +     e^{  i \phi_m} \,  |{\rm vac}\>\<{\rm vac}| \,,
\end{align} where $V_m$ is a unitary acting in the one-particle sector  $M$, and $\phi_m  \in  [0,2\pi)$ is a phase.  Physically, $\phi_m$ corresponds to the phase difference between states in the one-particle sector and the vacuum state. 

In the quantum optical example,   each unitary ${ U}_m$   can be realised by a Hamiltonian acting on the two polarisation modes associated to system $S_i$,  $i\in  \{1,2\}$.  For example, the unitary  $ Z \oplus e^{i \phi} \ketbra{\rm vac}{\rm vac}$ can be generated by the   Hamiltonian $H = \hbar [ (\xi + \theta/2) a_{H}^\dag  a_{ H} + (\xi - \theta/2) a_{V}^\dag  a_{V} ]$, where $a_{H}$ ($a_V$) are the annihilation operators for the appropriate modes with horizontal (vertical) polarisation, in suitable units.

\subsection{Effective evolution  with a control system}

The representation of a single particle in terms of abstract modes is equivalent  to a representation in terms of  
 a composite system  $M C$, consisting of a message-carrying system $M$  and a control system $C$, which determines the particle's time of transmission.  The change of representation is described by the mapping  
 \begin{align}
 \nonumber   |\psi\>_1 \otimes |{\rm vac}\>_2   &\longmapsto   |\psi\>_M \otimes |0\>_C \\
   |{\rm vac}\>_{1}           \otimes    |\psi\>_{2}    &  \longmapsto      |\psi\>_M\otimes |1\>_C \, ,
 \end{align}
 where $|\psi\>$ is an arbitrary state in the one-particle subspace.     If the control is in state $\ket{0}$, then the message is sent through the first application of the channel,  with the vacuum in the second application; {\em vice versa} if the control is in state $\ket{1}$.   If the control  is in a generic state $\omega$, 
 the  overall evolution   is described by an {\em effective channel}  $\map C_{\omega}$, which transforms a generic state $\rho$ of the message into the state  
\begin{align}\label{eq:rand_uni_full}
\map C_{\omega}   (\rho)    :=  \sum_{m,n}    p(m,n) \,    W_{mn} \,  (\rho \otimes \omega) \,  W_{mn}^\dag \, , 
\end{align}
where $W_{mn}$ is the unitary $W_{mn}   :=   V_m  \,e^{i\phi_n}    \otimes |0\>\<0|   +   e^{i\phi_m} \,   V_n  \otimes |1\>\<1|$.  The derivation of  Eq.~(\ref{eq:rand_uni_full}) is provided in  Appendix \ref{app:vacext}.
%

 When the probability distribution $p(m,n)$ is symmetric (that is, when $p(m,n) =  p(n,m)$ for every $m$ and $n$),  the effective channel  has the simple expression  
\begin{equation}\label{eq:outputsingle}
\map C_{\omega}  (\rho)
= \frac{   \map C (\rho)  + \map G (\rho)   }2  \otimes \omega  
+  \frac{ \map C  (\rho)  -  \map G (\rho) }2 \otimes   Z\omega  Z \,  , 
 \end{equation}
with      \begin{align}\label{C}
\map C  (\rho)  :  =   \sum_{m,n}  \,  p(m,n)  \,   V_m \rho  V_m^\dag
\end{align}       and 
\begin{align}\label{G} 
\map G (\rho) :=  \sum_{m,n} \!  p(m,n)  \,
e^{i \,  (\phi_{n} -  \phi_{m} ) }  
V_m   \rho   V_{n}^\dagger \, .
\end{align} 
(See Appendix \ref{app:vacext} for the derivation.)
Here, the map   $\map C$ is the 	quantum channel  representing the evolution of the message when it is sent at a definite time (either $t_1$ or $t_2$). The channel $\map C$ depends only on the marginal probability distribution $p_1  (m):  =\sum_n \,  p(m,n)$,  and it is independent of the correlations.   Instead,   the map $ \map G$ can generally depend on the correlations between the evolution of the particle at two mutually exclusive moments of time. We call $ \map G$  the {\em interference  term}. 


\section{Classical communication through correlated white noise}  
\label{sec:comms}

\subsection{Correlated white noise}

Consider  the  case where 
the evolution  at any definite time step is completely depolarising  on the message-carrying sector $M$, that is, 
 \begin{align}\label{depolarising}
 \map C_{|j\>\<j|}  (\rho  )     =    \frac {I}d    \otimes |j\>\<j|  \qquad \forall \rho \, , \forall  j  \in  \{0,1\}\, ,
 \end{align}
 where $\map C_{|j\>\<j|}$ is the quantum   channel obtained by plugging $\omega  =  |j\>\<j|$ into Eq.\ \eqref{eq:rand_uni_full}.   Eq.\ \eqref{depolarising}  implies that, whenever the particle is sent at a definite moment of time, the message is replaced by white noise.   Accordingly, the channel $\map C$ in Eq.\ \eqref{C} is depolarising.

 When the probability distribution $p(m,n)$ is symmetric,  Eq.\  \eqref{eq:outputsingle} becomes  
\begin{equation}\label{eq:outputsingle1}
\map C_{\omega}  (\rho)
= \frac{   I/d + \map G (\rho)   }2  \otimes \omega  
+  \frac{  I/d -  \map G (\rho) }2 \otimes   Z\omega  Z \,  . 
 \end{equation}
In the realisation of the random unitary channel,   we will take the unitaries $\{  V_m\}$ to be an orthogonal basis  for the space of $d\times d$ matrices. Accordingly,  the set $\{  V_m\}$ will contain $d^2$ unitaries, labelled by integers from $0$ to $d^2-1$. For qubits, we will take   $\{ V_m\}$ to be the four Pauli matrices $\{  I,  X,  Y,  Z\}$, labelled as $V_0  =  I$,  $V_1  =  X$,  $V_2=  Y$, and $V_3 = Z$.  

In terms of the probability distribution $p(m,n)$, the  condition  \eqref{depolarising} amounts to requiring  that the marginal probability distributions $p_1 (m)$ and $p_2 (n)$ be uniform, that is 
\begin{align}\label{localuniform}
p_1  (m) =  p_2 (n)  = \frac  1{d^2}  \qquad  \forall m,n  \in  \{0,...,d^2-1\} \,.
\end{align}  
  The probability distributions $p(m,n)$  satisfying Eq.\ (\ref{localuniform}) form a convex polytope whose  extreme points  are probability distributions of the form $p(m,n)=\delta_{m, \sigma(n) }/d^2$, where $\sigma$ is a permutation of the set $\{0,\dots, d^2-1\}$  \cite{birkhoff1946three}.

  For the identity permutation, satisfying $\sigma  (m)  =  m$ for all values of $m$,  the probability distribution $p(m,n)$ is symmetric, and     the interference term   \eqref{G} is the completely depolarising channel  $\map G (\rho) = I/d~\forall \rho$. Hence, the channel  $\map C_\omega$ in   Eq.\ \eqref{eq:outputsingle1} is  completely depolarising, and   no information can be transmitted through it, no matter what state $\omega$ is used.    In the following, we will show that, instead, other types of permutations  enable a  perfect transmission of classical information.

\subsection{Perfect communication through  correlated completely depolarising channels}\label{subsec:perfect}

Here we focus on the case where the message is a qubit ($d=  2$).  Let  $\sigma$ be a permutation that swaps two pairs of indices, for example mapping $(0,1,2,3)$ into $(1,0,3,2)$. In this case, the probability distribution $p(m,n)  =  \delta_{m,  \sigma (n)}/4$ is symmetric, and  the interference term is 
\begin{align}\label{latentissimo}
 \map G  (\rho)  =   \frac  {
  \rho  X e^{i\left( \phi _{1} -   \phi _{0} \right)} +Y \rho  Z e^{i\left( \phi _{3} -   \phi _{2} \right)}   +  {\rm h.c.}}4\, ,   
\end{align}
where ${\rm h.c.}$  denotes the Hermitian conjugate of the preceding matrices.  

 
Note that $\map G (\rho)$ depends only on the differences $\phi_1-  \phi_0$ and $\phi_3-\phi_2$.  We now show that, by suitably choosing the differences $\phi_1-  \phi_0$ and $\phi_3-\phi_2$, and the state $\omega$, it is possible to achieve a perfect transmission of classical information.      When $\phi_1\!-\!\phi_0 = 0$ and $\phi_3\!-\!\phi_2= \pi/2$,    the interference term becomes 
\begin{align}
\map G   (\rho)=  \frac{ \{  \rho,   X  \}  -      \{ Z\rho  Z   ,   X\}}4 \, ,
\end{align}  
where $\{A,B\}  =  AB+  BA$ denotes the anticommutator of two generic operators $A$ and $B$.      In particular, choosing $\rho  =  |\pm\>\<\pm|$, with $|\pm \>:  =  (|0\>  \pm  |1\>)/\sqrt 2$, we obtain  
\begin{align}
\map G  (|\pm\>\<\pm|)  =   \pm  \, \frac   I  2  \, .   
\end{align}
Combining this  relation with the depolarising  condition  $  \map  C   (  |\pm \>\<\pm|)  = I/2$, and inserting these two relations into  into Eq.\ \eqref{eq:outputsingle}, we obtain 
\begin{align} 
\map C_{\omega}    ( \ketbra{\pm}{\pm})=   \frac  I 2  \otimes \omega_\pm \, , 
\end{align} 
with $\omega_+   : = \omega$ and $\omega_-  :  =  Z\omega Z$.    In other words, the net effect of the superposition of correlated depolarising channels is to transfer information from the message  to the output state of the control.   

Putting the control in the  state $\omega=|+\>\<+|$, one obtains the orthogonal  output states $\omega_{\pm} =  |\pm \>\<\pm|$. Hence, a sender can encode a bit into the states  $|\pm\>$, and a  receiver  will be able to decode the bit in principle without error, by measuring the control system in the basis $\{|+\>,|-\>\}$.  

In summary, there exist time-correlated channels that look completely depolarising when the message is sent at any definite moment of time, and yet allow for a perfect transmission of classical information by sending messages at a coherent superposition of different times.

\subsection{Maximum capacity in the lack of correlations}\label{sec:maxcapacity_nocor}
We now show that correlations in the probability distribution $p(m,n)$ are essential in order to achieve the  perfect  communication task discussed in the previous subsection.  Specifically,  we prove that no perfect communication is possible in the lack of correlations,  that is, when the probability distribution factorises as $p(m,n)  =  p_1 (m)  \,  p_2(n)  = 1/d^4$  (cf.\ Eq.\ \eqref{localuniform}).  
For qubit messages ($d=2$), we show that, in the lack of correlations, 
\begin{enumerate}
\item  the classical  capacity of the channel $\map C_{{\omega}}$ is upper bounded by $0.5$ bits,  meaning that it is impossible to transmit more than $0.5$ bits per use of the channel, 
\item   the maximum  classical capacity of the channel $\map C_{{\omega}}$  over arbitrary states $\omega$ of the control system and over arbitrary (not necessarily random-unitary) realisations of the completely depolarising channel  is equal to 0.16 bits. 
\end{enumerate}
The first result follows from an analytical upper bound on the classical capacity, while the second result follows from numerical optimisation.

\subsubsection{Analytical  bound on the classical capacity}

  The derivation of the bound consists of three steps, whose details are provided in  Appendix \ref{app:bound_nocor}.

 The first step  is to prove that, in the lack of correlations and for message dimension $d=2$,     the channel $\map C_{\omega}$ is entanglement-breaking \cite{horodecki2003entanglement},  i.e.\ it transforms all entangled states into  separable states.    For entanglement-breaking channels, it is known that the classical capacity coincides with the {\em Holevo capacity} \cite{shor2002additivity}.  For a generic quantum  channel $\map E$,  the Holevo capacity is $\chi (\map E)   =  \max_{\{ p_x,  \rho_x\}}   \,  H   \left[  \sum_x  p_x\,  \map E (\rho_x) \right]   -   \sum_x \,  p_x \,  \map E(\rho_x)$, where the maximum is over all possible ensembles $\{  p_x\, , \rho_x  \}$ consisting of a probability distribution $\{p_x\}$ and a set of density matrices $\{  \rho_x\}$, and  $H(\rho): =  -\Tr[\rho\log \rho]$ is the von Neumann entropy of a generic state $\rho$, $\log$ denoting  the logarithm in base 2.  

 The second step is to observe that state of the control that maximises the Holevo capacity of the channel $\map C_{\omega}$  is  $\omega   =  { |+\>\<+|} $.   This result holds for arbitrary message dimension $d\ge 2$, and, in fact, it holds even in the presence of correlations, as long as the probability distribution $p(m,n)$ is symmetric.

Finally, the third step is to show that, in the lack of correlations and for arbitrary message dimension $d\ge 2$,   the Holevo capacity of the channel $\map C_{|+\>\<+|}$   is upper bounded by $1/d$.  

Putting the three steps together, we obtain that, in the lack of correlations and for qubit messages,  the classical capacity of the channel $\map C_\omega$ is upper bounded by $1/2$ for every possible state $\omega$.  Hence, the perfect transmission of 1 bit achieved in Subsection \ref{subsec:perfect} is impossible in the lack of correlations. 

\subsubsection{Numerical evaluation of the capacity}

The evaluation of the Holevo capacity  involves  an optimisation over all possible input ensembles. For quantum channels with $d$-dimensional input, the optimisation can be restricted to ensembles with  up to $d^2$ linearly independent pure states \cite{davies1978information}. In practice, however, the optimisation   is often  hard to carry out even in dimension $d=2$. To make the optimisation feasible, we first show  that in our case the optimisation can be reduced to an optimisation over   ensembles that  depend only on three real parameters  $q, p_0, p_1 \in [0,1]$.  The proof of this result is provided in Appendix \ref{app:numerical_proofs}.

Building on the above results, we can numerically evaluate the  largest value of the Holevo  capacity,   and therefore the classical capacity, for all possible qubit channels (i.e.\ $d=2$) of the form \eqref{eq:rand_uni_full} with  $p(m,n)  =1/16$.  We set the state of the control to $\omega=  |+\>\<+|$, which we know to guarantee the maximum Holevo information (cf. Lemma \ref{lem:plusismore} in Appendix \ref{app:bound_nocor}).  

\begin{figure}
	\centering
		\subfloat[\label{fig:fs-chi}]{%
		\includegraphics[width=0.55\linewidth]{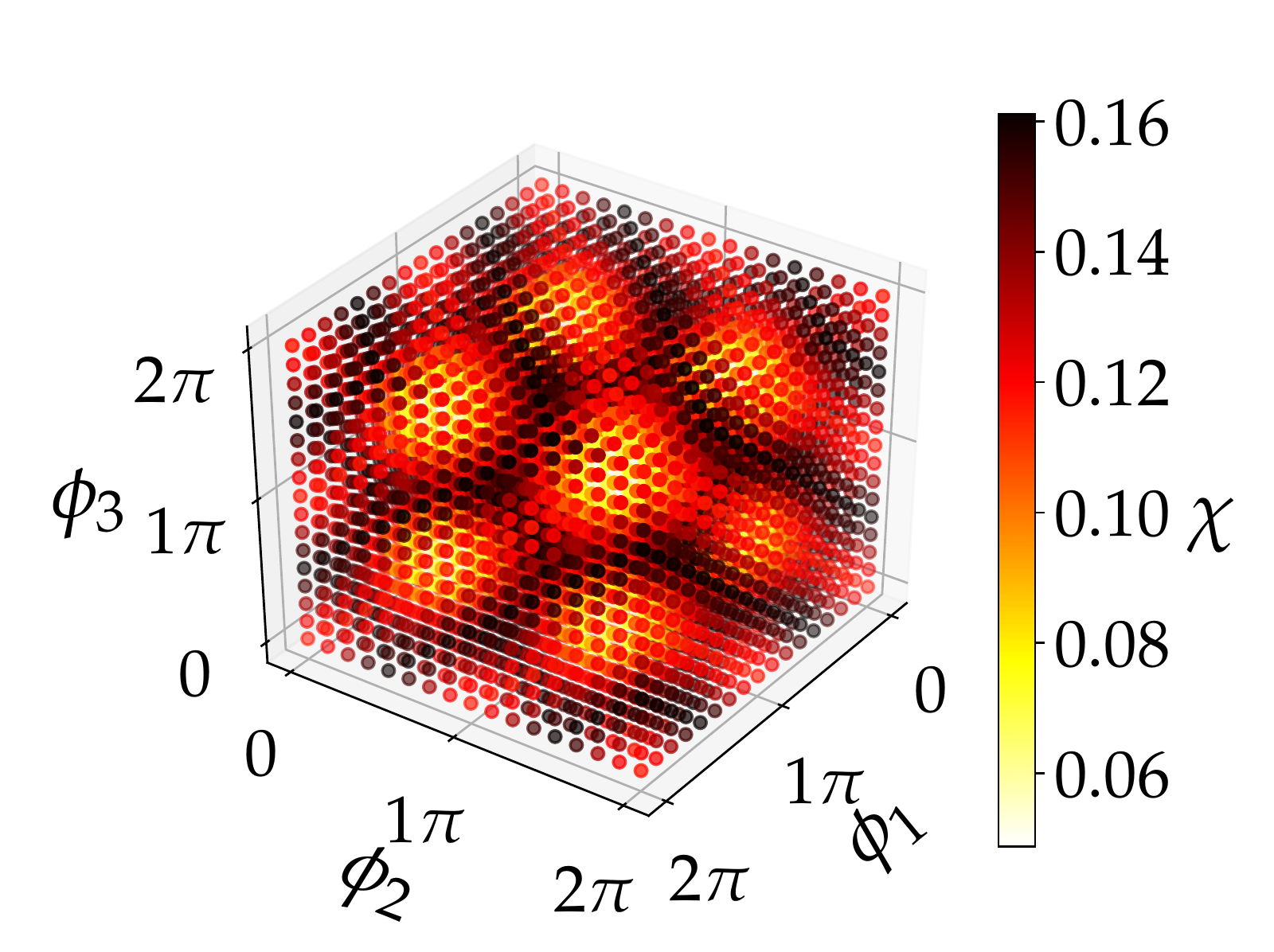}%
	}
	\subfloat[\label{fig:sup1032-chi}]{%
		\includegraphics[width=0.45\linewidth]{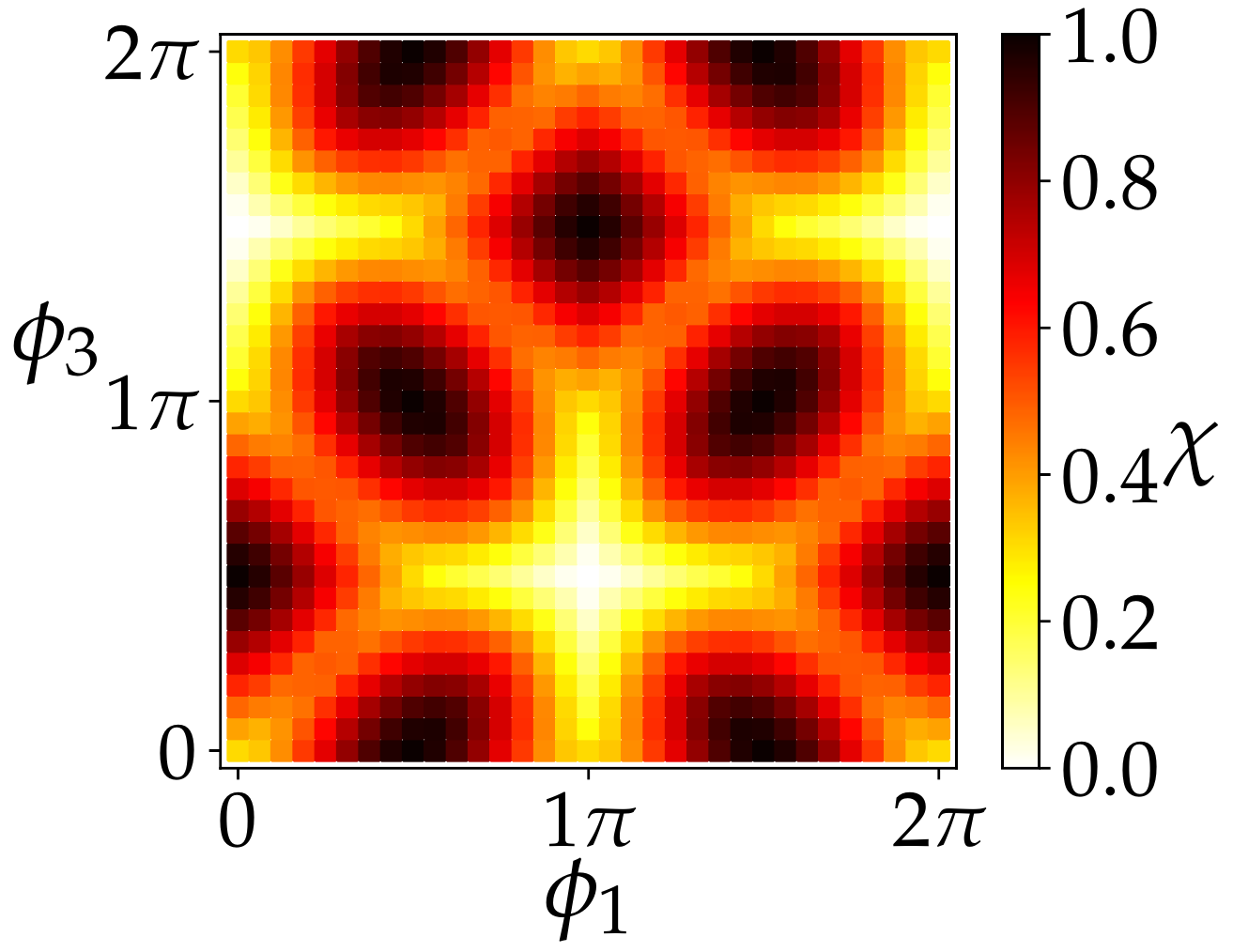}%
	}
	\caption{\label{fig:sup-chi} Performance in the transmission of a single particle through a correlated depolarising channel.  (a)  Classical capacity   in the lack of correlations.  Without loss of generality, $\phi_0=0$.  The maximum capacity is $0.016$ bits.   (b) Lower bound to the classical capacity  achieved with the correlated probability distribution $p(m,n) =  \delta_{n,\sigma (m)}/4$, where $\sigma$ is  the permutation that exchanges $0$ with $1$, and $2$ with $3$.    Without loss of generality, we set $\phi_0  = \phi_2=0$.   The maximum lower bound is 1 bit. }
\end{figure}

The resulting value of the Holevo capacity is a function of the phases $\{\phi_m\}_{m\in \{0,1,2,3\}}$ in Eq.\  \eqref{G}.  One phase, say $\phi_0$, can be set to $0$ without loss of generality, as it represents a global phase. 
In Figure \ref{fig:sup1032-chi}, we  provide a 3-dimensional plot showing the exact values of the Holevo information, and therefore by the arguments above, the classical capacity, for all possible values of the phases $\phi_1,  \phi_2,$ and $\phi_3$.
 The  maximum over all possible  choices of phases   is  $0.16$ bits. 
 
 In Appendix \ref{app:numerical_proofs} we also show that $0.16$ bits is the maximum capacity achievable with arbitrary    (not necessarily random unitary) channels that reduce to the depolarising channel in the  one-particle subspace  sector.    The value $0.16$ was previously found to be a lower bound to the classical capacity \  \cite{abbott2018communication}, and our result shows that the lower bound is actually tight:  0.16 is the best classical capacity one can obtain by sending a single particle through a superposition of paths traversing two identical, independent channels that are completely depolarising in the one-particle subspace.


\subsection{Lower bound to the classical capacity in the presence of  correlations}

In the correlated case, we do not have a proof that the classical capacity coincides with the Holevo capacity.   On top of that, the evaluation of the Holevo capacity  generally requires an optimisation over all possible ensembles of $d^2$ linearly independent pure states, which is computationally challenging. Here,  we  circumvent this problem by computing a lower bound to the Holevo capacity, obtained by restricting the optimisation to the set of all {\em orthogonal} ensembles, that is, input ensembles consisting of two orthogonal qubit states. In general,  this lower bound may not be tight   \cite{fuchs1997nonorthogonal,king2002qubit,hayashi2004qubit}, but it is nevertheless interesting as it quantifies the maximum performance of a natural set of encoding strategies. Since the Holevo capacity is always a lower bound to the classical capacity,  the above lower bound is also a lower bound to the classical capacity.

Here, we evaluate the  lower bound for the correlated channel with  $p(m,n)  =    \delta_{n,  \sigma  ( m )}/4 $, where $\sigma$  is the permutation that exchanges $0$ with $1$, and $2$ with $3$.   This particular choice is interesting because as  we have seen in Subsection \ref{subsec:perfect}, it can  reach the maximum capacity of 1 bit. 
We now inspect how the lower bound depends on the phases.  

  Since the interference term \eqref{latentissimo} depends only on the differences   $\phi_1-  \phi_0$ and $\phi_3-\phi_2$, we set $\phi_0 =  \phi_2=  0$ and scan the possible values of $\phi_1$ and $\phi_3$.    For the state of the control system,  we choose again $\omega=  |+\>\<+|$, as it maximises the Holevo capacity (cf.\ Lemma \ref{lem:plusismore} in Appendix \ref{app:bound_nocor}).     The lower bound to the  Holevo capacity     is shown in Figure \ref{fig:sup1032-chi} for all values of $\phi_1$ and $\phi_3$.

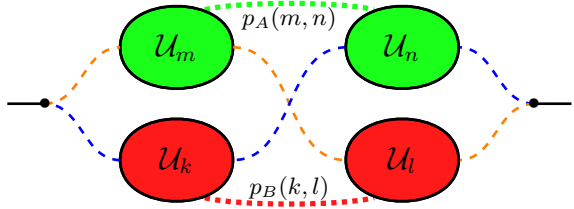
\begin{figure}
	\input{fig_freestyle_doublecorsup.tikz}
	\caption{\label{fig:doublecorsup} A single particle is sent through a superposition of two paths (orange and blue dashed lines), each traversing two independent channels (green and red ovals), each of which exhibits time correlations between successive uses.   The green and red dotted lines represent the correlations between the two subsequent uses of the same channel.}
\end{figure}

\section{Communication through multiple  time-correlated channels}\label{sec:network}

Time-correlated channels  can be used to mimic the use of ordinary quantum channels in a superposition of different causal orders \cite{oreshkov2019time,chiribella2019shannon2q}. 
  In this section we show that time correlations are a necessary resource for reproducing the benefits of the superposition of orders in quantum communication, and that, in fact, time correlations are an even more powerful resource than the ability to combine channels in a superposition of orders.

\subsection{A network of time-correlated channels} 

Suppose that two time-correlated channels $\map R_A$ and $\map R_B$, each of the form (\ref{correlatedRU}), are arranged as in  Figure \ref{fig:doublecorsup}, and that a single particle is sent through a superposition of two alternative paths visiting each of the two channels exactly  once.   When the control system is initialised in the state $\omega$,  the overall evolution of the message and the control is described by the {\em effective channel}  $\map E_\omega$ defined as
\begin{align}\label{eq:sup2pairs}
\map E_\omega  (\rho)   := \sum_{m,n,k,l}  p_A(m,n)  p_B(k,l)  \,  W_{mnkl}  (\rho \otimes \omega)  W_{mnkl}^\dag  \, ,
\end{align}   
with 
\begin{align}
\nonumber 
W_{mnkl}   :=   &V^{(B)}_{l}  V^{(A)}_{m}    \,e^{i  (\phi^{(B)}_{k}+ \phi^{(A)}_{n}) }   \otimes |0\>\<0| \\
  &  +    V^{(A)}_{n}  V^{(B)}_{k}  \,e^{i  (\phi^{(A)}_{m} + \phi^{(B)}_{l})}    \otimes |1\>\<1| \, .  \label{Wmnkl}
\end{align}
Here,  $p_A(m,n), p_B(k,l), \{  V^{(A)}_m\}, \{  V^{(B)}_l\}, \{\phi^{(A)}_m\}$, and $\{\phi^{(B)}_m\}$  are defined as in Equations (\ref{correlatedRU}) and (\ref{UVphi}).  The derivation of Eq.\  \eqref{eq:sup2pairs} is provided in Appendix  \ref{app:switch}.

An interesting special case occurs when  the probability distributions  $p_A (m,n)$ and $p_B (k,l)$ are  perfectly correlated, that is 
\begin{equation}\label{perfectcor} 
\begin{split}
  p_A(m,n)   &= p_{1A}  (m)  \delta_{mn} \quad  \forall m,n  \\
p_B(k,l)   &= p_{1B}  (k) \,  \delta_{kl} \quad~~ \forall k,l \, ,
\end{split}
\end{equation}
 where $p_{1A} (m)$ and $p_{1B} (k)$ are the marginal probability distributions of $p_A  (m,n)$ and $p_{B}  (k,l)$, respectively.  Under this condition,     the network in Figure \ref{fig:doublecorsup} reproduces the action of two random unitary channels  in a superposition of two alternative orders \cite{chiribella2019shannon2q}. 
 
 Mathematically, the operation of putting two quantum channels in a superposition of orders is described by the    quantum {\tt  SWITCH}  \cite{chiribella2009beyond,chiribella2013quantum}, a higher-order transformation that takes as inputs  two generic  channels $\map A$ and $\map B$ (with $d$-dimensional input and ouput systems)  and produces as output a new quantum channel $\map S  (\map A,\map B)$ with Kraus operators 
\begin{align}\label{krausswitch}
S_{mk}  :  =   A_m B_k \otimes |0\>\<0|  +  B_k A_m  \otimes |1\>\<1|    \, ,
\end{align} 
where $\{A_m\}$  ($\{ B_k\}$)  are  Kraus operators of $\map A$ ($\map B$), and $\{|0\>, |1\>\}$ is a basis for a control qubit that determines the relative order between $\map A$ and $\map B$.   Notably, the overall channel  $\map S  (\map A, \map B)$ is independent of the choice of Kraus representations for the input channels $\map A$ and $\map B$. 

 When  the control qubit is put in a fixed state $\omega$, the quantum   {\tt  SWITCH} of channels $\map A$ and $\map B$ yields the effective channel    
\begin{align}\label{effectiveswitch}
\map S_\omega  (\rho)  & :=  \sum_{m,k}   S_{mk}   (\rho\otimes \omega)  S_{mk}^\dag   \,,  
\end{align}   
with $S_{mk}$ as in Eq.\ \eqref{krausswitch}.  
In particular,  here we are interested in the case where  the channels $\map A$ and $\map B$ are random unitary, with  Kraus operators $A_m  := \sqrt{  p_{1A}  (m)}  \, V^{(A)}_m$  and $B_k:= \sqrt{  p_{1B}}  \,  V^{(B)}_k$.  With this choice,  the channel $\map S_\omega$  in Eq.\ \eqref{effectiveswitch} coincides with the channel $\map E_\omega$ in Eq.\  \eqref{eq:sup2pairs} under the condition that the probability distributions $p_A  (m,n)$ and $p_B (k,l)$ are perfectly correlated (cf.\ Eq.\  \eqref{perfectcor}).  

When the channels $\map A$ and $\map B$ are completely depolarising,  Ref.\ \cite{ebler2018enhanced} showed that the  channel $\map S_\omega$ resulting from  the quantum {\tt  SWITCH}  can transmit $0.049$ bits of classical information, provided that the control is initialised   in the state $\omega = |+\>\<+|$.  
Later,  the value $0.049$ was proven to be exactly equal to the classical capacity \cite{chiribella2020quantum}.    Since the channels $\map E_\omega$ and  $\map S_\omega$ coincide,  we conclude that the time-correlated network in Figure \ref{fig:doublecorsup} can achieve a capacity of $0.049$ bits.

 In the following, we provide two new results: 
\begin{enumerate}
\item We show that time correlations are strictly necessary in order to achieve the quantum {\tt SWITCH}  capacity of 0.049 bits. Specifically, we show numerically that the maximum classical capacity in the uncorrelated case is 0.018 bits for random-unitary realisations of the completely depolarising channel, and  0.024 bits for arbitrary realisations.  This result shows that, when the quantum {\tt SWITCH} is  reproduced by the network in Figure \ref{fig:doublecorsup}, the  origin of the communication enhancement is not just the interference of paths, but rather   the combined effect of  the interference of paths {\em and} of the time correlations. 
\item We show that 
there exist time correlations that achieve a classical capacity of at least $0.31$ bits. This result shows that the access to time correlations  is generally a stronger resource 
than the ability to combine ordinary channels in a superposition of orders.  
\end{enumerate}

\begin{figure}
	\centering
		\subfloat[\label{fig:fsqs-chi}]{%
		\includegraphics[width=0.55\linewidth]{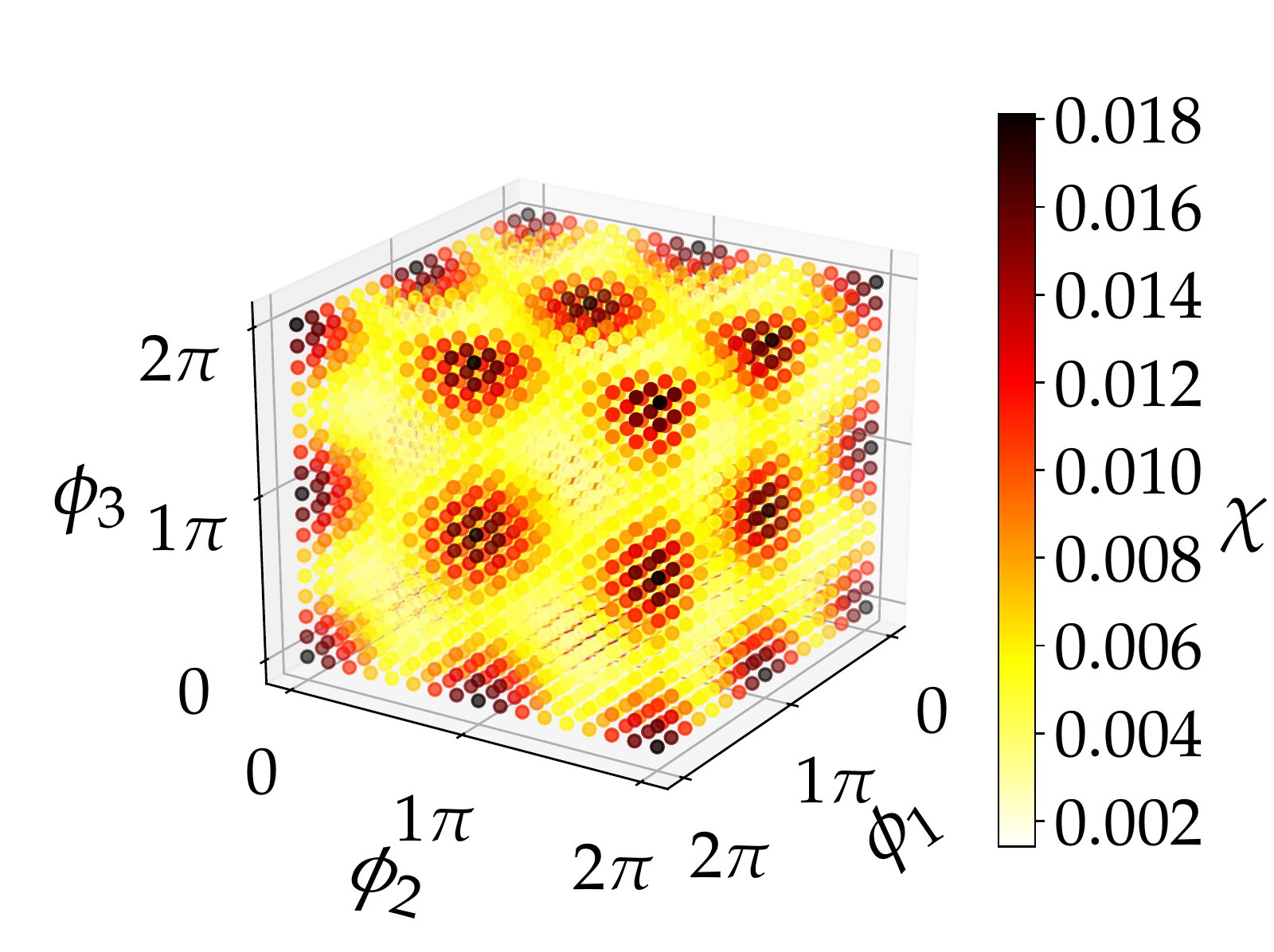}%
	}
	\subfloat[\label{fig:sup10321032-chi}]{%
		\includegraphics[width=0.45\linewidth]{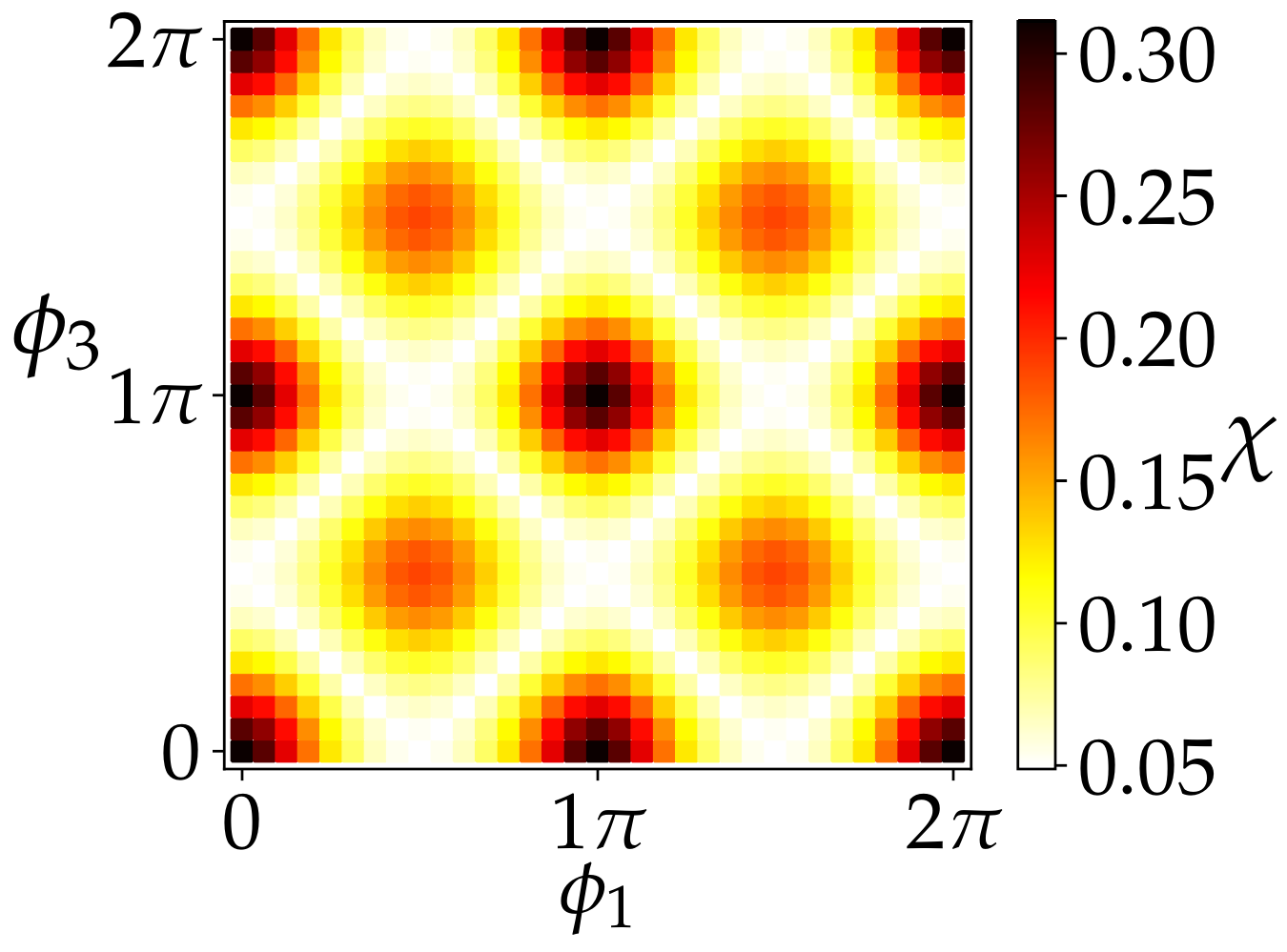}
	}
	\caption{\label{fig:sup-net-chi} Performance in the transmission of a single particle through a network of correlated depolarising channels, arranged as in Figure \ref{fig:doublecorsup}. (a) Classical capacity in the lack of correlations.  Without loss of generality, $\phi_0=0$.   The maximum capacity is $0.018$ bits.  (b) Lower bound to the classical capacity achieved with maximal correlations corresponding to the probability distributions $p_A  (m,n)  =  p_B (m,n)   =  \delta_{n,\sigma (m)}/4$,  where $\sigma $ is the permutation that exchanges 0 with 1, and 2 with 3.   Without loss of generality, $\phi_0  =  \phi_2=0$. The maximum lower bound is $0.31$ bits.  }	 
\end{figure}

\subsection{Maximum capacity in the lack of correlations}\label{subsec:maxcapacitydouble}
Here we evaluate the maximum  amount of classical information that can be transmitted through the network in Figure \ref{fig:doublecorsup} when the channels are completely depolarising and no  correlation is present, that is, when $p_A(m,n) =p_B(k,l)= 1/16 ~ \forall m,n , k,l \in \{0,1,2,3\}$.

The evaluation of the maximum  capacity follows the same steps as  in Subsection \ref{sec:maxcapacity_nocor}. The main  observations are: 
\begin{enumerate}
\item in the lack of correlations, the channel $\map E_\omega$ in Eq. \eqref{eq:sup2pairs}  is entanglement-breaking, and therefore its classical capacity coincides with the Holevo capacity
\item the control state $\omega$ that maximises the Holevo capacity of the channel $\map E_\omega$ is $\omega  =  |+\>\<+|$
\item without loss of generality, the maximisation of the Holevo information can be reduced to ensembles that depend only on three real paramters  $q, p_0$, and  $p_1$   in  $[0,1]$. 
\end{enumerate} 
 The derivation of these results is provided in Appendix \ref{app:double}. 
 
Building on  the above observations, we evaluate the  capacity of the channel  $\map E_\omega$ in Eq.\ \eqref{eq:sup2pairs} by scanning all possible values of the phases $\{  \phi_m\}_{m=0}^3$.   The result is  the plot shown in    Figure \ref{fig:fsqs-chi}.  The largest classical capacity over all random unitary realisations is  0.018 bits, which is strictly smaller than the value $0.049$ bits achieved by the superposition of orders.

Furthermore, we also extend the optimisation from random unitary realisations to arbitrary realisations of the completely depolarising channel. For this broader class of realisations, we numerically obtain   that the maximum capacity  is $0.024$ bits.

Summarising,  the best classical capacity one can obtain by sending a single particle through the network in Figure  \ref{fig:doublecorsup},  in the lack  of   correlations between the two paths, is $0.018$ bits, and the capacity can be increased to $0.024$ bits by replacing the random unitary channels with more general realisations of the completely depolarising channel. 


Note that both values $0.018$ and $0.024$  are below the 0.049 bits of classical capacity achieved by the quantum {\tt SWITCH}.    This result shows that,  when the  quantum {\tt SWITCH} is reproduced by the correlated network in Figure \ref{fig:doublecorsup}, it offers a communication advantage over all communication protocols where a single particle travels in a superposition of two paths on which it experiences uncorrelated  noisy processes. Hence, we conclude that, in this scenario,  the origin of the communication advantages of the quantum {\tt SWITCH} is not merely the superposition of paths, but rather the non-trivial interplay between the superposition of paths and the time correlations in the noise.     

Our results also imply a {\em caveat}  about  terminology.   The quantum  {\tt SWITCH} of two channels $\map A$ and $\map B$  is sometimes described informally as a {\em ``superposition of channels $\map A  \map B$ and $\map B \map A$."}  While this  expression  may be formally correct (at least according to a broad notion of superposition \cite{chiribella2019shannon2q}),   
 it can be misleading if taken at face value,  because it does not  mention explicitly the requirement of correlations between the channels $\map A$ and $\map B$ in the two branches of the superposition.  

\subsection{Time correlations surpassing the quantum {\tt SWITCH} capacity}  
We now show that the classical capacity of $0.049$ bits, achieved by the quantum {\tt SWITCH}, can be surpassed using  more general  time correlations.   We prove this result explicitly, by exhibiting a pair of time-correlated channels that achieve a capacity at least $0.31$ bits.

Our choice of channels corresponds to   $p_A(m,n)=p_B(m,n) =  \delta_{n,\sigma (m)}/4$, where $\sigma$ is the permutation that exchanges $0$ with 1, and 2 with 3.  This choice is motivated by the fact that  the  permutation $\sigma$ guarantees the maximum communication capacity in the case where a single time-correlated channel is used (cf.\ Subsection \ref{subsec:perfect}).

 With the above choice,    the effective channel   describing the transmission of the message is
\begin{equation}\label{eq:outputdouble}
\map E_{\omega}  (\rho)
= \frac{ \frac I 2  + \map K (\rho)   }2  \otimes \omega  
+  \frac{  \frac I 2  -  \map K (\rho) }2 \otimes Z\omega Z \,  , 
\end{equation}
with
\begin{equation}\label{eq:g2_1032}
\map K (\rho)
:= \frac{1}{8} \! \left\{ \! \left[  \cos 2(\phi_1-\phi_0 ) + \cos 2(\phi_3-\phi_2) \right] \rho + \! 2 X \rho X \! \right\} .
\end{equation}
The derivation of this formula is provided in Appendix \ref{app:switch}.   Note that the channel $\map E_\omega$ depends only on the phase differences $\phi_1\!-\!\phi_0$ and  $\phi_3\!-\!\phi_2$, via Eq.\ \eqref{eq:g2_1032}.

We now provide a lower bound to the classical capacity of the channel  $\map E_\omega$.  As we did earlier in the paper,  we lower bound the classical capacity by the Holevo capacity, and, in turn, we lower bound the Holevo capacity by restricting the maximisation to orthogonal input ensembles.  For the state of the control qubit,  we pick  $\omega= |+\>\<+|$, which is the  choice  that maximises the Holevo capacity (cf.\ Lemma \ref{lem:plusismore} in Appendix \ref{app:bound_nocor}). 

The lower bound to the classical capacity is shown  in Figure \ref{fig:sup10321032-chi} for all possible values of the phase differences $\phi_1\!-\!\phi_0$ and  $\phi_3\!-\!\phi_2$. The highest lower bound  over all combinations of phases $\{\phi_{m}\}_{m=0}^3$ is given by $0.31$ bits.   
This value is larger than the classical capacity  of 0.049 bits achieved by the quantum {\tt SWITCH}, corresponding to perfect correlations $p_A(m,n) = p_B(m,n)  =  \delta_{m,n}/4$.  This result implies that  not only can  time correlations reproduce the superposition of causal orders, but  they can also surpass its advantages.

\section{Noise on the control degree of freedom}\label{sec:noisypath}

So far we have assumed that the message-carrying  degree of freedom of the particle undergoes noise during transmission, while the control degree of freedom is noiseless. However, in practical scenarios, this will only be an approximation to the actual physics.   We now briefly discuss the effect of noise on the control system,  focussing in particular on dephasing noise, of the form  	
\begin{equation}
	\map P (\omega) = s Z \omega Z + (1-s) \omega \, ,
	\end{equation}
where	 $s \in ~ [0, 1/2]$ is a probability and $\omega$ is the initial state of the control. 
For a more detailed investigation into the effects of noise on the control system, we refer the reader to a recent  related work \cite{kristjansson2021network}.

\begin{figure}
	\vspace{1ex}
	\includegraphics[width=0.8\linewidth]{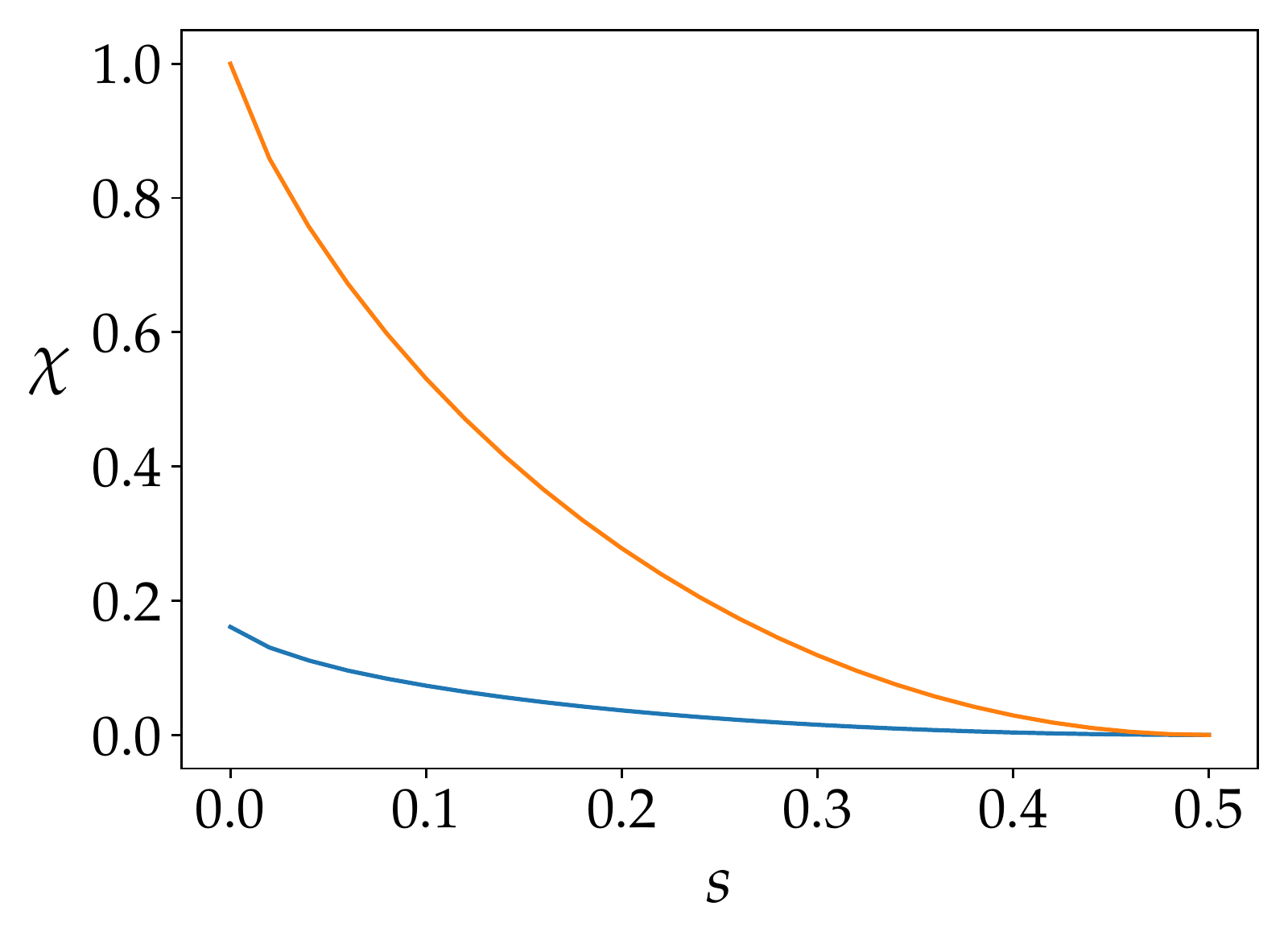}
	\caption{\label{fig:s-chi}  \textit{Blue}: Maximum classical capacity in the absence of correlations, as a function of the  dephasing parameter $s$.  The maximum is computed over all  realisations of the completely depolarising channel, and is achieved  by the random unitary realisation with the choice of phases $\{\phi_m\}$ that give the maximum capacity of 0.16 bits when $s=0$.   \textit{Orange}: Lower bound to the maximal classical capacity in the presence of correlations, as a function of  the dephasing parameter $s$.   The lower bound  is computed by considering   the correlated probability distribution $p(m,n) =  \delta_{n,\sigma (m)}/4$, where $\sigma$ is  the permutation that exchanges $0$ with $1$, and $2$ with $3$, and $\phi_0=\phi_1=\phi_2=0, \phi_3=\pi/2$. 
	}  
\end{figure}

For simplicity, here we focus on the communication scenario involving  a single transmission line, as in Figure \ref{fig:freestyle}.    In this setting,  the evolution experienced by a single particle is described by the channel 
\begin{align}
\map C_\omega'   :  =  (\map I_M\otimes \map P)  \map C_\omega \,,
\end{align}
 obtained by dephasing the control system at the output of the  channel $\map C_\omega$ in Eq.\  \eqref{eq:outputsingle}.   By inserting the expression \eqref{eq:outputsingle} into the above equation, it is immediate to see that the effect of  dephasing is to dampen the  interference term  $\map G$  in the effective channel \eqref{eq:outputsingle}:   specifically,  the interference term changes  from $\map G$ to $ (1-2s) \, \map G$.    
 
 In the case of completely depolarising channels on the message degree of freedom, the presence of a non-zero interference term  means that, as long as the dephasing of the control is not complete  ($s\not  =  1/2$), the superposition  of evolutions can still allow   for a non-zero amount of classical information to be transmitted, thereby offering an advantage over the transmission at a definite moment of time.    
 
 Figure \ref{fig:s-chi} shows the behaviour of the classical capacity  as a function of  the dephasing parameter $s$.      The figure shows that  correlations between two uses of the channel offer an enhancement of the classical capacity. To make this point,  we first evaluate  numerically  the maximum capacity achievable  in the lack of correlations, with arbitrary realisations of the completely depolarising channel (blue curve). Notably, the capacity for every fixed value of  $s$ is   achieved by the same realisation of the completely depolarising channel that achieves the maximum capacity in the ideal $s=0$ case.      We then show that a higher capacity can be achieved with the  correlated channel  described in Subsection \ref{subsec:perfect}.  To this purpose, we numerically evaluate a lower bound to the Holevo capacity (and therefore to classical capacity), obtained by  restricting the maximisation to orthogonal input ensembles  (orange curve).      Note that both the blue and orange curves are above 0 for every non-maximal amount of dephasing ($s\not  =  1/2$), meaning that the single particle transmission at a coherent superposition of times offers an advantage over the transmission at a definite time.

\section{Conclusions}\label{sec:concl}
We have shown that a single quantum particle can sense the correlations between noisy processes at different moments of time. By sending the particle at a superposition of different times, one can take advantage of these correlations and  boost the communication rate to values that would be impossible if the moment of transmission were a classical, well-defined variable.

An important avenue for future research is the experimental realisation of our protocols, as well as the experimental exploration of their noise robustness to timing errors and decoherence between the two different modes used to create the superposition.  On the theoretical side, it is interesting to apply our framework for  single-particle communication to more complex scenarios, e.g.\ involving the transmission  of a single particle at more than two times, or even in continuous time. It is also interesting to analyse other communication tasks,  such as the   two-way communication   proposed in Ref.\ \cite{del2018two}.   Moreover, the extension from single particle communication to other communication protocols with a finite number of particles is a natural next step of this research.

At the foundational level, time-correlated channels provide an insight into the resources used by the  existing experiments on the superposition of causal order.      We analysed a basic setup that reproduces  the overall result of  the quantum {\tt SWITCH}   by sending a single particle in a superposition of paths through time-correlated channels.  In this setup, we showed that time-correlations are a necessary resource to reproduce the communication advantages of the quantum {\tt SWITCH}.  Moreover, we observed that, with more elaborate patterns of correlations, one can achieve an even greater enhancement than the one found for the superposition of orders.   This result establishes time-correlated channels as an appealing resource, which can be used as a testbed for foundational results on causal order,  and, at the same time, as a building block for new  communication protocols.

\section*{Acknowledgements}
We acknowledge discussions with Robert Spekkens, Sandu Popescu, Paul Skrzypczyk,  Debbie Leung, Aephraim Steinberg,  Philippe Grangier, Philip Walther, Giulia Rubino, Caslav Brukner,  David Schmid,  Chiara Macchiavello, Massimiliano F.\ Sacchi, Santiago Sempere Llagostera, Robert Gardner, Kwok Ho Wan, Raj Patel, Ian Walmsley, Giulio Amato, Kavan Modi, Guillaume Boisseau, Alastair Abbott, Marco T\'ulio Quintino, Fabio Costa, Daniel Ebler,   Sina Salek, and Carlo Sparaciari. The numerical simulations presented in this paper were written using the Python software package QuTiP and the circuit diagrams were drawn using TikZiT. This work is supported by  the National Natural Science Foundation of China through grant 11675136,  the Hong Research Grant Council through grant 17307719, the Croucher Foundation,  the HKU Seed Funding for Basic Research,  the UK Engineering and Physical Sciences Research Council (EPSRC) through grant EP/R513295/1, and the Perimeter Institute for Theoretical Physics.  Research at the Perimeter Institute is supported by the Government of Canada through the Department of Innovation, Science and Economic Development Canada and by the Province of Ontario through the Ministry of Research, Innovation and Science. This publication was made possible through the support of the grants  60609 `Quantum Causal Structures' and 61466 `The Quantum Information Structure of Spacetime (QISS)' (qiss.fr) from the John Templeton Foundation. The opinions expressed in this publication are those of the authors and do not necessarily reflect the views of the John Templeton Foundation.


%


\setcounter{secnumdepth}{2}

\begin{appendix}

	\section{Transmission of a single particle through a superposition of multiple ports}\label{app:vacext}

Here we provide a mathematical framework for describing the transmission of a single particle at a superposition of different times, and, more generally, for describing the transmission of the particle on a superposition of different trajectories, each passing through one of the ports of a multiport quantum device.

\subsection{Multiport quantum devices and their vacuum extensions}

A transmission line with a single input port   is described by a  quantum channel, that is, a completely positive trace-preserving map  transforming density matrices on the particle's Hilbert space.  In the following we will denote by $\Chan  (S \to S')$ the set of quantum channels with input system $S$ and (possibly different) output system $S'$.   When $S=S'$ we  will use the shorthand $\Chan(S)$. The action of a quantum channel $\map A$ on a density matrix $\rho$   can be conveniently  written in the Kraus representation $\map A\left( \rho \right) =\sum_{i} A_{i} \rho A_{i}^{\dagger}$, where $\left\{A_{i}\right\}$ is a (non-unique) set of  operators, satisfying $ \sum_{i} A_{i}^{\dagger} A_{i} = I$.   
	
A transmission line with $k$ input/output ports is described by a $k$-partite quantum channel  $\map B \in  \Chan\left(S^{(1)} \otimes \cdots  \otimes  S^{(k)}    \to S'^{(1)}  \otimes \cdots\otimes S'^{(k)}\right)$ with $k$ input-output pairs $(S^{(i)},S'^{(i)})_{i=1}^{k}$.  

	\begin{figure}
		\input{fig_cor.tikz}
		\caption{\label{fig:cor}  The left-hand side depicts a 2-step correlated quantum channel $\map B$ taking two input states on systems $S^{(1)}$ and $S^{(2)}$, in succession. 
			The right-hand side shows the physical implementation of the 2-step channel via two unitary channels $\map W_{1}$ and $\map W_{2}$ \cite{chiribella2008quantum,chiribella2009theoretical} where the memory between the two uses of the channel is realised by  an environment $E$, which is inaccessible to the communicating parties.}  
	\end{figure}
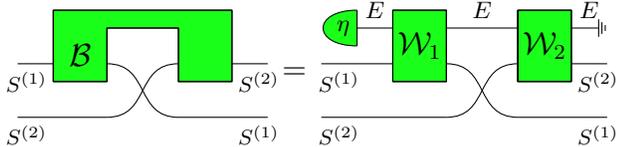

A transmission line that can be used $k$ times in succession is described by \textit{$k$-step quantum channel} \cite{macchiavello02correlations} (also known as a quantum $k$-comb \cite{chiribella2008quantum,chiribella2009theoretical}).   A $k$-step quantum channel is a special type of $k$-partite channel $\map B$ with  the additional  property that  no signal propagates from an input $S^{(i)}$ to any group of  outputs $S'^{(j)}$ with $j<i$  \cite{chiribella2008quantum}.  We will denote the set of $k$-step quantum channels as $\Chan(S^{(1)}  \to S'^{(1)},     \dots  ,   S^{(k)}  \to S'^{(k)})$, or simply  $\Chan(S^{(1)},      \dots  ,   S^{(k)})$ when the input and output of each pair coincide.      For $k=2$, an example of  2-step  quantum channel is illustrated in Figure \ref{fig:cor}.

	The possibility that no particle is sent through a port of a device  can be described using the notion of \textit{vacuum extension} \cite{chiribella2019shannon2q}. Consider first a single-port device,  described by an ordinary quantum channel $\map A \in \Chan(S)$. When no particle is sent through the device, we describe the input as  the vacuum state $\ket{\textrm{vac}}$, that is,  a state in a {\em vacuum sector}  $\textrm{Vac}$ \cite{aaberg2004operations,zhou2011adding,
		chiribella2019shannon2q,dong2019controlled}, which is orthogonal to the one-particle sector $S$. Overall, the device acts on an extended system $\widetilde{S} := S \oplus \textrm{Vac}$, which is associated with the Hilbert space given by $\spc{H}_S \oplus \spc{H}_{\rm Vac}$, where $\spc{H}_{\rm Vac}$ is the  vacuum Hilbert space, here assumed to be one-dimensional.  
	
	Given a quantum  channel $\map{A}$, a vacuum extension $\widetilde {\map{A}}$ of $\map{A}$ is any channel which acts as $\map{A}$ (respectively, $\map I_\textrm{Vac}$) when the input is a state in sector $S$ (respectively, $\textrm{Vac}$).  The Kraus operators of  $\widetilde {\map{A}}$ are $\widetilde A_i  =  A_i  \oplus  \alpha_i \,  |\rm vac \>\<\rm vac  |  $,
	where    $\{A_i\}_{i=0}^{r-1}$ is a Kraus representation of $\map A$, and $\{\alpha_i\}_{i=0}^{r-1}$ are \textit{vacuum amplitudes} satisfying
	$\sum_{i=0}^{r-1} \, |   \alpha_i|^2   =  1$.

	A given channel has infinitely many possible vacuum extensions. In an actual communication scenario, the vacuum extension can be determined by probing the action of the channel on superpositions of the vacuum and one-particle states. Physically, the choice of vacuum extension is determined by the Hamiltonian of the field describing the vacuum and the one-particle sector.

The notion of vacuum extension can be easily extended  to the case of $k$-partite channels, which include  $k$-step channels as a special case.   For simplicity, we focus on the $k=2$ case, but the extension to $k\ge 2$ is straightforward.

 Consider a transmission line described by a bipartite  channel $\map B \in \Chan(S^{(1)}\otimes S^{(2)})$. A vacuum extension of the channel  $\map B$ is another bipartite channel $\widetilde{\map B}   \in \Chan(\widetilde S^{(1)} \otimes \widetilde S^{(2)})$, acting on the extended systems $\widetilde S^{(1)}  :  =  S^{(1)}   \oplus {\rm Vac}^{(1)}$  and $\widetilde S^{(2)}    :=   S^{(2)}   \oplus {\rm Vac}^{(2)}$.  
 In general, the systems $S^{(1)},S^{(2)}$ can represent the systems accessible at the same location at two consecutive moment of time,  or it can represent the systems accessible at different locations at the same time (as considered in Refs.\ \cite{abbott2018communication,chiribella2019shannon2q}), or more generally, they can represent any pair of independently aderressable systems, representing the input/output ports of our multiport device.

\subsection{A single particle travelling through multiple ports}

	 In order to be able to send the same quantum particle to either of the ports of the device, we require the isomorphism $S^{(1)} \cong S^{(2)} \cong M$, where $M$ is the message-carrying degree of freedom of the particle.    In this case,   the tensor product  $\widetilde S^{(1)}   \otimes \widetilde S^{(2)}  $ contains a no-particle sector ${\rm Vac}^{(1)} \otimes {\rm Vac}^{(2)}$, a one-particle sector  $( S^{(1)}   \otimes {\rm Vac}^{(2)} )  \oplus ( {\rm Vac}^{(1)} \otimes S^{(2)}  )$, and a two-particle sector $ S^{(1)}  \otimes S^{(2)}$.  
 The one-particle sector is isomorphic to $M\otimes C$, where $C$ is a qubit system, representing the degree of freedom of the particle that controls its time of transmission. 
	When the control is in state $\ket{0}$, the message is sent through the first application of the channel and the vacuum is sent in the second application; vice versa for the control in state $\ket{1}$.

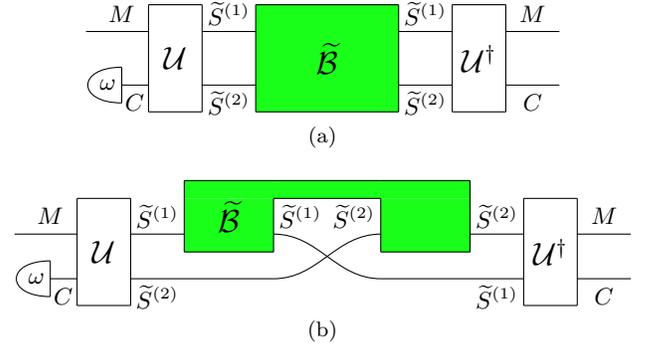
\begin{figure}		
			\centering
		\subfloat[\label{fig:sup_space}]{%
			\input{fig_singlecorsup_space.tikz}%
		}
	
		\subfloat[\label{fig:sup}]{%
		\input{fig_singlecorsup.tikz}%
		}
		\caption{
			 (a) Transmission of a single particle through a bipartite quantum channel $\widetilde {\map B}$ (green).   
			(b) Transmission of a single particle through a 2-step quantum channel $\widetilde {\map B}$  (green).  In both caes, the particle is represented by a composite system $M\otimes C$, where $M$ represents the degrees of freedom used as the message, and $C$ represents the degrees of freedom used as the control.    The isomorphism $\map U$ converts the composite system $M \otimes C$ into the one-particle sector $(S^{(1)} \otimes {\rm Vac} )  \oplus ( {\rm Vac} \otimes S^{(2)} )$ of $\widetilde S^{(1)} \otimes \widetilde S^{(2)}$. The inverse map $\map U^\dagger$ converts the output state back into $M \otimes C$.   For the  applications in this paper, we take the input of the control system $C$ to be fixed in the state $\omega$ whilst the message system $M$ is accessible to the sender.}
	\end{figure}

	We now define the situation in which a single particle is sent at a superposition of two different ports.   We call the process experienced by the particle the   {\em superposition channel}  $\map S (  \widetilde{\map B} )$, and define it  as the restriction of
	$\widetilde{\map B}$  to the one-particle sector, regarded as isomorphic to the composite system ``message + control."   Explicitly, the action of the superposition channel is defined as 
		\begin{align}\label{eq:sup}
	\map S (  \widetilde{\map B} )   :=   \map U^\dag \circ  \widetilde{\map B}  \circ \map U \,,
	\end{align}
	where  $\map U (\cdot) := U (\cdot) U^\dagger$ is the isomorphism between $M\otimes C$ and the one-particle sector $(S^{(1)}\otimes {\rm Vac})  \oplus ({\rm Vac}\otimes S^{(2)} )$, with 
	\begin{align}\label{eq:iso}
	\nonumber U   (  |\psi\>_M  \otimes |0\>_C)    &:=  |\psi\>_{\widetilde{S}^{(1)}}  \otimes |{\rm vac}\>_{\widetilde{S}^{(2)}}  \\
	U   (  |\psi\>_M  \otimes |1\>_C)    &:=  |{\rm vac}\>_{\widetilde{S}^{(1)}} \otimes  |\psi\>_{\widetilde{S}^{(2)}}  \, .
	\end{align} 
Mathematically, the transformation $\map S : \Chan(\widetilde{S}^{(1)} \otimes  \widetilde{S}^{(2)}) \rightarrow \Chan(M \otimes C)$ is a quantum supermap, that is, a transformation from quantum channels to quantum channels satisfying appropriate consistency requirements \cite{chiribella2008transforming,chiribella2009theoretical,chiribella2013quantum}.  
An illustration of the supermap $\map S$ is provided in subfigure \ref{fig:sup_space}.

Note that definition \eqref{eq:sup} can be applied in particular to $k$-step quantum channels, which are a special case of $k$-partite channels.    The illustration of the  supermap $\map S$ in this special case  is provided in subfigure \ref{fig:sup}.

 The same definition can be adopted for the transmission of a single particle through a $k$-partite multiport device.  In this case, the device is represented by a $k$-partite quantum channel $\map B \in  \Chan(S^{(1)}\otimes \dots  \otimes  S^{(k)})$,   with  $S^{(1)}  \cong  S^{(2)}  \cong \cdots  \cong  S^{(k)}$, and with vacuum extension   $\widetilde {\map B} \in  \Chan(\widetilde S^{(1)}\otimes \dots  \otimes \widetilde  S^{(k)})$. The superposition channel is then defined as the  restriction of   $\widetilde {\map B}$ to the one-particle sector
\begin{align} \nonumber 
&
\bigoplus_{j=1}^k    {\rm Vac}^{(1)} \! \otimes \! \cdots \!  \otimes \!  {\rm Vac}^{(j-1)}  \otimes S^{(j)}   \otimes \!  {\rm Vac}^{(j+1)} \!  \otimes \! \cdots  \! \otimes \! {\rm Vac}^{(k)}\\
& \qquad \qquad \qquad  \cong  M\otimes  C \, , 
 \end{align} 
where $C$ is now a $k$-dimensional control system.

	\subsection{Derivation of Eq.\ \eqref{eq:rand_uni_full} in the main text}  
	
	We now specialise to the case of correlated  channels of the random unitary form
	\begin{align}\label{eq:correlatedRU_app}
	\map R    =  \sum_{m,n}   p(m,n)   \,    \map V_m \otimes \map V_n  \in \Chan(S^{(1)} , S^{(2)} ) \, ,
	\end{align} 
	where $\map V_m (\cdot ) := V_m (\cdot) V_m^\dagger$ is a unitary channel, $\{ V_m\}$ is a set of unitary gates, and  $p(m,n)$ is a joint probability distribution. 
	The vacuum extension of each unitary $V_m$ is taken to be another unitary $U_m$, which we write as 
	\begin{align}\label{aaaa}
	\widetilde{ V}_m := U_m = V_m \oplus e^{i \phi_m} \ketbra{\rm vac}{\rm vac} \,,
	\end{align} where the vacuum amplitude is given by a complex phase, representing the coherent action of each possible noisy process on the one-particle and vacuum sectors. 
	This leads to the  vacuum extension
	\begin{align}\label{eq:vacext_correlatedRU_app}
	\widetilde{\map R}    =  \sum_{m,n}   p(m,n)   \,    \widetilde{\map V}_m ,  \widetilde{\map V}_n \in \Chan(\widetilde{S}^{(1)} , \widetilde{ S}^{(2)} ) \, ,
	\end{align} 
	with $\widetilde{\map V}_m  (\cdot ) := \widetilde{V}_m (\cdot )  \widetilde{V}_m^\dagger$,
	which is equivalent to Equation \eqref{correlatedRU} in the main text, with $U_m = \widetilde{ V}_m$.
	
	The use of the channel $\map R$, specified by the vacuum extension $\widetilde{\map R}$, at a superposition of times  is given by:
	\begin{align}\label{eq:sup1}
	\map S (  \widetilde{\map R})   = \! \!  \sum_{m,n=0}^{r-1} p(m,n)  \, \map U^\dag \! \circ \! \left(
	\widetilde{\map V}_m \otimes \widetilde{\map V}_n \right)
	\! \circ  \map U  \, .
	\end{align}
	
	Explicitly, we have the expression  
	\begin{align}\label{bbbb}
	\map S (  \widetilde{\map R})     (\rho\otimes \omega)  =  \sum_{m,n}  C_{mn}     \,  (\rho\otimes \omega)  \,  C_{mn}^\dag \, ,
	\end{align}
	where $\rho$  (respectively, $\omega$) is an arbitrary state of the message  (respectively, control), and   
\begin{align}	
\nonumber C_{mn}    &:= \sqrt{p(m,n)} e^{i \phi_n}   V_{m}  \,   \otimes |0\>\<0| \\
   &~~ +  \sqrt{p(m,n)}  V_{n }  e^{i \phi_m} \,    \otimes |1\>\<1| \, ,
\end{align} 
$e^{i \phi_m}$ being the vacuum amplitude in Eq. \eqref{aaaa}.  
	Eq. \eqref{bbbb}   coincides with  Equation \eqref{eq:rand_uni_full} in the main text, with  $\map C:  =\map S (  \widetilde{\map R})$   and $W_{mn} := C_{mn}/\sqrt{p(m,n)}$.

\subsection{Derivation of Eq.\ \eqref{eq:outputsingle}--\eqref{G} in the main text}

 It is useful to consider the case where the probability distribution $p(m,n)$ is symmetric, that is, $  p(m,n)  =  p(n,m)$ for every $m$ and $n$.  In this case, the superposition channel has the simple expression 
\begin{align}\label{eq:outputsingle_full}
\map S  (\widetilde  {\map R})    =  \frac{   \map   R_1    +     \map G}2  \otimes \map I    +   \frac{\map R_1  -  \map G}2 \otimes \map Z  \, , 
\end{align}
where $\map Z$   is the unitary channel associated  to the Pauli matrix $Z$,   $\map R_1$  is the reduced channel  defined by 
\begin{align}
\map R_1  (\rho)  :  =  \sum_m\,  p_1 (m) \,  V_m \rho  V_m^\dag  \qquad p_1 (m):  =  \sum_n  \,  p(m,n) \, ,   
\end{align}  
and $\map G$ is the linear map defined by 
\begin{align}
\map G   (\rho) :  =    \sum_{m,n}  \,  p(m,n)  ~ e^{i   (  \phi_n-\phi_m)}  \,  V_m\rho  V_n^\dag \, .  
\end{align}

\section{Analytical bound on the classical capacity in the lack of correlations}\label{app:bound_nocor}
This section refers to the scenario where the  message is transmitted at a superposition of two possible times, experiencing  independent noisy processes that are completely depolarising in the one-particle subspace.      This section   makes use of the notation  introduced in Appendix \ref{app:vacext}.

\subsection{Proof that the superposition of uncorrelated completely depolarising channels is entanglement-breaking}\label{app:eb}

Let  $\map A (\cdot) = \sum_{m=0}^{r-1} A_m (\cdot) A_m^\dagger \in \Chan(S)$ be a generic quantum channel, and let  $\widetilde{ \map {A} }  \in \Chan(\widetilde{ S})$ be a vacuum extension of $\map A$.   Using  Eq.  (\ref{eq:sup}), we obtain  
	\begin{align} \label{eq:Fsup_gen}
	\nonumber
	\map S & ( \widetilde {\map A} \otimes \widetilde{\map A} )   \\
	&=    \frac{ \map A\left( \rho \right)  + F  \rho  F^\dag   }2  \otimes \map I  
	+  \frac{  \map A\left( \rho \right)     -  F\rho F^\dag}2 \otimes  \map Z   \, ,
	\end{align}  
	where $\map I$  (respectively, $\map Z$) is the identity channel  (respectively, Pauli channel corresponding to the Pauli matrix $Z$), and 
	\begin{align}\label{F}  F :  = \sum_{m}  \,  \overline \alpha_m \,  A_m
	\end{align} is the {\em  vacuum interference operator}   defined in Ref.\ \cite{chiribella2019shannon2q}. 

Now, let $\map A$ be the completely depolarising channel $\map D:  \rho  \mapsto  I/d$, with    vacuum extension $\widetilde{ \map {D}  }$.  For a fixed state $\omega$ of the control system,  consider the effective channel   defined by  
	\begin{align}
	\nonumber    \map S  (\widetilde  {\map D} \! \otimes  
	\! \widetilde {\map D})   (  \rho  \otimes \omega)  &=     \frac{   I/d  + F  \rho  F^\dag   }2  \otimes \map I  
	+  \frac{  I/d    -  F\rho F^\dag}2 \otimes  \map Z \\
	  &  =:  \map C_{\omega,  F } (\rho)  \, .   \label{comega}
	\end{align} 
	
	For $d =  2$, we have the following result: 
	\begin{prop}\label{prop:comegaentbreak}    The channel    $\map C_{\omega,  F} $  in Eq.\ \eqref{comega} is entanglement-breaking for $d= 2$.
	\end{prop} 
	
	The proof uses the following lemma:  
	\begin{lem}\label{lemma:F}
	Let $\map D$ be a completely depolarising channel with vacuum extension $\widetilde{\map D}$ and vacuum interference operator $F$. Then, the operator norm of $F$ satisfies the inequality $||F||_\infty \le \frac{1}{\sqrt{d}}$.
\end{lem}
\Proof 
Let the Kraus operators and vacuum amplitudes of $\map D$ be given by $\{A_i\}, \{\alpha_i\}$, respectively.
By definition, 
\begin{equation}
||  F ||_\infty   =  \max_{ \{ |v\> :  ||  |v\> ||=1 \} } ~ \max_{\{ |w\> :  || |w\> ||  =1 \} }   \<v |  F |w\>   
\end{equation}
 and  
 \begin{equation}
 \begin{split}
  \left| \<v | F |  w \> \right|  &=  \left|  \sum_i \overline{ \alpha_i}  \<v|  A_i  |w \> \right|   \\ 
  &\le \sqrt{ \left( \sum_i  |\alpha_i|^2 \right)      \left(\sum_j  \<v | A_j  |w  \>\<  w |  A_j^\dag |v\>  \right)   }   \\
 &= \sqrt{ \<v  |   \map D  \left(|w\>\<w| \right)   |v\>  }   
 \end{split}
 \end{equation}
If $\map D$ is the completely depolarising channel, then  $\map D ( |w\>\<w| )  =  I/d  $
and therefore the bound becomes
$| \<v | F |  w \> |  \le  \sqrt{1/d}$   
which implies 
$|| F ||_\infty \le  \sqrt{1/d}$.
\qed
  
\medskip  

We are now ready to provide the proof of Proposition \ref{prop:comegaentbreak}.

 {\bf Proof of Proposition \ref{prop:comegaentbreak}.}      To prove that a channel is  entanglement-breaking, it is sufficient show that it transforms a maximally entangled state into a separable state \cite{horodecki2003entanglement}.  
	Let $|\Phi^+\>  =  \sum_{k=0}^{d-1}  \,   |k\>\otimes |k\>/\sqrt{d}$ be the canonical maximally entangled state.  When the channel $\map C_{\omega,  F}$ is applied, the output state is 
	\begin{align}
\nonumber (\map C_{\omega,  F} \otimes \map I)   (|\Phi^+\>\<\Phi^+|)     =  &    \left(  \frac  {   I\otimes I }{d^2}   +   G_F\right)  \otimes \frac \omega 2\\
\label{outputuncor}& +    \left(  \frac  {   I\otimes I }{d^2}   -   G_F\right)  \otimes \frac {Z\omega Z} 2 \, ,
 	\end{align}
with $  G_F:  =   (F \otimes I)   (|\Phi^+\>\<\Phi^+|)  (F\otimes I)^\dag$. 	

We now show that the operators $\frac{I  \otimes I}{d^2}  \pm   G_F$ are proportional to states with positive partial transpose. To this purpose, note that the partial transpose of $G_F$ on the second space is  
\begin{align}
G_F^{\tau_2}   =     (F\otimes I) \frac{{\tt SWAP}}d  (F \otimes I)^\dag  \, .  
\end{align}  
Hence, for every unit vector $|\Psi\>$ we have the bound,
\begin{align}
\nonumber \<\Psi|   G_F^{\tau_2}  |\Psi\>  &  \le       \frac{\< \Psi|   (F F^\dag \otimes I)  | \Psi\>}{d} \\
  \nonumber  &  \le    \frac{ \|    F  F^\dag  \|_{\infty}}{d}  \\
 \nonumber   & =  \frac{\|   F\|_{\infty}^2}{d}   \\
   &  \le  \frac 1{d^2} \, .  \label{boundonGnorm}
  \end{align}  	
where the first inequality follows from Schwarz' inequality, and the last inequality follows from  Lemma \ref{lemma:F}.   

Using Eq.\ (\ref{boundonGnorm}), we obtain the relation  
\begin{align}
\<  \Psi  |     \left(  \frac{I\otimes I}{d^2}  \pm  G_F\right)^{\tau_2} \,  |\Psi\>    \ge   \frac 1{d^2}  -  \<\Psi  |  G_F^{\tau_2}  |\Psi\>   \ge 0  \, .  
\end{align} 
Since $|\Psi\>$ is an arbitrary vector, we conclude that the operator $\left(\frac{I\otimes I}{d^2}  \pm  G_F\right)^{\tau_2}$ has positive partial transpose.   For $d=2$, the Peres-Horodecki criterion \cite{peres1996separability,horodecki2001separability}, guarantees that $\frac {I\otimes I}4   \pm  G_F$ is proportional to a separable state.  Hence, the whole output state \eqref{outputuncor} is separable. \qed

	\subsection{Optimal control state for maximizing the Holevo capacity}
Proposition \ref{prop:comegaentbreak} implies that the classical capacity of the channel $\map C_{\omega, F}$ is equal to its Holevo capacity (see \cite{shor2002additivity}).  Here we show that the Holevo capacity is maximised by the state $\omega  =  |+\>\<+|$.    In fact, we prove a more general result:  
\begin{lem}\label{lem:plusismore}
Let  $\map C_\omega$ be an arbitrary channel  of the form 
\begin{align}\label{generalcomega}
\map  C_{\omega}  (\rho) :  =   \map L_+  (\rho) \otimes \omega  +    \map L_-  (\rho)  \otimes Z\omega Z  \,   ,
\end{align}
where $\map L_\pm$ are arbitrary linear maps.    Then,  for every density matrix $\omega$,  the Holevo capacity satisfies the bound   $\chi\left(\map C_{\omega}\right) \le \chi\left( \map C_{|+\>\<+|} \right)$.  
\end{lem}

\Proof   The Holevo capacity  is  known to be monotonically decreasing under the adtion of quantum channels,  namely $\chi (\map E) \ge \chi  (\map F \circ \map E)$  for every pair of channels $\map E$ and $\map F$.   For every channel $\map C_{\omega}$ of the form  \eqref{generalcomega}, we have     the relation 
\begin{align}
\map C_{\omega}  =        (\map I_M \otimes \map P_\omega )    \circ \map C_{|+\>\<+|}  \, ,
\end{align} 
where $\map P_\omega$ is the quantum channel  defined by 
\begin{align}
\map P_\omega(  \gamma)    :=    \<  +  |\gamma  |+\>  \,   \omega  +  \< -  |  \gamma  |-\>   \, Z\omega Z    
\end{align}
 for an arbitrary state $\gamma$.  Hence, we have  $\chi    (\map C_{\omega})   =  \chi  \left[ (\map I_M \otimes \map P_\omega )    \circ \map C_{|+\>\<+|}    \right]   \le \chi    (\map C_{|+\>\<+|})$.    \qed  
 
 \medskip  
Lemma \ref{lem:plusismore} holds in particular for 
\begin{enumerate}
\item the channel $\map  C_{\omega,  F} $ defined in Eq.\ \eqref{comega}
\item the  channel $\map  C_{\omega} $ defined in Eq.\ \eqref{eq:outputsingle1} of the main text
\item the channel $\map  E_{\omega,  F} $ defined in Eq.\ \eqref{eq:outputdouble} of the main text.  
\end{enumerate}
	
	\subsection{Bound on the Holevo capacity}

	\begin{prop}\label{prop:F1}
The Holevo capacity of the channel  $\map C_{\omega,  F} $ defined in Eq.\ \eqref{comega} is upper bounded as  		
	\begin{equation}\label{holevoboundF1}
		\begin{split}
		\chi   (\map  C_{\omega, F} ) \le \frac{ \log (2 d)}{d}   &+ \frac{\frac{1}{d}+||F||_\infty^{~2}}2 \log  \frac{\frac{1}{d}+||F||_\infty^{~2}}2 \\ 
		&+ \frac{\frac{1}{d}-||F||_\infty^{~2}}2 \log  \frac{\frac{1}{d}-||F||_\infty^{~2}}2 		\, ,
		\end{split}
		\end{equation}
		where $F$ is the vacuum interference operator defined in  Eq.\ \eqref{F}.  
			\end{prop}

\Proof    For a fixed vacuum extension, and therefore for a fixed vacuum interference operator $F$,  the Holevo capacity of the channel $\map C_\omega$ is  upper bounded by the Holevo capacity of the channel  $\map C_{|+\>\<+|, F}$  (Lemma \ref{lem:plusismore}). Hence,      it is enough to prove the bound for the channel $\map C_{|+\>\<+|}$.     

Note that the output of channel $\map C_{|+\>\<+|,  F}$ has dimension $2 d$.       For a generic channel  $\map E$ with $(2d)$-dimensional output, the Holevo capacity is  upper bounded as  \cite{holevo2002remarks}
\begin{align}\label{minimumoutputentropy}
\chi (\map E)   \le   \log  (2d)     -  \min_{\rho }    H      \left [\map E  (\rho) \right]  \, ,  
\end{align}
where $ H(\rho) :  = -  \Tr [\rho  \log \rho] $ is the von Neumann entropy, and the minimisation can be restricted without loss of generality to pure states.

We now upper bound the right-hand-side of Eq.\ \eqref{minimumoutputentropy} for $\map E  = \map C_{|+\>\<+|, F}$. The  action of the channel $\map C_{|+\>\<+|,  F}$ on a generic input state $\rho$ is 
\begin{align}
\map C_{|+\>\<+|,  F} (\rho)   =    \frac  {  \frac  Id  \! + 
	\!  F\rho  F^\dag}2  \otimes |\!+\!\>\<\!+\!|  +      \frac  {  \frac Id \!  - \!   F\rho  F^\dag}2  \otimes |\!-\!\>\<\!-\!| \, ,
\end{align}
as one can deduce from Eqs.\ \eqref{comega}  and \eqref{eq:Fsup_gen}.      
 
In the case of a pure state $\rho  =  |\psi\>\<\psi|$,  we write $F|\psi\>  =  k \, |\varphi\>$, where $|\varphi\>$ is a unit vector and $k$ is a normalisation constant.  
With this notation,   we obtain  
\begin{equation} \label{eq:Fsup_gen_D}
\begin{split}
 \map C_{|+\>\<+|,  F} (|\psi\>\<\psi|) &=    \frac{ (\frac{1}{d}+k^2) \ketbra{\varphi}{\varphi}  + \frac{1}{d}    P_\perp  }2  \otimes |+\>\<+|  \\
&~+   \frac{ (\frac{1}{d}-k^2) \ketbra{\varphi}{\varphi}  + \frac{1}{d}   P_\perp }2 \otimes |-\>\<-|  \, ,
\end{split}
\end{equation}  
with $P_\perp:  =  I -  |\varphi\>\<\varphi|$.  
The von Neumann entropy of this state is 
\begin{align}
\begin{split} 	
H &\left[  \map C_{|+\>\<+|,  F} (|\psi\>\<\psi|) \right]
\\
&= - \frac{\frac{1}{d}+k^2}2 \log  \frac{\frac{1}{d}+k^2}2  
	       -  \frac{d-  1}{2d}  \log  \frac{1}{2d} \\
	      & \quad -  \frac{\frac{1}{d}-k^2}2 \log  \frac{\frac{1}{d}-k^2}2 
	      -  \frac{d-1}{2d}  \log   \frac{1}{2d} \\
	 &= \frac{d-1}{d}  \log (2d)    \\  
	   &\quad   - \frac{\frac{1}{d}+k^2}2 \log  \frac{\frac{1}{d}+k^2}2  
	     -  \frac{\frac{1}{d}-k^2}2 \log  \frac{\frac{1}{d}-k^2}2 
  \end{split}  \label{eq:entropy_min}
\end{align}
Now, note that one has 
\begin{align}
k     =  \|    F\,  |\psi\>  \|   \le  \|   F\|_\infty     \le  \frac  1{\sqrt d} \, ,        
\end{align}  
where the last inequality follows from Lemma \ref{lemma:F}.   The  expression (\ref{eq:entropy_min}) is monotonically decreasing for  $k$  in the interval $[0,1/\sqrt d]$.    Hence, one has the lower bound 
\begin{align}
\begin{split}
	H \left[  \map C_{|+\>\<+|,  F} (|\psi\>\<\psi|) \right]&  \ge  \frac{d-1}{d}  \log (2d)    \\  
	   &\quad   - \frac{\frac{1}{d}+  \|   F\|_\infty^2}2 \log  \frac{\frac{1}{d}+\|  F\|_{\infty}^2}2  \\
	     &\quad  -  \frac{\frac{1}{d}-\|  F\|_\infty^2}2 \log  \frac{\frac{1}{d}-\|  F\|_\infty^2}2 
  \end{split}  \, .
\end{align}
Inserting this expression into Eq.\  \eqref{minimumoutputentropy} with $\map E  =  \map C_{|+\>\<+|,  F}$, we then obtain Eq.\ (\ref{holevoboundF1}). \qed

\begin{cor}\label{cor:F2}
The Holevo capacity of the channel  $\map C_{\omega,  F} $ defined in Eq. \eqref{comega} is upper bounded as  $\chi   (\map  C_{\omega,  F} )  \le 1/d$.    In particular, for 		 $d=2$, one has the bound $		\chi   (\map  C_{\omega,  F} )  \le 0.5$. 
		\end{cor}

\Proof  Immediate from the fact that the right-hand-side of Eq. \eqref{holevoboundF1} is monotonically decreasing with $\|F\|_\infty$, and that  $\|  F\|_\infty$ is upper bounded by $1/\sqrt  d$  (Lemma \ref{lemma:F}). \qed

\section{Maximisation of the Holevo information for the superposition of  independent depolarising channels}\label{app:numerical_proofs}

Here we prove a series of results that enable a complete numerical maximisation of the Holevo information of the channel \eqref{comega} 
\begin{align}
\map  C_{\omega, F}  :  \rho \mapsto     \frac { I/d  +  F\rho F^\dag}2  \otimes \omega   +  \frac{I/d  -  F\rho F^\dag}2  \otimes  Z\omega Z
\end{align}
 over all input ensembles, over all states of the control system, and over all vacuum extensions of the completely depolarising channel. This Appendix makes use of notation introduced in the previous appendices.

 Let us start from the maximisation over the vacuum extensions, which are in one-to-one correspondence with the possible operators $F$.     
  \begin{lem}\label{lem:ensembles}
Without loss of generality, the operator $F$ that maximises the Holevo information of the channel $\map C_{\omega,  F} $ can be taken to be of the form  $F  =  a  \, |0\>\<0|   + b  |1\>\<1|  $, with $a^2  +  b^2  \le  1/d$, ~$a,b\ge 0$.   
  \end{lem}    

\Proof   Using the singular value decomposition, $F$ can be written as $ F  =    U  F'  V $, where  $U$ and $V$ are suitable unitary matrices, and $F'$ is diagonal in the basis $\{ |0\>,  |1\>\}$.  Now the capacity of the channel  $\map C_{\omega, F} $ is equal to the capacity of the channel $\map C_{\omega,  F'}   =  (  \map U\otimes \map I_C)^\dag    \circ  \map C_{\omega ,  F} \circ  \map V^{\dag}$, where $\map U^\dag$ and $\map V^\dag$ are the inverses of the unitary channels associated to the unitary matrices $U$ and $V$, respectively, and $\map I_C$ is the identity channel on the control system.   Notice that $F'$ is also a vacuum interference operator associated to the completely depolarising channel.   Hence, the maximisation of the Holevo capacity can be restricted to channels with diagonal vacuum interference operator.  
    
Next, we note that, for a vacuum extension of the completely depolarising channel, the vacuum interference operator $F$ must satisfy the condition  $\Tr F^\dag F\le 1/d$ \cite{abbott2018communication}.  For an operator of the form $F =  a \,  |0\>\<0|  +  b \,  |1\>\<1|$, this implies the inequality $|a|^2+  |b|^2 \le 1/d$. Finally, we show  that $a,b$ can restricted to positive numbers. Let $W=a' \ketbra{0}{0} + b' \ketbra{1}{1}$, where $ a' = \bar{a}/|a|, b' = \bar{b}/|b|$. Then $ F'' := W F = F W =|a | \ketbra{0}{0} + 
|b| \ketbra{1}{1}$. The capacity of the channel $\map C_{\omega,  F''}   =  (  \map  W \otimes \map I_C)    \circ  \map C_{\omega ,  F} $ (where $\map W$ is the unitary channel associated with the unitary $W$)  is equal to the capacity of the channel  $\map C_{\omega, F} $. Therefore, a maximisation of the Holevo capacity can be restricted to vacuum interference operators with positive coefficients in the computational basis.
\qed 

Let us consider now the maximisation over all possible ensembles.  The key result here is that the maximisation can be reduced to the optimisation of $d$ vectors with positive coefficients in the computational basis.  
\begin{lem}\label{lem:diagonalF}
When the operator $F$ is diagonal in the computational basis, the input ensemble that maximises the Holevo information after application of  the channel $\map C_{\omega,  F}$ can be chosen without loss of generality to be of the form  
\begin{align}
\left\{  \frac {p_{x}}d,   \,  M^j  |\psi_x\>\<\psi_x|  M^{j  \dag} \right\}_{x\in  \{ 0,\dots,  d-1 \},   \,  j\in \{0,  \dots,  d-1\}} \,,
\end{align} 
where $ (p_x)_{ x\in  \{  0,\dots,  d-1\} }$ is a probability distribution,  $M$ is the unitary operator $M  :  =  \sum_{m  =  0}^{d-1}\,  \omega^m   \,  |m\>\<m|$, $\omega:  =  e^{2\pi  i  /d}$, and   $|\psi_x\>$ is a unit vector with positive coefficients in the computational basis $\{  |m\>\}_{m=0}^{d-1}$. 
\end{lem}    

\Proof   When $F$ is diagonal, the channel $\map C_{\omega,  F}$ has the covariance property 
\begin{align}
\map C_{\omega,  F}    \circ \map U_{\bs \theta}    =    (\map U_{\bs \theta} \otimes \map I_C)  \circ   \map C_{\omega,  F} \qquad \,  \forall {\bs \theta}\,   ,  
\end{align}
where  ${\bs \theta}  =  (\theta_0, \theta_1, \dots ,  \theta_{d-1})$ is a vector of $d$ phases, and  $\map U_\theta$ is the unitary channel associated to  the unitary matrix $U_{\bs \theta}  =  \sum_{m=0}^{d-1}   \,  e^{i  \theta_m  }  \,  |m\>\<m|$.  Note that, in particular, we have
\begin{align}
\map C_{\omega ,  F}   \circ \map M^j    =    (\map M^j \otimes \map I_C)  \circ   \map C_{\omega,  F} \qquad \,  j\in  \{0,
\dots,  d-1\} \, ,
\end{align}
where $\map M$ is the unitary channel associated to the unitary operator  $M$  defined in the statement of the lemma.  

For covariant channels,    Davies \cite{davies1978information} showed that the optimal input ensembles can be chosen without loss of generality to be covariant. In our case, this means that the optimal ensemble can be chosen to be of the form 
\begin{align}\label{E}
{\sf E}  :  = \left\{  \frac {p_{x}}d,   \,  \map M^j  (  \rho_x)  \right\}_{x\in   X,   \,  j\in \{0,  \dots,  d-1\}} \,,
\end{align} 
for some finite set $X$, some probability distribution $(p_x)_{x\in X}$ and some set of density matrices $(\rho_x)_{x \in  X}$.     In the same paper, Davies   also showed that the ensemble  can be chosen without loss of generality to consist of {\em pure} states, possibly at the price of increasing the size of the set $X$.

 We now show that one can choose $|X|\le d$ without loss of generality. 
 Let $\sf E$ be an optimal  covariant ensemble, and let  
 \begin{align}
 \<\rho\>   =  \frac 1  d   \,  \sum_{j=0}^{d-1}   \sum_{x  \in  X}   \,  p_x \,     \map M^j  (\rho_x )      
 \end{align}
  be its average state.     Fixing $X$, the set of covariant ensembles with average state $\<\rho\>$ is a convex set.   Since the Holevo information is a convex function of the ensemble  \cite{davies1978information},   the maximisation can be restricted without loss of generality to the extreme points.  
  
Now, note that the covariant ensembles $\sf E$ are in one-to-one  correspondence with covariant positive-operator-valued-measures  (POVMs)  $(  P_{x,j})_{x\in \set X,  j\in  \{0,\dots,  d-1\}}$, via the correspondence  
\begin{align}
  P_{x,j}   :   =  \frac  {\map M^j      (\xi_x)  }{d}\qquad  \xi_x  :  =    \<  \rho  \>^{-\frac 12}  \,     p_x\,   \rho_x  \,   \<\rho\>^{-\frac 12}  \, .     
\end{align}   
Since the correspondence is linear, the extreme ensembles are in one-to-one correspondence with the extreme POVMs.  The latter have been characterised by one of us in Ref.\ \cite{chiribella2006extremal}, where it was shown that a necessary condition for extremality is that the ranks of  the operators  $\xi_{x}$, denoted by $r_x$, satisfy the condition  
\begin{align}
\sum_{x\in X}   r_x^2 \le \sum_{\mu   }    \,  m_\mu^2 \, ,      
\end{align}    
  where the sum on the right-hand-side runs over the irreducible representations  (irreps) contained in the decomposition of the representation $\{   M^j\}_{j=0}^{d-1}$,   and $m_{\mu}$ is the multiplicity of the irrep $\mu$.   Now, the representation $\{   M^j\}_{j=0}^{d-1}$ has $d$ irreps, each with unit multiplicity.   Hence, the bound becomes  
\begin{align}
\sum_{x\in X}   r_x^2 \le d \, .      
\end{align}    
    In particular, this means that the number of non-zero operators  $\xi_x$ is at most $d$.  
    
In terms of the ensemble $\sf E$, this means that the number of values of $x$ with $p_x\not  =  0$   is at most $d$.     Hence, the maximisation of the Holevo information can be restricted without loss of generality to covariant ensembles with $|X| \le d$.  
    
 Recall that the optimal ensemble can be chosen without loss of generality to consist of pure states. The final step is to guarantee that these pure states have non-negative coefficients in the computational basis.    
 For a covariant ensemble   ${\sf E}  =  \{  p_x/d  \, ,   \,   M^j   |\psi_x\>\<  \psi_x |  M^{  j \dag}
 \}$, let us expand  each state as  $|\psi_x\>  =   \sum_m \,     |c_m|  \,  e^{i\theta_{x,m}}  |m\>$, where $\{\theta_{x, m}\}$ are suitable phases.    Then, we can define the new states   $|\psi_x'\>  :  =  U_{-{\bs \theta}_x} |\psi_x\>$, with ${\bs \theta}_x : =  (\theta_{x,0}, \dots,  \theta_{x,d-1})$.  By  construction, these states have positive coefficients in the computational basis,  and the corresponding ensemble ${\sf E'} :  =\{  p_x/d  \, ,   \,   M^j   |\psi_x'\>\<  \psi_x' |  M^{  j \dag}
 \}$ gives rise to the same Holevo information  as $\sf E$, when fed into the channel $\map C_{\omega,  F}  $.   \qed

 \begin{cor}\label{cor:chiexplicit}
When the operator $F$ is diagonal in the computational basis, the Holevo capacity of the channel $\map C_{\omega, F}$ is given by 
 \begin{align}
\nonumber  \chi (\map C_{\omega, F})    & =    \!
\max_{  \{  p_x  \, ,  |\psi_x\> \}} \! \left\{    \!   H \!   \left[ \!  \map C_{\omega, F} \! \left(  \!     \sum_{x,  m}  p_x\, |\<m|  \psi_x\> |^2  \, |m\>\<m| \right) \right] \right.   \\
   & \qquad\qquad   \qquad -   \left.  \sum_x   p_x\,    H \left[\map C_{\omega, F}  (|\psi_x\>\<\psi_x|  )\right]\right\} \, ,
 \end{align}  
 where the maximum is over the ensembles of $d$ pure states with positive coefficients in the computational basis. 
 \end{cor}   
 \Proof  Immediate from the definition of the Holevo information for the ensemble obtained by applying channel   $\map C_{\omega, F}$  to the pure state ensemble in Lemma \ref{lem:diagonalF}, using the relations,   
 \begin{align}
 H [  \map C_{\omega, F}   (  M^j  |\psi\>\<\psi|  M^{j\, \dag}) ]   =  H [  \map C_{\omega, F}   ( |\psi\>\<\psi| ) ] \, ,\\
 \frac 1 d \, \sum_{j  =  0}^{d-1}    \,    M^j  |\psi\>\<\psi|   M^{j \dag}    =    \sum_{m  = 0}^{d-1}  \,  |\<  m| \psi\>|^2 \,   |m\>\<m|  \, ,       
 \end{align}
 valid for every vector  $|\psi\>$.  \qed  
 
 \medskip  
 
For qubit messages ($d=  2$), we finally obtain an upper bound on the classical capacity: 
 \begin{theo}\label{theorem_numerics}
 For every vacuum extension of the completely depolarising channel and for every state of the control qubit,  the classical capacity of the  channel resulting from the superposition of two independent depolarising qubit channels is upper bounded as     
 
  \begin{align}
 \nonumber
C  (\map C_{\omega,  F})   \le   \max_{\substack   {a\ge 0,   b\ge 0 \\ a^2  + b^2  \le 1/2 }}\, \max_{0\le q,  p_0,  p_1\le 1}   
H \!   \left[   \map C_{\omega, F} \! \left(     
\rho_{q}
 \right) \right]   \\
  -   q  H \left[\map C_{\omega, F}  (|\psi_0\>\<\psi_0|  ) \right] 
 - (1 \!-\! q) H \left[\map C_{\omega, F}  (|\psi_1\>\<\psi_1|  ) \right] \, , \\
 \begin{cases}
  |\psi_0\>    &=  \sqrt{  p_0} \, |0\>  +  \sqrt{1-p_0}\,  |1\>  \, ,\\
  |\psi_1\>  &=  \sqrt{p_1} \,  |1\>  + \sqrt{  1-p_1} \,  |1\>  \, ,\\
 \rho_q &= [q p_0 + (1-q) p_1 ]\ketbra{0}{0} \\
 &\quad + [q (1-p_0) + (1-q) (1-p_1)] \ketbra{1}{1}
  \end{cases}
  \end{align}    
\end{theo}   
 \Proof For $d=2$,     Proposition \ref{prop:comegaentbreak} guarantees that the channel $\map C_{\omega,  F}$ is entanglement breaking, and therefore its classical capacity is equal to the Holevo capacity.  Lemma \ref{lem:plusismore} guarantees that the maximum of the Holevo capacity is attained by the state $\omega =  |+\>\<+|$.   Then, Lemma \ref{lem:diagonalF} guarantees the maximum of the Holevo capacity of the channel $\map C_{ |+\>\<+|,  F}$ can be obtained with a diagonal operator $F  =  a\,  |0\>\<0|  + b \,|1\>\<1|$, $a,b \ge 0$.  The Holevo capacity  of $\map C_{ |+\>\<+|,  F}$  can be computed explicitly using Corollary \ref{cor:chiexplicit}, with   
 \begin{align}
\nonumber  |\psi_0\>   &:  =  \sqrt{  p_0} \, |0\>  +  \sqrt{1-p_0}\,  |1\>  \\
|\psi_1\>  &: =  \sqrt{p_1} \,  |1\>  + \sqrt{  1-p_1} \,  |1\>  \, .
 \end{align}   
 Finally, an upper bound is obtained by relaxing the constraint on $a$ and $b$ to $a^2 + b^2 \le 1/d$  (Lemma \ref{lem:diagonalF}).  
 \qed  
		
	\section{Transmission of a single particle through a network of two-step channels}\label{app:switch}
	
	In the following  we  will use  the notation  introduced in Appendix \ref{app:vacext}.
	
	\subsection{Derivation of Eq.\ \eqref{eq:sup2pairs} in the main text}
	
Let $\map A$ and $\map B$ be two-step 	 channels, with vacuum extensions   $\widetilde{\map A}  \in \Chan(\widetilde{A}^{(1)} ,  \widetilde{A}^{(2)})$ and $\widetilde{\map B} \in \Chan(\widetilde{B}^{(1)} ,  \widetilde{B}^{(2)})$.  For simplicity, here we take all the systems $\widetilde A^{(1)},  \widetilde A^{(2)},  \widetilde B^{(1)}  , \widetilde B^{(2)}$  to be isomorphic.

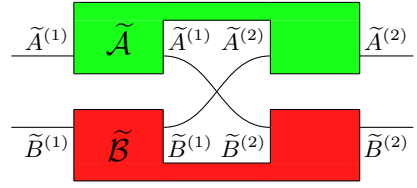
\begin{figure}
	\input{fig_doublecor.tikz}
	\caption{\label{fig:doublecor}  The  channel $   \map Z ( \widetilde{\map A} ,  \widetilde{\map B}  ) $, obtained by  connecting two vacuum-extended 2-step channels $\widetilde {\map A}$  (green) and $\widetilde {\map B}$ (red) such that output of the first use of each channel is connected to  the  input of the second use of the other channel. 
	}  
\end{figure}

We now connect the 2-step channels  $\widetilde{\map A}$ and $\widetilde{\map B} $  in such a way that the output of the first use of each channel is fed into the  input of the second use of the other channel, as in Figure  \ref{fig:doublecor}.  This particular composition of two 2-step channels is described by a supermap $\map Z$ that maps pairs of channels in  $\Chan(\widetilde A^{(1)} , \widetilde A^{(2)}) \times \Chan(\widetilde B^{(1)},  \widetilde   B^{(2)})$ into  bipartite channels in  $\Chan(\widetilde A^{(1)} \otimes  \widetilde B^{(1)} \to \widetilde {B}^{(2)} \otimes 
	,  \widetilde{A}^{(2)})$. 

\medskip  	
	
	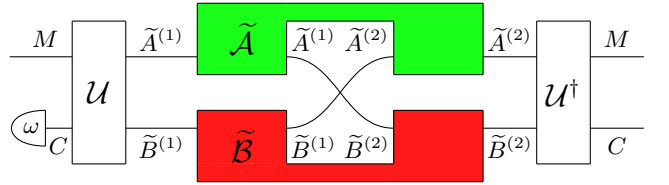
\begin{figure}
		\input{fig_doublecorsup.tikz}
		\caption{\label{fig:doublesup}  The superposition channel $\map S  [ \map Z ( \widetilde{\map A} ,  \widetilde{\map B}  ) ] $ of two 2-step channels ${ \map A}$ and $\map {B}$, specified by the vacuum extensions $ \widetilde {\map A}$ (green) and $  \widetilde {\map B}$ (red), where the alternative paths traverse the two correlated channels in the opposite order. For the applications in this paper, the input of the control system $C$ is fixed in the state $\omega$, whilst the message system $M$ is accessible to the sender.
		}  
	\end{figure}
  \medskip

We can now consider the scenario in which a single particle is sent in a superposition of going through the $A$-port and the $B$-port of the channel  $\map Z  ( \widetilde {\map R}_A,  \widetilde  {\map R}_B)$.  Following Eq. \eqref{eq:sup},  the evolution of the particle is  described by the superposition channel  
\begin{align}
\map S   \left[\map Z  ( \widetilde {\map A},  \widetilde  {\map B}) \right]   :=  \map U^\dag \circ      \map Z  ( \widetilde {\map A},  \widetilde  {\map B}) \circ \map U \, ,
\end{align} 
with $\map U$ defined as in Eqs. \eqref{eq:sup} and \eqref{eq:iso}.    The 	superposition channel $\map S   \left[\map Z  ( \widetilde {\map A},  \widetilde  {\map B}) \right]$  is illustrated in Figure \ref{fig:doublesup}.

Let us apply the above construction  to the special case where the channels  $\widetilde{\map A}$ and $\widetilde{\map B}$  are of the random unitary form
	\begin{align}
\nonumber  \widetilde {\map A} = 	\widetilde{\map R}_A    &:=  \sum_{m,n}   p_A(m,n)   \,    \widetilde{\map V}^{(A)}_m \otimes  \widetilde{\map V}^{(A)}_n  \\
 \widetilde {\map B}
=	  \widetilde{\map R}_B    &:=  \sum_{k,l}   p_B(k,l)   \,    \widetilde{\map V}^{(B)}_k \otimes   \widetilde{\map V}^{(B)}_l  \, ,
	\end{align} 
	where $\widetilde{\map V}_m^{(A)}$ and   $\widetilde{\map V}_k^{(B)}$ are the unitary channels corresponding to the unitary operators 
	\begin{align}
\nonumber	\widetilde{V}^{(A)}_m   &:=  V^{(A)}_m  \oplus  e^{i\phi_m^{(A)}} \,  |{\rm vac}\>\<{\rm vac} | \\
     \widetilde{V}^{(B)}_k &:=  V^{(B)}_k  \oplus  e^{i\phi_k^{(B)}} \,  |{\rm vac}\>\<{\rm vac} | \,,
     \end{align}
      respectively. With this choice, we have
      	\begin{align}
		\nonumber &\map Z  (\widetilde{\map R}_A,  \widetilde{\map R}_B)  \\
		&  =  \!\!\! \sum_{m,n,k,l} \!\!\!  p_A(m,n) p_B(k,l)   \,   ( \widetilde{\map V}^{(B)}_l \circ \widetilde{\map V}^{(A)}_m ) \otimes ( \widetilde{\map V}^{(A)}_n \circ \widetilde{\map V}^{(B)}_k ) \, .
	\end{align} 
	and 
\begin{align}
\nonumber &\map S   \left[ \! \map Z \! \left( \widetilde{\! \map {R}}_A ,  \widetilde{ \map {R} \! }_B \right) \right]    \,  ( \cdot )    \\
&  = \sum_{m,n,k,l}  \,  p_A (m,n)  \,  p_{B}  (k,l)  \,  W_{mnkl}       \,  (\cdot)  \,    W_{mnkl}^\dag \, , 
\end{align}	
with  
\begin{align}
\nonumber   W_{mnkl} 
  &:= V^{(B)}_{l}  V^{(A)}_{m}    \,e^{i  (\phi_{k}^{(B)}+ \phi_{n}^{(A)}) }   \otimes |0\>\<0|  \\
    &  \quad  +    V^{(A)}_{n}  V^{(B)}_{k}  \,e^{i  (\phi_{m}^{(A)} + \phi_{l}^{(B)})}    \otimes |1\>\<1| \, .
\end{align}
   This proves  Equation \eqref{eq:sup2pairs} in the main text.

	\subsection{Derivation of Eqs.\ \eqref{eq:outputdouble}--\eqref{eq:g2_1032} in the main text}

	For the control (in this case the path of the particle) initialised in the  state $\omega$, the superposition channel  specified by the vacuum extension $\map Z (\widetilde{ \map R}_{A}  , \widetilde{ \map R}_{B})$ is given by
	\begin{widetext}
	\begin{align}\label{eq:outputdouble_full}
\nonumber 		 \map S \! \left[ \! \map Z \! \left( \widetilde{\! \map {R}}_A ,  \widetilde{ \map {R} \! }_B \right) \right]  \!  (\rho \! \otimes \!\omega) \!    &   =   \sum_{m,n,k,l}  p_A (m,n)\,  p_B  (k,l)  \,  W_{mnkl}    (\rho \otimes \omega)  W_{mnkl}^\dag \\  
\nonumber &     =  \sum_{m,n,k,l}  \Big\{   \,    p_A (m,n)\,  p_B  (k,l)    ~       V_l^{(B)}    V_m^{(A)}    \rho    V_m^{(A) \dag}  V_l^{(B) \dag} \otimes \omega_{00}\,   |0\>\<0|      \\      
\nonumber &     \qquad   \qquad+     p_A (m,n)\,  p_B  (k,l)    ~       V_n^{(A)}    V_k^{(B)}    \rho    V_k^{(B) \dag}  V_n^{(A) \, \dag} \otimes \omega_{11}\,   |1\>\<1|      \\      
\nonumber &     \qquad   \qquad   +     p_A (m,n)\,  p_B  (k,l)    ~       V_l^{(B)}    V_m^{(A)}    \rho    V_k^{(B) \,\dag}  V_n^{(A)  \dag}\, e^{i  \left[  \phi_n^{(A)}  +   \phi_k^{(B)}   -  \phi_m^{(A)}   -  \phi_l^{(B)}\right]} \otimes \omega_{01}\,   |0\>\<1|   +  {\rm h.c.} \Big\}\\
  \nonumber &  =      \map R_B \map R_A      (\rho)  \otimes \omega_{00} \,  |0\>\<0|      + \map R_A \map R_B      (\rho)  \otimes \omega_{11} \,  |1\>\<1|   +  \map K  (\rho)   \otimes \omega_{01} \,  |0\>\<1|   +   \left[  \map K  (\rho)\right]^\dag   \otimes \omega_{10} \,  |1\>\<0| \,,
\end{align}
where $\map K$ is the linear map defined by  
\begin{align}
\map K (\rho):  =  \sum_{m,n,k,l} \,   p_A (m,n)\,  p_B  (k,l)    ~       V_l^{(B)}    V_m^{(A)}    \rho    V_k^{(B) \,\dag}  V_n^{(A)  \dag}\, e^{i  \left[  \phi_n^{(A)}  +   \phi_k^{(B)}   -  \phi_m^{(A)}   -  \phi_l^{(B)}\right]} \, .
\end{align}	
\end{widetext}
We now restrict our attention to  the case where
\begin{enumerate}
\item the two channels   $ \widetilde{\! \map {R}}_A$ and $  \widetilde{ \map {R} \! }_B$ are identical   (this implies that one  can  choose  without loss of generality $p_A   (m,n)  = p_B  (m,n):=  p(m,n)$ for every $m$ and $n$,  $V_m^{(A)}  =  V_m^{(B)}  :=  V_m$, and  $\phi_m^{(A)}  =  \phi_m^{(B)}  =:  \phi_m$ for every $m$),       
\item the probability distribution  $p (m,n)$ is symmetric, namely $p(m,n)=  p(n,m)$ for every $m,n$.  
\end{enumerate}
Under these conditions, the operator $\map K(\rho)$ is self-adjoint for every density matrix $\rho$, and the effective channel can be rewritten as 
\begin{align}\label{basta}
 \nonumber \map S \! \left[ \! \map Z \! \left( \widetilde{\! \map {R} }_A ,  \widetilde{ \map {R} }_B \right) \right]  \!  (\rho \! \otimes \!\omega) \!    = &    \frac  {\map  R^2    (\rho)  +  \map K(\rho)}2  \otimes \omega  \\
 &   +  \frac  {\map  R^2    (\rho)  -  \map K(\rho)}2  \otimes  Z\omega Z    \, ,   
\end{align}
with 
\begin{align}
\map R (\rho) :  =  \sum_{m,n}  \,  p(m,n) \,  V_m \rho  V_m^\dag  \, .
\end{align}

In particular, suppose that the unitaries $\{  V_m\}_{m=0}^{d^2-1}$ form an orthogonal basis, and that   the probability $p(m,n)$ has the form $p(m,n)  =  \delta_{n,\sigma (m)}/d^2$, for a permutation $\sigma$ that makes $p(m,n)$ symmetric.   In this case, Eq.\ \eqref{basta} becomes
 \begin{align}
 \nonumber \map S \! \left[ \! \map Z \! \left( \widetilde{\! \map {R} }_A ,  \widetilde{ \map {R} }_B \right) \right]  \!  (\rho \! \otimes \!\omega) \!    = &    \frac  {I/d  +  \map K(\rho)}2  \otimes \omega  \\
 &   +  \frac  {  I/d  -  \map K(\rho)}2  \otimes  Z\omega Z    \, ,     
\end{align}
with 
\begin{align}
\map K (\rho) \!  =  \!  \frac 1{d^4} \!\! \sum_{m,k}  \! V_{\sigma (k)}^{(\!B\!)}   V_m^{(\!A\!)} \rho  V_k^{(\!B\!) \dag}  V_{\sigma(m)}^{(\!A\!) \dag}  \,    e^{i  \left[  \phi_{\sigma(\! m  \! )}^{(\!A\!)}  \!  +   \phi_k^{(\!B\!)}  \!   -  \phi_m^{(\!A\!)}  \!   -  \phi_{\sigma(\!k \!)}^{(\!B\!)}    \right]}   .
\end{align}
Setting $d=2$ and choosing $\sigma$ to be the permutation that exchanges 0 with 1, and 2 with 3, we obtain Eqs.\ \eqref{eq:outputdouble}--\eqref{eq:g2_1032}  of the main text.  

 \section{Proofs of the statements in Subsection \ref{subsec:maxcapacitydouble}}\label{app:double}

 Here we consider the scenario of Figure \ref{fig:doublesup}, in the special case where the 2-step channels $\widetilde {\map A}$ and $\widetilde {\map B}$  are of the product form  $\widetilde {\map A}  =  \widetilde {\map A}_{1}  \otimes \widetilde {\map A}_{2}$ and  $\widetilde {\map B}  =  \widetilde {\map B}_{1}  \otimes \widetilde {\map B}_{2}$, respectively.  In this case, the combination of the channels  in the network of Figure \ref{fig:doublecor} gives the bipartite channel
 \begin{align}
\map Z  (\widetilde {\map A} \otimes \widetilde {\map B})    =     \widetilde {\map B}_{2} \widetilde {\map A}_{1} \otimes \widetilde {\map A}_{2}  \widetilde {\map B}_{1}\, .  
 \end{align}
When a single particle is sent into one of the two ports of this channel, the resulting evolution is described by the superposition channel
\begin{align}
\map S  \left[  \map Z  (\widetilde {\map A} \otimes \widetilde {\map B})   \right]  &  =   \map S  ( \widetilde {\map B}_{2} \widetilde {\map A}_{1} \otimes \widetilde {\map A}_{2}  \widetilde {\map B}_{1} ) \, ,
\end{align}
where $\map S$ is the supermap defined in Eq.\ \eqref{eq:sup}.

\begin{figure}
	\vspace{1ex}
	\includegraphics[width=0.9\linewidth]{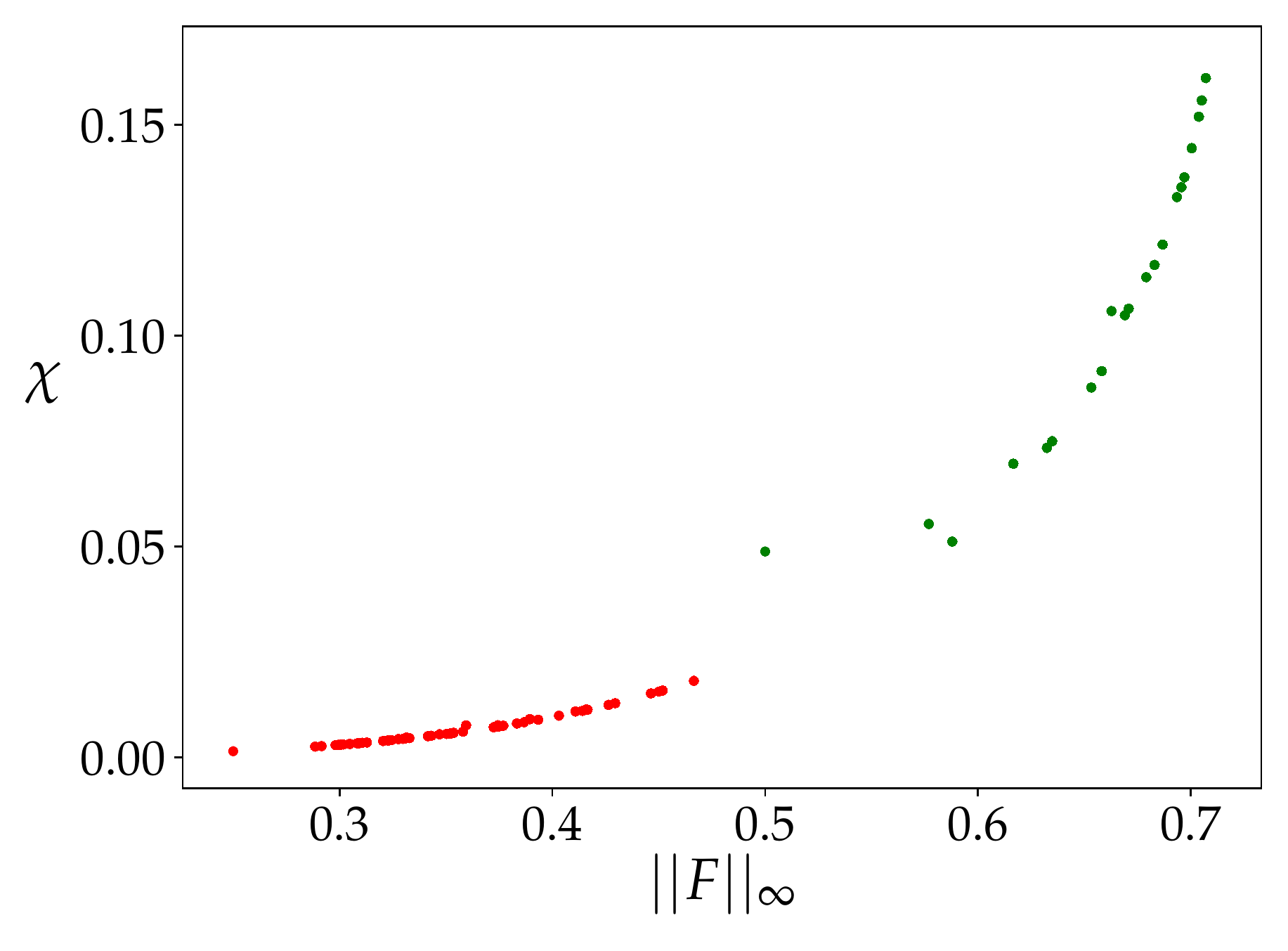}
	\caption{\label{fig:fnorm-chi}  \textit{Green}: A plot of the classical capacity against $||F||_{\infty} $ for the channel $\map C_{\omega,  F}$ . \textit{Red}: A plot of the classical capacity against $||F||^2_{\infty} $   for the  channel $\map C_{\omega, { F^2}}$. In both cases $F =  \sum _{m = 0}^{3} \frac{1}4 e^{ - i  \phi _{m}} V_{ m}$ and  is sampled over the phase parameters $\{\phi_{1},\phi_{2},\phi_{3}\}$ with a numerical precision of $\pi/8$ for each parameter. We set $\phi_{0}=0$ without loss of generality, as $F \rho F^\dagger$ is invariant under the phase group $U(1)$. The classical capacity is here equal to the Holevo capacity (see Appendix \ref{app:bound_nocor}) and the Holevo capacity was calculated using the methods outlined in Appendix \ref{app:numerical_proofs}.
	}  
\end{figure}

We now restrict our attention to the case where the channels  $\widetilde {\map A}_{1},   \widetilde {\map A}_{2},  \widetilde {\map B}_{1}$, and $\widetilde {\map B}_{2}$ are all equal to each other, and are all equal to $\map{\widetilde   D}$, a vacuum extension of the completely depolarising channel.   In this case, the action of the  superposition channel on a generic product state $\rho\otimes \omega$ is   
\begin{align}
\map S  &(\map{\widetilde   D}^2  \otimes \map{ \widetilde   D}^2 )   (\rho\otimes \omega) \\
&=     \frac{   I/d  + F^2  \rho  F^{2 \,\dag}   }2  \otimes \omega  
	+  \frac{  I/d    -  F^2\rho F^{2\dag}}2 \otimes  Z\omega Z  \, ,
 	  \end{align}
 where $F$ is the vacuum interference operator associated to channel   $\map{\widetilde   D}$.   The above equation follows from Eq.\ \eqref{comega} and from the observation that the vacuum interference operator of $\map{\widetilde   D}^2$ is $F^2$.

Note that  one has the equality
\begin{align}
\map S  (\map{\widetilde   D}^2  \otimes \map{ \widetilde   D}^2 )    (\rho\otimes \omega) \equiv \map  C_{\omega,  F^2}  (\rho) \, ,
\end{align}
using the notation of Eq.\  \eqref{comega}.  
That is, in the lack of correlations   the configuration of channels depicted in Figure \ref{fig:doublesup}  gives rise to   the  effective channel    in Equation \eqref{eq:Fsup_gen}, with $F$ replaced by $F^2$.
This means that all of the results in Appendices \ref{app:bound_nocor}--\ref{app:numerical_proofs} apply to this scenario as well, with $F$ replaced by $F^2$. In particular, the classical capacity can be determined numerically using Theorem \ref{theorem_numerics}, with the maximisation constraint now being that for the vacuum interference operator $F^2 = g \ketbra{0}{0} + h \ketbra{1}{1}$, $g+h \leq 1/d$, where $g,h \geq 0$.

The classical capacity of the channels $\map C_{\omega,  F}$ and $\map C_{\omega, F^2}$ can be evaluated numerically.  For the cases where each completely depolarising channel is implemented by a random unitary channel (cf.\ Eqs.\ \eqref{eq:rand_uni_full} and \eqref{eq:sup2pairs}, respectively, in the main text), Figure \ref{fig:fnorm-chi} show a scatter plot with the capacities of both  channels  in the same graph against the norm of the corresponding vacuum interference operator,  $F$ or $F^2$, for same combination of phases $\phi_1,\phi_2,\phi_3$ as shown in Figs.\ \ref{fig:sup-chi} and \ref{fig:sup-net-chi}.

\end{appendix}

\end{document}

%% file: figure_styles.tikzstyles
\tikzstyle{dotted line}=[-, style=dotted, tikzit draw=brown]
\tikzstyle{dashed line}=[-, style=dashed, tikzit draw=cyan]
\tikzstyle{green fill line}=[-, fill={green!90!black}, tikzit draw=green]

%% file: fig_freestyle_corsup.tikz
\begin{tikzpicture}[scale=1]
	\begin{pgfonlayer}{nodelayer}
		\node [style=none] (0) at (-4.5, 0) {};
		\node [style=none] (1) at (-1.5, 0) {};
		\node [style=none] (6) at (4.5, 0) {};
		\node [style=none] (7) at (-6.5, -1) {};
		\node [style=none] (8) at (-7.5, -1) {};
		\node [style=none] (9) at (-4.5, -2) {};
		\node [style=none] (10) at (-1.5, -2) {};
		\node [style=none] (11) at (1.5, -2) {};
		\node [style=none] (12) at (4.5, -2) {};
		\node [style=none] (13) at (6.5, -1) {};
		\node [style=none] (14) at (1.5, -2) {};
		\node [style=none] (15) at (4.5, -2) {};
		\node [style=none] (16) at (-6.5, -1) {};
		\node [style=none] (17) at (-4.5, 0) {};
		\node [style=none] (18) at (-1.5, 0) {};
		\node [style=none] (19) at (6.5, -1) {};
		\node [style=none] (20) at (7.5, -1) {};
		\node [style=none] (21) at (6.5, -1) {};
		\node [style=none] (22) at (-1.5, -2) {};
		\node [style=none] (23) at (1.5, 0) {};
		\node [style=none] (24) at (1.5, 0) {};
		\node [style=none] (25) at (4.5, 0) {};
		\node [style=none] (26) at (-2.25, 1) {};
		\node [style=none] (27) at (2.25, 1) {};
		\node [style=none] (30) at (0, 0.75) {$p(m,n)$};
		\draw [fill=green!90!white, line width=1pt,bend right=90, looseness=1.25] (0.center) to (1.center) to (0.center);
		\draw [fill=green!90!white, line width=1pt,bend right=90, looseness=1.25] (24.center) to (25.center) to (24.center);
		\draw [green!90!white,line width=2pt, style=dotted line, in=165, out=15, looseness=0.50] (26.center) to (27.center);
	\end{pgfonlayer}
	\begin{pgfonlayer}{edgelayer}
	\node [style=none, font={\large}] (28) at (-3, 0) {$\map U_m$};
	\node [style=none, font={\large}] (29) at (3, 0) {$\map U_n$};
		\draw [line width=1pt, bend left=90, looseness=1.25] (0.center) to (1.center);
		\draw [line width=1pt, bend right=90, looseness=1.25] (0.center) to (1.center);
		\draw [line width=1pt] (8.center) to (7.center);
		\draw [blue,line width=1pt, style=dashed line, in=165, out=0] (7.center) to (9.center);
		\draw [blue,line width=1pt, style=dashed line, bend right=15, looseness=0.50] (9.center) to (10.center);
		\draw [blue,line width=1pt, style=dashed line, in=0, out=-180] (13.center) to (6.center);
		\draw [red,line width=1pt, style=dashed line, in=-180, out=0] (16.center) to (17.center);
		\draw [red,line width=1pt, style=dashed line, in=165, out=0] (18.center) to (14.center);
		\draw [red,line width=1pt, style=dashed line, bend right=15, looseness=0.50] (14.center) to (15.center);
		\draw [red,line width=1pt, style=dashed line, in=15, out=180] (19.center) to (15.center);
		\draw [line width=1pt] (21.center) to (20.center);
		\draw [blue,line width=1pt, style=dashed line, in=15, out=180] (23.center) to (22.center);
		\draw [line width=1pt, bend left=90, looseness=1.25] (24.center) to (25.center);
		\draw [line width=1pt, bend right=90, looseness=1.25] (24.center) to (25.center);
		\node [style=none] (88) at (-6.5, -1) {$\bullet$};
		\node [style=none] (89) at (6.5, -1) {$\bullet$};
	\end{pgfonlayer}
\end{tikzpicture}

%% file: fig_freestyle_doublecorsup.tikz
\begin{tikzpicture}[scale=1]
	\begin{pgfonlayer}{nodelayer}
		\node [style=none] (0) at (-4.5, 0) {};
		\node [style=none] (1) at (-1.5, 0) {};
		\node [style=none] (6) at (4.5, 0) {};
		\node [style=none] (7) at (-6.5, -1.5) {};
		\node [style=none] (8) at (-7.5, -1.5) {};
		\node [style=none] (9) at (-4.5, -3) {};
		\node [style=none] (13) at (6.5, -1.5) {};
		\node [style=none] (14) at (1.5, -3) {};
		\node [style=none] (15) at (4.5, -3) {};
		\node [style=none] (16) at (-6.5, -1.5) {};
		\node [style=none] (17) at (-4.5, 0) {};
		\node [style=none] (18) at (-1.5, 0) {};
		\node [style=none] (19) at (6.5, -1.5) {};
		\node [style=none] (20) at (7.5, -1.5) {};
		\node [style=none] (21) at (6.5, -1.5) {};
		\node [style=none] (22) at (-1.5, -3) {};
		\node [style=none] (23) at (1.5, 0) {};
		\node [style=none] (24) at (1.5, 0) {};
		\node [style=none] (25) at (4.5, 0) {};
		\node [style=none] (26) at (-2.25, 1) {};
		\node [style=none] (27) at (2.25, 1) {};
		\node [style=none] (31) at (-4.5, -3) {};
		\node [style=none] (32) at (-1.5, -3) {};
		\node [style=none] (33) at (4.5, -3) {};
		\node [style=none] (34) at (-4.5, -3) {};
		\node [style=none] (35) at (-1.5, -3) {};
		\node [style=none] (36) at (1.5, -3) {};
		\node [style=none] (37) at (1.5, -3) {};
		\node [style=none] (38) at (4.5, -3) {};
		\node [style=none] (39) at (-2.25, -4) {};
		\node [style=none] (40) at (2.25, -4) {};
		\draw [green!90!white,line width=2pt, style=dotted line, in=165, out=15, looseness=0.50] (26.center) to (27.center);
		\draw [red!90!white,line width=2pt, style=dotted line, in=-165, out=-15, looseness=0.50] (39.center) to (40.center);
		\draw [fill=green!90!white, line width=1pt,bend right=90, looseness=1.25] (0.center) to (1.center) to (0.center);
		\draw [fill=green!90!white, line width=1pt,bend right=90, looseness=1.25] (24.center) to (25.center) to (24.center);
		\draw [fill=red!90!white, line width=1pt,bend right=90, looseness=1.25] (31.center) to (32.center) to (31.center);
		\draw [fill=red!90!white, line width=1pt,bend right=90, looseness=1.25] (37.center) to (38.center) to (37.center);
		\node [style=none] (98) at (0, 0.75) {$p_A(m,n)$};
		\node [style=none] (99) at (0, -3.75) {$p_B(k,l)$};
	\end{pgfonlayer}
	\begin{pgfonlayer}{edgelayer}
	\node [style=none, font={\large}] (41) at (-3, -3) {$\map U_{k}$};
	\node [style=none, font={\large}] (44) at (3, -3) {$\map U_{l}$};
	\node [style=none, font={\large}] (28) at (-3, 0) {$\map U_{m}$};
	\node [style=none, font={\large}] (29) at (3, 0) {$\map U_{n}$};
		\draw [line width=1pt, bend left=90, looseness=1.25] (0.center) to (1.center);
		\draw [line width=1pt, bend right=90, looseness=1.25] (0.center) to (1.center);
		\draw [line width=1pt] (8.center) to (7.center);
		\draw [blue, line width=1pt, style=dashed line, in=0, out=-180] (13.center) to (6.center);
		\draw [orange, line width=1pt, style=dashed line, in=-180, out=0] (16.center) to (17.center);
		\draw [orange, line width=1pt, style=dashed line, in=180, out=0] (18.center) to (14.center);
		\draw [orange, line width=1pt, style=dashed line, in=0, out=180] (19.center) to (15.center);
		\draw [line width=1pt] (21.center) to (20.center);
		\draw [blue, line width=1pt, style=dashed line, in=0, out=180] (23.center) to (22.center);
		\draw [line width=1pt, bend left=90, looseness=1.25] (24.center) to (25.center);
		\draw [line width=1pt, bend right=90, looseness=1.25] (24.center) to (25.center);
		\draw [line width=1pt, bend right=90, looseness=1.25] (31.center) to (32.center);
		\draw [line width=1pt, bend left=90, looseness=1.25] (31.center) to (32.center);
		\draw [line width=1pt, bend right=90, looseness=1.25] (37.center) to (38.center);
		\draw [line width=1pt, bend left=90, looseness=1.25] (37.center) to (38.center);
		\draw [blue,line width=1pt,style=dashed line, in=0, out=-180] (34.center) to (16.center);
		\node [style=none] (88) at (-6.5, -1.5) {$\bullet$};
		\node [style=none] (89) at (6.5, -1.5) {$\bullet$};
	\end{pgfonlayer}
\end{tikzpicture}

%% file: fig_cor.tikz
\begin{tikzpicture}[scale=0.95, circuit ee IEC]
	\begin{pgfonlayer}{nodelayer}
	\draw[fill=green!90!white] (-7.5,1) -- (-7.5,3) -- (-2.5,3) -- (-2.5,1) -- (-4,1) -- (-4,2.5) -- (-6,2.5) -- (-6,1) -- (-7.5,1) -- (-7.5,3) ;
	\draw[fill=green!90!white] (2,1) -- (2,3) -- (3.5,3) -- (3.5,1) -- (2,1) -- (2,3);
	\draw[fill=green!90!white] (5.5,1) -- (5.5,3) -- (7,3) -- (7,1) -- (5.5,1) -- (5.5,3);
		\node [style=none] (0) at (-4, 3) {};
		\node [style=none] (1) at (-4, 1) {};
		\node [style=none] (2) at (-2.5, 3) {};
		\node [style=none] (3) at (-2.5, 1) {};
		\node [style=none] (4) at (-6, 3) {};
		\node [style=none] (5) at (-6, 1) {};
		\node [style=none] (6) at (-7.5, 1) {};
		\node [style=none] (7) at (-7.5, 3) {};
		\node [style=none] (8) at (-7.5, 2.5) {};
		\node [style=none] (9) at (-7.5, 1.5) {};
		\node [style=none] (10) at (-6, 2.5) {};
		\node [style=none] (11) at (-6, 1.5) {};
		\node [style=none] (12) at (-4, 2.5) {};
		\node [style=none] (13) at (-2.5, 2.5) {};
		\node [style=none] (14) at (-2.5, 1.5) {};
		\node [style=none, font={\footnotesize}] (15) at (-8.25, 1) {$S^{(1)}$};
		\node [style=none, font={\footnotesize}] (16) at (-1.75, -0.5) {$S^{(1)}$};
		\node [style=none] (17) at (-1.5, 1.5) {};
		\node [style=none, font={\large}] (18) at (-3.25, 1.75) {};
		\node [style=none, font={\large}] (19) at (-6.75, 1.75) {$\map{B}$};
		\node [style=none, font={\large}] (20) at (2.75, 2) {$\map{W}_{1}$};
		\node [style=none] (21) at (3.5, 3) {};
		\node [style=none] (22) at (2, 2.5) {};
		\node [style=none] (23) at (5.5, 3) {};
		\node [style=none] (24) at (7, 3) {};
		\node [style=none] (25) at (2, 1.5) {};
		\node [style=none] (26) at (7, 1.5) {};
		\node [style=none] (27) at (7, 2.5) {};
		\node [style=none] (28) at (2, 1) {};
		\node [style=none] (29) at (2, 3) {};
		\node [style=none] (30) at (3.5, 3) {};
		\node [style=none] (31) at (3.5, 1) {};
		\node [style=none] (32) at (5.5, 3) {};
		\node [style=none] (33) at (8, 1.5) {};
		\node [style=none] (34) at (5.5, 1) {};
		\node [style=none] (35) at (7, 1) {};
		\node [style=none] (37) at (1, 3) {};
		\node [style=none] (38) at (1, 2.5) {};
		\node [style=none] (39) at (1, 2) {};
		\node [style=none, font={\large}] (40) at (6.25, 2) {$\map{W}_{2}$};
		\node [style=none] (41) at (3.5, 2.5) {};
		\node [style=none] (42) at (5.5, 2.5) {};
		\node [style=none] (43) at (7.75, 2.5) {};
		\node [style=none, font={\large}] (44) at (-0.75, 1.25) {$=$};
		\node [style=none, ground=none, xshift=-0.45em] (45) at (8, 2.5) {};
		\node [style=none] (46) at (7.5, 3) {$E$};
		\node [style=none] (47) at (4.5, 3) {$E$};
		\node [style=none] (48) at (-8.5, 1.5) {};
		\node [style=none] (49) at (0, 1.5) {};
		\node [style=none] (50) at (-4, 0) {};
		\node [style=none] (51) at (-1.5, 0) {};
		\node [style=none] (52) at (-8.5, 0) {};
		\node [style=none] (53) at (-6, 0) {};
		\node [style=none] (54) at (-6, 0) {};
		\node [style=none] (55) at (-4, 1.5) {};
		\node [style=none, font={\footnotesize}] (56) at (-1.75, 1) {$S^{(2)}$};
		\node [style=none, font={\footnotesize}] (57) at (-8.25, -0.5) {$S^{(2)}$};
		\node [style=none] (58) at (0, 0) {};
		\node [style=none] (59) at (3.5, 1.5) {};
		\node [style=none] (60) at (3.5, 0) {};
		\node [style=none, font={\footnotesize}] (61) at (0.5, -0.5) {$S^{(2)}$};
		\node [style=none, font={\footnotesize}] (62) at (7.75, -0.5) {$S^{(1)}$};
		\node [style=none] (63) at (5.5, 0) {};
		\node [style=none] (64) at (5.5, 1.5) {};
		\node [style=none] (65) at (3.5, 0) {};
		\node [style=none] (66) at (8, 0) {};
		\node [style=none, font={\footnotesize}] (67) at (0.5, 1) {$S^{(1)}$};
		\node [style=none, font={\footnotesize}] (68) at (7.75, 1) {$S^{(2)}$}; : 
		\node [style=none] (69) at (1.5, 3) {$E$};
		\draw [fill={green!90!white}, bend right=90, looseness=3.25] (37.center) to (39.center);
	\end{pgfonlayer}
	\begin{pgfonlayer}{edgelayer}
		\node [style=none] (36) at (0.5, 2.5) {$\, \, \eta$};
		\draw (7.center) to (4.center);
		\draw (5.center) to (6.center);
		\draw (6.center) to (7.center);
		\draw (0.center) to (2.center);
		\draw (2.center) to (3.center);
		\draw (3.center) to (1.center);
		\draw (14.center) to (17.center);
		\draw (5.center) to (10.center);
		\draw (10.center) to (12.center);
		\draw (12.center) to (1.center);
		\draw (4.center) to (0.center);
		\draw (29.center) to (21.center);
		\draw (31.center) to (28.center);
		\draw (28.center) to (29.center);
		\draw (32.center) to (24.center);
		\draw (24.center) to (35.center);
		\draw (35.center) to (34.center);
		\draw (26.center) to (33.center);
		\draw (31.center) to (30.center);
		\draw (23.center) to (34.center);
		\draw (37.center) to (39.center);
		\draw (41.center) to (42.center);
		\draw (38.center) to (22.center);
		\draw (27.center) to (43.center);
		\draw (48.center) to (9.center);
		\draw (49.center) to (25.center);
		\draw [in=180, out=0, looseness=1.25] (11.center) to (50.center);
		\draw (50.center) to (51.center);
		\draw (52.center) to (53.center);
		\draw [in=180, out=0, looseness=1.25] (54.center) to (55.center);
		\draw [in=180, out=0, looseness=1.25] (59.center) to (63.center);
		\draw (63.center) to (66.center);
		\draw (58.center) to (65.center);
		\draw [in=180, out=0, looseness=1.25] (60.center) to (64.center);
	\end{pgfonlayer}
\end{tikzpicture}

%% file: fig_singlecorsup_space.tikz
\begin{tikzpicture}[scale=0.95]
	\begin{pgfonlayer}{nodelayer}
	\draw[style=dashed,fill=green!90!white, draw=none] (-4,-1.5) -- (-4,1.5) -- (0,1.5) -- (0,-1.5) -- (-4,-1.5) -- (-4,1.5) ;
		\node [style=none] (0) at (1.5, 1.5) {};
		\node [style=none] (1) at (1.5, -1.5) {};
		\node [style=none] (2) at (3, 1.5) {};
		\node [style=none] (3) at (3, -1.5) {};
		\node [style=none] (4) at (-5.5, 1.5) {};
		\node [style=none] (5) at (-5.5, -1.5) {};
		\node [style=none] (6) at (-7, -1.5) {};
		\node [style=none] (7) at (-7, 1.5) {};
		\node [style=none] (8) at (-7, 0.75) {};
		\node [style=none] (9) at (-7, -0.75) {};
		\node [style=none] (10) at (-7.75, -0.75) {};
		\node [style=none] (11) at (-5.5, 0.75) {};
		\node [style=none] (12) at (-4, -0.75) {};
		\node [style=none] (13) at (3, 0.75) {};
		\node [style=none] (14) at (3, -0.75) {};
		\node [style=none] (15) at (-8.75, 0.75) {};
		\node [style=none] (16) at (-7.75, 1.25) {$M$};
		\node [style=none] (17) at (-4.75, 1.25) {$\widetilde{S}^{(1)}$};
		\node [style=none] (91) at (-4.75, -1.25) {$\widetilde{S}^{(2)}$};
		\node [style=none] (18) at (3.75, 1.25) {$M$};
		\node [style=none] (19) at (-7.5, -1.25) {$~C$};
		\node [style=none] (21) at (3.75, -1.25) {$C$};
		\node [style=none, font={\large}] (22) at (-6.25, 0) {$\map{U}$};
		\node [style=none, font={\large}] (23) at (2.25, 0) {$\map{U^{\dagger}}$};
		\node [style=none] (24) at (-7.75, -0.25) {};
		\node [style=none] (25) at (-7.75, -1.25) {};
		\node [style=none] (26) at (-8.25, -0.75) {$\, \, \omega$};
		\node [style=none] (27) at (4.5, 0.75) {};
		\node [style=none] (28) at (4.5, -0.75) {};
		\node [style=none] (32) at (1.5, 0.75) {};
		\node [style=none] (34) at (0, 0.75) {};
		\node [style=none] (37) at (-4, 0.75) {};
		\node [style=none] (38) at (-4, -1.5) {};
		\node [style=none] (41) at (0, -1.5) {};
		\node [style=none] (42) at (0, 1.5) {};
		\node [style=none] (43) at (0, 1.5) {};
		\node [style=none] (44) at (-4, 1.5) {};
		\node [style=none, font={\large}] (48) at (-2, 0) {$\map{\widetilde{B}}$};
		\node [style=none] (70) at (0.75, 1.25) {$\widetilde{S}^{(1)}$};
		\node [style=none] (71) at (0.75, -1.25) {$\widetilde{S}^{(2)}$};
		\node [style=none] (73) at (-5.5, -0.75) {};
		\node [style=none] (92) at (0, -0.75) {};
		\node [style=none] (93) at (1.5, -0.75) {};
	\end{pgfonlayer}
	\begin{pgfonlayer}{edgelayer}
		\draw (7.center) to (4.center);
		\draw (4.center) to (5.center);
		\draw (5.center) to (6.center);
		\draw (6.center) to (7.center);
		\draw (0.center) to (2.center);
		\draw (2.center) to (3.center);
		\draw (3.center) to (1.center);
		\draw (1.center) to (0.center);
		\draw (9.center) to (10.center);
		\draw [bend right=90, looseness=3.25] (24.center) to (25.center);
		\draw (24.center) to (25.center);
		\draw (15.center) to (8.center);
		\draw (13.center) to (27.center);
		\draw (14.center) to (28.center);
		\draw (44.center) to (43.center);
		\draw (41.center) to (38.center);
		\draw (38.center) to (44.center);
		\draw (41.center) to (42.center);
		\draw (11.center) to (37.center);
		\draw (73.center) to (12.center);
		\draw (34.center) to (32.center);
		\draw (92.center) to (93.center);
	\end{pgfonlayer}
\end{tikzpicture}

%% file: fig_singlecorsup.tikz
\begin{tikzpicture}[scale=0.95]
	\begin{pgfonlayer}{nodelayer}
		\draw[style=dashed,fill=green!90!white, draw=none] (-4,0) -- (-4,1.5) -- (-1.5,1.5) -- (-1.5,0) -- (-4,0) -- (-4,1.5) ;
		\draw[style=dashed,fill=green!90!white,draw=none] (1.5,0) -- (1.5,1.5) -- (4,1.5) -- (4,0) -- (1.5,0) -- (1.5,1.5) ;
		\draw[style=dashed,fill=green!90!white,draw=none]  (-4,1.5) -- (-4,2)  -- (4,2) -- (4,1.5) -- (-4,1.5) -- (-4,2)   ;
		\node [style=none] (0) at (5.5, 1.5) {};
		\node [style=none] (1) at (5.5, -1.5) {};
		\node [style=none] (2) at (7, 1.5) {};
		\node [style=none] (3) at (7, -1.5) {};
		\node [style=none] (4) at (-5.5, 1.5) {};
		\node [style=none] (5) at (-5.5, -1.5) {};
		\node [style=none] (6) at (-7, -1.5) {};
		\node [style=none] (7) at (-7, 1.5) {};
		\node [style=none] (8) at (-7, 0.5) {};
		\node [style=none] (9) at (-7, -0.75) {};
		\node [style=none] (10) at (-7.75, -0.75) {};
		\node [style=none] (11) at (-5.5, 0.5) {};
		\node [style=none] (12) at (-1.5, -0.75) {};
		\node [style=none] (13) at (7, 0.5) {};
		\node [style=none] (14) at (7, -0.75) {};
		\node [style=none] (15) at (-8.75, 0.5) {};
		\node [style=none] (16) at (-7.75, 1) {$M$};
		\node [style=none] (17) at (-4.75, 1) {$\widetilde{S}^{(1)}$};
		\node [style=none] (91) at (-4.75, -1.25) {$\widetilde{S}^{(2)}$};
		\node [style=none] (18) at (7.75, 1) {$M$};
		\node [style=none] (19) at (-7.5, -1.25) {$~C$};
		\node [style=none] (21) at (7.75, -1.25) {$C$};
		\node [style=none, font={\large}] (22) at (-6.25, 0) {$\map{U}$};
		\node [style=none, font={\large}] (23) at (6.25, 0) {$\map{U^{\dagger}}$};
		\node [style=none] (24) at (-7.75, -0.25) {};
		\node [style=none] (25) at (-7.75, -1.25) {};
		\node [style=none] (26) at (-8.25, -0.75) {$\, \, \omega$};
		\node [style=none] (27) at (8.5, 0.5) {};
		\node [style=none] (28) at (8.5, -0.75) {};
		\node [style=none] (32) at (5.5, 0.5) {};
		\node [style=none] (33) at (1.5, -0.75) {};
		\node [style=none] (34) at (-1.5, 0.5) {};
		\node [style=none] (35) at (4, 2) {};
		\node [style=none] (37) at (-4, 0.5) {};
		\node [style=none] (38) at (-4, 0) {};
		\node [style=none] (39) at (1.5, 2) {};
		\node [style=none] (40) at (1.5, 1.5) {};
		\node [style=none] (41) at (-1.5, 0) {};
		\node [style=none] (42) at (-1.5, 1.5) {};
		\node [style=none] (43) at (-1.5, 2) {};
		\node [style=none] (44) at (-4, 2) {};
		\node [style=none] (45) at (4, 1.5) {};
		\node [style=none] (46) at (4, 0) {};
		\node [style=none] (47) at (-4, 1.5) {};
		\node [style=none, font={\large}] (48) at (-2.75, 1) {$\map{\widetilde{B}}$};
		\node [style=none] (49) at (1.5, 0) {};
		\node [style=none] (50) at (4, 0.5) {};
		\node [style=none] (68) at (1.5, 0.5) {};
		\node [style=none] (70) at (-0.75, 1) {$\widetilde{S}^{(1)}$};
		\node [style=none] (71) at (4.75, 1) {$\widetilde{S}^{(2)}$};
		\node [style=none] (72) at (0.75, 1) {$\widetilde{S}^{(2)}$};
		\node [style=none] (92) at (4.75, -1.25) {$\widetilde{S}^{(1)}$};
		\node [style=none] (73) at (-5.5, -0.75) {};
		\node [style=none] (74) at (5.5, -0.75) {};
	\end{pgfonlayer}
	\begin{pgfonlayer}{edgelayer}
		\draw (7.center) to (4.center);
		\draw (4.center) to (5.center);
		\draw (5.center) to (6.center);
		\draw (6.center) to (7.center);
		\draw (0.center) to (2.center);
		\draw (2.center) to (3.center);
		\draw (3.center) to (1.center);
		\draw (1.center) to (0.center);
		\draw (9.center) to (10.center);
		\draw [bend right=90, looseness=3.25] (24.center) to (25.center);
		\draw (24.center) to (25.center);
		\draw (15.center) to (8.center);
		\draw (13.center) to (27.center);
		\draw (14.center) to (28.center);
		\draw (44.center) to (43.center);
		\draw (41.center) to (38.center);
		\draw (38.center) to (44.center);
		\draw (39.center) to (35.center);
		\draw (35.center) to (46.center);
		\draw (46.center) to (49.center);
		\draw (41.center) to (42.center);
		\draw (42.center) to (40.center);
		\draw (40.center) to (49.center);
		\draw (43.center) to (39.center);
		\draw (11.center) to (37.center);
		\draw (50.center) to (32.center);
		\draw [in=180, out=0] (12.center) to (68.center);
		\draw [in=-180, out=0] (34.center) to (33.center);
		\draw (73.center) to (12.center);
		\draw (74.center) to (33.center);
	\end{pgfonlayer}
\end{tikzpicture}

%% file: fig_doublecor.tikz
\begin{tikzpicture}[scale=0.95]
	\begin{pgfonlayer}{nodelayer}
	\draw[style=dashed,fill=green!90!white, draw=none] (-4,0) -- (-4,1.5) -- (-1.5,1.5) -- (-1.5,0) -- (-4,0) -- (-4,1.5) ;
	\draw[style=dashed,fill=green!90!white,draw=none] (1.5,0) -- (1.5,1.5) -- (4,1.5) -- (4,0) -- (1.5,0) -- (1.5,1.5) ;
	\draw[style=dashed,fill=green!90!white,draw=none]  (-4,1.5) -- (-4,2)  -- (4,2) -- (4,1.5) -- (-4,1.5) -- (-4,2)  ;
	\draw[style=dashed,fill=red!90!white, draw=none] (-4,-2.5) -- (-4,-1) -- (-1.5,-1) -- (-1.5,-2.5) -- (-4,-2.5) -- (-4,-1) ;
	\draw[style=dashed,fill=red!90!white,draw=none] (1.5,-2.5) -- (1.5,-1) -- (4,-1) -- (4,-2.5) -- (1.5,-2.5) -- (1.5,-2.5) ;
	\draw[style=dashed,fill=red!90!white,draw=none]  (-4,-3) -- (-4,-2.5)  -- (4,-2.5) -- (4,-3) -- (-4,-3) -- (-4,-2.5)   ;
		\node [style=none] (11) at (-5.75, 0.5) {};
		\node [style=none] (12) at (-1.5, -1.5) {};
		\node [style=none] (17) at (-4.75, 1) {$\widetilde{A}^{(1)}$};
		\node [style=none] (32) at (5.5, 0.5) {};
		\node [style=none] (33) at (1.5, -1.5) {};
		\node [style=none] (34) at (-1.5, 0.5) {};
		\node [style=none] (35) at (4, 2) {};
		\node [style=none] (37) at (-4, 0.5) {};
		\node [style=none] (38) at (-4, 0) {};
		\node [style=none] (39) at (1.5, 2) {};
		\node [style=none] (40) at (1.5, 1.5) {};
		\node [style=none] (41) at (-1.5, 0) {};
		\node [style=none] (42) at (-1.5, 1.5) {};
		\node [style=none] (43) at (-1.5, 2) {};
		\node [style=none] (44) at (-4, 2) {};
		\node [style=none] (45) at (4, 1.5) {};
		\node [style=none] (46) at (4, 0) {};
		\node [style=none] (47) at (-4, 1.5) {};
		\node [style=none, font={\large}] (48) at (-2.75, 1) {$\map{\widetilde{A}}$};
		\node [style=none] (49) at (1.5, 0) {};
		\node [style=none] (50) at (4, 0.5) {};
		\node [style=none] (68) at (1.5, 0.5) {};
		\node [style=none] (70) at (-0.75, 1) {$\widetilde{A}^{(1)}$};
		\node [style=none] (71) at (4.75, 1) {$\widetilde{A}^{(2)}$};
		\node [style=none] (72) at (0.75, 1) {$\widetilde{A}^{(2)}$};
		\node [style=none] (74) at (5.5, -1.5) {};
		\node [style=none] (75) at (-1.5, -1.5) {};
		\node [style=none] (76) at (4, -3) {};
		\node [style=none] (78) at (-4, -1.5) {};
		\node [style=none] (79) at (-4, -1) {};
		\node [style=none] (80) at (1.5, -3) {};
		\node [style=none] (81) at (1.5, -2.5) {};
		\node [style=none] (82) at (-1.5, -1) {};
		\node [style=none] (83) at (-1.5, -2.5) {};
		\node [style=none] (84) at (-1.5, -3) {};
		\node [style=none] (85) at (-4, -3) {};
		\node [style=none] (86) at (4, -2.5) {};
		\node [style=none] (87) at (4, -1) {};
		\node [style=none] (88) at (-4, -2.5) {};
		\node [style=none, font={\large}] (89) at (-2.75, -2) {$\map{\widetilde{B}}$};
		\node [style=none] (90) at (1.5, -1) {};
		\node [style=none] (91) at (4, -1.5) {};
		\node [style=none] (92) at (1.5, -1.5) {};
		\node [style=none] (93) at (-0.75, -2) {$\widetilde{B}^{(1)}$};
		\node [style=none] (94) at (0.75, -2) {$\widetilde{B}^{(2)}$};
		\node [style=none] (95) at (-4.75, -2) {$\widetilde{B}^{(1)}$};
		\node [style=none] (96) at (4.75, -2) {$\widetilde{B}^{(2)}$};
		\node [style=none] (97) at (-5.75, -1.5) {};
	\end{pgfonlayer}
	\begin{pgfonlayer}{edgelayer}
		\draw (44.center) to (43.center);
		\draw (41.center) to (38.center);
		\draw (38.center) to (44.center);
		\draw (39.center) to (35.center);
		\draw (35.center) to (46.center);
		\draw (46.center) to (49.center);
		\draw (41.center) to (42.center);
		\draw (42.center) to (40.center);
		\draw (40.center) to (49.center);
		\draw (43.center) to (39.center);
		\draw (11.center) to (37.center);
		\draw (50.center) to (32.center);
		\draw [in=180, out=0] (12.center) to (68.center);
		\draw [in=-180, out=0] (34.center) to (33.center);
		\draw (85.center) to (84.center);
		\draw (82.center) to (79.center);
		\draw (79.center) to (85.center);
		\draw (80.center) to (76.center);
		\draw (76.center) to (87.center);
		\draw (87.center) to (90.center);
		\draw (82.center) to (83.center);
		\draw (83.center) to (81.center);
		\draw (81.center) to (90.center);
		\draw (84.center) to (80.center);
		\draw (91.center) to (74.center);
		\draw (97.center) to (78.center);
	\end{pgfonlayer}
\end{tikzpicture}

%% file: fig_doublecorsup.tikz
\begin{tikzpicture}[scale=0.95]
	\begin{pgfonlayer}{nodelayer}
	\draw[style=dashed,fill=green!90!white, draw=none] (-4,0) -- (-4,1.5) -- (-1.5,1.5) -- (-1.5,0) -- (-4,0) -- (-4,1.5) ;
	\draw[style=dashed,fill=green!90!white,draw=none] (1.5,0) -- (1.5,1.5) -- (4,1.5) -- (4,0) -- (1.5,0) -- (1.5,1.5) ;
	\draw[style=dashed,fill=green!90!white,draw=none]  (-4,1.5) -- (-4,2)  -- (4,2) -- (4,1.5) -- (-4,1.5) -- (-4,2)  ;
	\draw[style=dashed,fill=red!90!white, draw=none] (-4,-2.5) -- (-4,-1) -- (-1.5,-1) -- (-1.5,-2.5) -- (-4,-2.5) -- (-4,-1) ;
	\draw[style=dashed,fill=red!90!white,draw=none] (1.5,-2.5) -- (1.5,-1) -- (4,-1) -- (4,-2.5) -- (1.5,-2.5) -- (1.5,-2.5) ;
	\draw[style=dashed,fill=red!90!white,draw=none]  (-4,-3) -- (-4,-2.5)  -- (4,-2.5) -- (4,-3) -- (-4,-3) -- (-4,-2.5)   ;
		\node [style=none] (0) at (5.5, 1.5) {};
		\node [style=none] (1) at (5.5, -2.5) {};
		\node [style=none] (2) at (7, 1.5) {};
		\node [style=none] (3) at (7, -2.5) {};
		\node [style=none] (4) at (-6, 1.5) {};
		\node [style=none] (5) at (-6, -2.5) {};
		\node [style=none] (6) at (-7.5, -2.5) {};
		\node [style=none] (7) at (-7.5, 1.5) {};
		\node [style=none] (8) at (-7.5, 0.5) {};
		\node [style=none] (9) at (-7.5, -1.5) {};
		\node [style=none] (10) at (-8.25, -1.5) {};
		\node [style=none] (11) at (-6, 0.5) {};
		\node [style=none] (12) at (-1.5, -1.5) {};
		\node [style=none] (13) at (7, 0.5) {};
		\node [style=none] (14) at (7, -1.5) {};
		\node [style=none] (15) at (-9.25, 0.5) {};
		\node [style=none] (16) at (-8.25, 1) {$M$};
		\node [style=none] (17) at (-5, 1) {$\widetilde{A}^{(1)}$};
		\node [style=none] (18) at (7.75, 1) {$M$};
		\node [style=none] (19) at (-8, -2) {$~C$};
		\node [style=none] (21) at (7.75, -2) {$C$};
		\node [style=none, font={\large}] (22) at (-6.75, -0.5) {$\map{U}$};
		\node [style=none, font={\large}] (23) at (6.25, -0.5) {$\map{U^{\dagger}}$};
		\node [style=none] (24) at (-8.25, -1) {};
		\node [style=none] (25) at (-8.25, -2) {};
		\node [style=none] (26) at (-8.75, -1.5) {$\, \, \omega$};
		\node [style=none] (27) at (8.5, 0.5) {};
		\node [style=none] (28) at (8.5, -1.5) {};
		\node [style=none] (32) at (5.5, 0.5) {};
		\node [style=none] (33) at (1.5, -1.5) {};
		\node [style=none] (34) at (-1.5, 0.5) {};
		\node [style=none] (35) at (4, 2) {};
		\node [style=none] (37) at (-4, 0.5) {};
		\node [style=none] (38) at (-4, 0) {};
		\node [style=none] (39) at (1.5, 2) {};
		\node [style=none] (40) at (1.5, 1.5) {};
		\node [style=none] (41) at (-1.5, 0) {};
		\node [style=none] (42) at (-1.5, 1.5) {};
		\node [style=none] (43) at (-1.5, 2) {};
		\node [style=none] (44) at (-4, 2) {};
		\node [style=none] (45) at (4, 1.5) {};
		\node [style=none] (46) at (4, 0) {};
		\node [style=none] (47) at (-4, 1.5) {};
		\node [style=none, font={\large}] (48) at (-2.75, 1) {$\map{\widetilde{A}}$};
		\node [style=none] (49) at (1.5, 0) {};
		\node [style=none] (50) at (4, 0.5) {};
		\node [style=none] (68) at (1.5, 0.5) {};
		\node [style=none] (70) at (-0.75, 1) {$\widetilde{A}^{(1)}$};
		\node [style=none] (71) at (4.75, 1) {$\widetilde{A}^{(2)}$};
		\node [style=none] (72) at (0.75, 1) {$\widetilde{A}^{(2)}$};
		\node [style=none] (73) at (-6, -1.5) {};
		\node [style=none] (74) at (5.5, -1.5) {};
		\node [style=none] (75) at (-1.5, -1.5) {};
		\node [style=none] (76) at (4, -3) {};
		\node [style=none] (78) at (-4, -1.5) {};
		\node [style=none] (79) at (-4, -1) {};
		\node [style=none] (80) at (1.5, -3) {};
		\node [style=none] (81) at (1.5, -2.5) {};
		\node [style=none] (82) at (-1.5, -1) {};
		\node [style=none] (83) at (-1.5, -2.5) {};
		\node [style=none] (84) at (-1.5, -3) {};
		\node [style=none] (85) at (-4, -3) {};
		\node [style=none] (86) at (4, -2.5) {};
		\node [style=none] (87) at (4, -1) {};
		\node [style=none] (88) at (-4, -2.5) {};
		\node [style=none, font={\large}] (89) at (-2.75, -2) {$\map{\widetilde{B}}$};
		\node [style=none] (90) at (1.5, -1) {};
		\node [style=none] (91) at (4, -1.5) {};
		\node [style=none] (92) at (1.5, -1.5) {};
		\node [style=none] (93) at (-0.75, -2) {$\widetilde{B}^{(1)}$};
		\node [style=none] (94) at (0.75, -2) {$\widetilde{B}^{(2)}$};
		\node [style=none] (95) at (-5, -2) {$\widetilde{B}^{(1)}$};
		\node [style=none] (96) at (4.75, -2) {$\widetilde{B}^{(2)}$};
	\end{pgfonlayer}
	\begin{pgfonlayer}{edgelayer}
		\draw (7.center) to (4.center);
		\draw (4.center) to (5.center);
		\draw (5.center) to (6.center);
		\draw (6.center) to (7.center);
		\draw (0.center) to (2.center);
		\draw (2.center) to (3.center);
		\draw (3.center) to (1.center);
		\draw (1.center) to (0.center);
		\draw (9.center) to (10.center);
		\draw [bend right=90, looseness=3.25] (24.center) to (25.center);
		\draw (24.center) to (25.center);
		\draw (15.center) to (8.center);
		\draw (13.center) to (27.center);
		\draw (14.center) to (28.center);
		\draw (44.center) to (43.center);
		\draw (41.center) to (38.center);
		\draw (38.center) to (44.center);
		\draw (39.center) to (35.center);
		\draw (35.center) to (46.center);
		\draw (46.center) to (49.center);
		\draw (41.center) to (42.center);
		\draw (42.center) to (40.center);
		\draw (40.center) to (49.center);
		\draw (43.center) to (39.center);
		\draw (11.center) to (37.center);
		\draw (50.center) to (32.center);
		\draw [in=180, out=0] (12.center) to (68.center);
		\draw [in=-180, out=0] (34.center) to (33.center);
		\draw (85.center) to (84.center);
		\draw (82.center) to (79.center);
		\draw (79.center) to (85.center);
		\draw (80.center) to (76.center);
		\draw (76.center) to (87.center);
		\draw (87.center) to (90.center);
		\draw (82.center) to (83.center);
		\draw (83.center) to (81.center);
		\draw (81.center) to (90.center);
		\draw (84.center) to (80.center);
		\draw (73.center) to (78.center);
		\draw (91.center) to (74.center);
	\end{pgfonlayer}
\end{tikzpicture}

%% file: correlations_final_final_arxiv.bbl
\begin{thebibliography}{75}%
	\makeatletter
	\providecommand \@ifxundefined [1]{%
		\@ifx{#1\undefined}
	}%
	\providecommand \@ifnum [1]{%
		\ifnum #1\expandafter \@firstoftwo
		\else \expandafter \@secondoftwo
		\fi
	}%
	\providecommand \@ifx [1]{%
		\ifx #1\expandafter \@firstoftwo
		\else \expandafter \@secondoftwo
		\fi
	}%
	\providecommand \natexlab [1]{#1}%
	\providecommand \enquote  [1]{``#1''}%
	\providecommand \bibnamefont  [1]{#1}%
	\providecommand \bibfnamefont [1]{#1}%
	\providecommand \citenamefont [1]{#1}%
	\providecommand \href@noop [0]{\@secondoftwo}%
	\providecommand \href [0]{\begingroup \@sanitize@url \@href}%
	\providecommand \@href[1]{\@@startlink{#1}\@@href}%
	\providecommand \@@href[1]{\endgroup#1\@@endlink}%
	\providecommand \@sanitize@url [0]{\catcode `\\12\catcode `\$12\catcode
		`\&12\catcode `\#12\catcode `\^12\catcode `\_12\catcode `\%12\relax}%
	\providecommand \@@startlink[1]{}%
	\providecommand \@@endlink[0]{}%
	\providecommand \url  [0]{\begingroup\@sanitize@url \@url }%
	\providecommand \@url [1]{\endgroup\@href {#1}{\urlprefix }}%
	\providecommand \urlprefix  [0]{URL }%
	\providecommand \Eprint [0]{\href }%
	\providecommand \doibase [0]{http://dx.doi.org/}%
	\providecommand \selectlanguage [0]{\@gobble}%
	\providecommand \bibinfo  [0]{\@secondoftwo}%
	\providecommand \bibfield  [0]{\@secondoftwo}%
	\providecommand \translation [1]{[#1]}%
	\providecommand \BibitemOpen [0]{}%
	\providecommand \bibitemStop [0]{}%
	\providecommand \bibitemNoStop [0]{.\EOS\space}%
	\providecommand \EOS [0]{\spacefactor3000\relax}%
	\providecommand \BibitemShut  [1]{\csname bibitem#1\endcsname}%
	\let\auto@bib@innerbib\@empty
	\bibitem [{\citenamefont {Bennett}\ and\ \citenamefont
		{Brassard}(1984)}]{BennettCh1984}%
	\BibitemOpen
	\bibfield  {author} {\bibinfo {author} {\bibfnamefont {C.~H.}\ \bibnamefont
			{Bennett}}\ and\ \bibinfo {author} {\bibfnamefont {G.}~\bibnamefont
			{Brassard}},\ }in\ \href@noop {} {\emph {\bibinfo {booktitle} {Int. Conf. on
				Computers, Systems and Signal Processing (Bangalore, India, Dec. 1984)}}}\
	(\bibinfo {year} {1984})\ pp.\ \bibinfo {pages} {175--9}\BibitemShut
	{NoStop}%
	\bibitem [{\citenamefont {Ekert}(1991)}]{ekert1991quantum}%
	\BibitemOpen
	\bibfield  {author} {\bibinfo {author} {\bibfnamefont {A.~K.}\ \bibnamefont
			{Ekert}},\ }\href@noop {} {\bibfield  {journal} {\bibinfo  {journal}
			{Physical Review Letters}\ }\textbf {\bibinfo {volume} {67}},\ \bibinfo
		{pages} {661} (\bibinfo {year} {1991})}\BibitemShut {NoStop}%
	\bibitem [{\citenamefont {Shor}(1995)}]{shor95error}%
	\BibitemOpen
	\bibfield  {author} {\bibinfo {author} {\bibfnamefont {P.~W.}\ \bibnamefont
			{Shor}},\ }\href@noop {} {\bibfield  {journal} {\bibinfo  {journal} {Physical
				Review A}\ }\textbf {\bibinfo {volume} {52}},\ \bibinfo {pages} {R2493}
		(\bibinfo {year} {1995})}\BibitemShut {NoStop}%
	\bibitem [{\citenamefont {Gottesman}(2010)}]{gottesman2010introduction}%
	\BibitemOpen
	\bibfield  {author} {\bibinfo {author} {\bibfnamefont {D.}~\bibnamefont
			{Gottesman}},\ }in\ \href@noop {} {\emph {\bibinfo {booktitle} {Quantum
				information science and its contributions to mathematics, Proceedings of
				Symposia in Applied Mathematics}}},\ Vol.~\bibinfo {volume} {68}\ (\bibinfo
	{year} {2010})\ pp.\ \bibinfo {pages} {13--58}\BibitemShut {NoStop}%
	\bibitem [{\citenamefont {Lidar}\ and\ \citenamefont
		{Brun}(2013)}]{lidar2013quantum}%
	\BibitemOpen
	\bibfield  {author} {\bibinfo {author} {\bibfnamefont {D.~A.}\ \bibnamefont
			{Lidar}}\ and\ \bibinfo {author} {\bibfnamefont {T.~A.}\ \bibnamefont
			{Brun}},\ }\href@noop {} {\emph {\bibinfo {title} {Quantum Error
				Correction}}}\ (\bibinfo  {publisher} {Cambridge University Press},\ \bibinfo
	{year} {2013})\BibitemShut {NoStop}%
	\bibitem [{\citenamefont {Macchiavello}\ and\ \citenamefont
		{Palma}(2002)}]{macchiavello02correlations}%
	\BibitemOpen
	\bibfield  {author} {\bibinfo {author} {\bibfnamefont {C.}~\bibnamefont
			{Macchiavello}}\ and\ \bibinfo {author} {\bibfnamefont {G.~M.}\ \bibnamefont
			{Palma}},\ }\href@noop {} {\bibfield  {journal} {\bibinfo  {journal}
			{Physical Review A}\ }\textbf {\bibinfo {volume} {65}},\ \bibinfo {pages}
		{050301(R)} (\bibinfo {year} {2002})}\BibitemShut {NoStop}%
	\bibitem [{\citenamefont {Kretschmann}\ and\ \citenamefont
		{Werner}(2005)}]{kretschmann2005quantum}%
	\BibitemOpen
	\bibfield  {author} {\bibinfo {author} {\bibfnamefont {D.}~\bibnamefont
			{Kretschmann}}\ and\ \bibinfo {author} {\bibfnamefont {R.~F.}\ \bibnamefont
			{Werner}},\ }\href@noop {} {\bibfield  {journal} {\bibinfo  {journal}
			{Physical Review A}\ }\textbf {\bibinfo {volume} {72}},\ \bibinfo {pages}
		{062323} (\bibinfo {year} {2005})}\BibitemShut {NoStop}%
	\bibitem [{\citenamefont {Caruso}\ \emph {et~al.}(2014)\citenamefont {Caruso},
		\citenamefont {Giovannetti}, \citenamefont {Lupo},\ and\ \citenamefont
		{Mancini}}]{caruso2014quantum}%
	\BibitemOpen
	\bibfield  {author} {\bibinfo {author} {\bibfnamefont {F.}~\bibnamefont
			{Caruso}}, \bibinfo {author} {\bibfnamefont {V.}~\bibnamefont {Giovannetti}},
		\bibinfo {author} {\bibfnamefont {C.}~\bibnamefont {Lupo}}, \ and\ \bibinfo
		{author} {\bibfnamefont {S.}~\bibnamefont {Mancini}},\ }\href@noop {}
	{\bibfield  {journal} {\bibinfo  {journal} {Reviews of Modern Physics}\
		}\textbf {\bibinfo {volume} {86}},\ \bibinfo {pages} {1203} (\bibinfo {year}
		{2014})}\BibitemShut {NoStop}%
	\bibitem [{\citenamefont {Pollock}\ \emph {et~al.}(2018)\citenamefont
		{Pollock}, \citenamefont {Rodr\'{\i}guez-Rosario}, \citenamefont
		{Frauenheim}, \citenamefont {Paternostro},\ and\ \citenamefont
		{Modi}}]{pollock2018nonmarkov}%
	\BibitemOpen
	\bibfield  {author} {\bibinfo {author} {\bibfnamefont {F.~A.}\ \bibnamefont
			{Pollock}}, \bibinfo {author} {\bibfnamefont {C.}~\bibnamefont
			{Rodr\'{\i}guez-Rosario}}, \bibinfo {author} {\bibfnamefont {T.}~\bibnamefont
			{Frauenheim}}, \bibinfo {author} {\bibfnamefont {M.}~\bibnamefont
			{Paternostro}}, \ and\ \bibinfo {author} {\bibfnamefont {K.}~\bibnamefont
			{Modi}},\ }\href@noop {} {\bibfield  {journal} {\bibinfo  {journal} {Physical
				Review A}\ }\textbf {\bibinfo {volume} {97}},\ \bibinfo {pages} {012127}
		(\bibinfo {year} {2018})}\BibitemShut {NoStop}%
	\bibitem [{\citenamefont {Ball}\ and\ \citenamefont
		{Banaszek}(2005)}]{ball2005hybrid}%
	\BibitemOpen
	\bibfield  {author} {\bibinfo {author} {\bibfnamefont {J.~L.}\ \bibnamefont
			{Ball}}\ and\ \bibinfo {author} {\bibfnamefont {K.}~\bibnamefont
			{Banaszek}},\ }\href@noop {} {\bibfield  {journal} {\bibinfo  {journal}
			{Journal of Physics A: Mathematical and General}\ }\textbf {\bibinfo {volume}
			{39}},\ \bibinfo {pages} {L1} (\bibinfo {year} {2005})}\BibitemShut {NoStop}%
	\bibitem [{\citenamefont {Bonato}\ \emph {et~al.}(2006)\citenamefont {Bonato},
		\citenamefont {Aspelmeyer}, \citenamefont {Jennewein}, \citenamefont
		{Pernechele}, \citenamefont {Villoresi},\ and\ \citenamefont
		{Zeilinger}}]{bonato2006influence}%
	\BibitemOpen
	\bibfield  {author} {\bibinfo {author} {\bibfnamefont {C.}~\bibnamefont
			{Bonato}}, \bibinfo {author} {\bibfnamefont {M.}~\bibnamefont {Aspelmeyer}},
		\bibinfo {author} {\bibfnamefont {T.}~\bibnamefont {Jennewein}}, \bibinfo
		{author} {\bibfnamefont {C.}~\bibnamefont {Pernechele}}, \bibinfo {author}
		{\bibfnamefont {P.}~\bibnamefont {Villoresi}}, \ and\ \bibinfo {author}
		{\bibfnamefont {A.}~\bibnamefont {Zeilinger}},\ }\href@noop {} {\bibfield
		{journal} {\bibinfo  {journal} {Optics Express}\ }\textbf {\bibinfo {volume}
			{14}},\ \bibinfo {pages} {10050} (\bibinfo {year} {2006})}\BibitemShut
	{NoStop}%
	\bibitem [{\citenamefont {Chiribella}\ \emph {et~al.}(2011)\citenamefont
		{Chiribella}, \citenamefont {Dall’Arno}, \citenamefont {D’Ariano},
		\citenamefont {Macchiavello},\ and\ \citenamefont
		{Perinotti}}]{chiribella2011quantum}%
	\BibitemOpen
	\bibfield  {author} {\bibinfo {author} {\bibfnamefont {G.}~\bibnamefont
			{Chiribella}}, \bibinfo {author} {\bibfnamefont {M.}~\bibnamefont
			{Dall’Arno}}, \bibinfo {author} {\bibfnamefont {G.~M.}\ \bibnamefont
			{D’Ariano}}, \bibinfo {author} {\bibfnamefont {C.}~\bibnamefont
			{Macchiavello}}, \ and\ \bibinfo {author} {\bibfnamefont {P.}~\bibnamefont
			{Perinotti}},\ }\href@noop {} {\bibfield  {journal} {\bibinfo  {journal}
			{Physical Review A}\ }\textbf {\bibinfo {volume} {83}},\ \bibinfo {pages}
		{052305} (\bibinfo {year} {2011})}\BibitemShut {NoStop}%
	\bibitem [{\citenamefont {Giovannetti}\ and\ \citenamefont
		{Fazio}(2005)}]{giovannetti2005information}%
	\BibitemOpen
	\bibfield  {author} {\bibinfo {author} {\bibfnamefont {V.}~\bibnamefont
			{Giovannetti}}\ and\ \bibinfo {author} {\bibfnamefont {R.}~\bibnamefont
			{Fazio}},\ }\href@noop {} {\bibfield  {journal} {\bibinfo  {journal}
			{Physical Review A}\ }\textbf {\bibinfo {volume} {71}},\ \bibinfo {pages}
		{032314} (\bibinfo {year} {2005})}\BibitemShut {NoStop}%
	\bibitem [{\citenamefont {Macchiavello}\ \emph {et~al.}(2004)\citenamefont
		{Macchiavello}, \citenamefont {Palma},\ and\ \citenamefont
		{Virmani}}]{macchiavello2004transition}%
	\BibitemOpen
	\bibfield  {author} {\bibinfo {author} {\bibfnamefont {C.}~\bibnamefont
			{Macchiavello}}, \bibinfo {author} {\bibfnamefont {G.~M.}\ \bibnamefont
			{Palma}}, \ and\ \bibinfo {author} {\bibfnamefont {S.}~\bibnamefont
			{Virmani}},\ }\href@noop {} {\bibfield  {journal} {\bibinfo  {journal}
			{Physical Review A}\ }\textbf {\bibinfo {volume} {69}},\ \bibinfo {pages}
		{010303(R)} (\bibinfo {year} {2004})}\BibitemShut {NoStop}%
	\bibitem [{\citenamefont {Ruggeri}\ \emph {et~al.}(2005)\citenamefont
		{Ruggeri}, \citenamefont {Soliani}, \citenamefont {Giovannetti},\ and\
		\citenamefont {Mancini}}]{ruggeri2005information}%
	\BibitemOpen
	\bibfield  {author} {\bibinfo {author} {\bibfnamefont {G.}~\bibnamefont
			{Ruggeri}}, \bibinfo {author} {\bibfnamefont {G.}~\bibnamefont {Soliani}},
		\bibinfo {author} {\bibfnamefont {V.}~\bibnamefont {Giovannetti}}, \ and\
		\bibinfo {author} {\bibfnamefont {S.}~\bibnamefont {Mancini}},\ }\href@noop
	{} {\bibfield  {journal} {\bibinfo  {journal} {EPL (Europhysics Letters)}\
		}\textbf {\bibinfo {volume} {70}},\ \bibinfo {pages} {719} (\bibinfo {year}
		{2005})}\BibitemShut {NoStop}%
	\bibitem [{\citenamefont {Cerf}\ \emph {et~al.}(2005)\citenamefont {Cerf},
		\citenamefont {Clavareau}, \citenamefont {Macchiavello},\ and\ \citenamefont
		{Roland}}]{cerf2005quantum}%
	\BibitemOpen
	\bibfield  {author} {\bibinfo {author} {\bibfnamefont {N.~J.}\ \bibnamefont
			{Cerf}}, \bibinfo {author} {\bibfnamefont {J.}~\bibnamefont {Clavareau}},
		\bibinfo {author} {\bibfnamefont {C.}~\bibnamefont {Macchiavello}}, \ and\
		\bibinfo {author} {\bibfnamefont {J.}~\bibnamefont {Roland}},\ }\href@noop {}
	{\bibfield  {journal} {\bibinfo  {journal} {Physical Review A}\ }\textbf
		{\bibinfo {volume} {72}},\ \bibinfo {pages} {042330} (\bibinfo {year}
		{2005})}\BibitemShut {NoStop}%
	\bibitem [{\citenamefont {Giovannetti}\ and\ \citenamefont
		{Mancini}(2005)}]{giovannetti2005bosonic}%
	\BibitemOpen
	\bibfield  {author} {\bibinfo {author} {\bibfnamefont {V.}~\bibnamefont
			{Giovannetti}}\ and\ \bibinfo {author} {\bibfnamefont {S.}~\bibnamefont
			{Mancini}},\ }\href@noop {} {\bibfield  {journal} {\bibinfo  {journal}
			{Physical Review A}\ }\textbf {\bibinfo {volume} {71}},\ \bibinfo {pages}
		{062304} (\bibinfo {year} {2005})}\BibitemShut {NoStop}%
	\bibitem [{\citenamefont {Ball}\ \emph {et~al.}(2004)\citenamefont {Ball},
		\citenamefont {Dragan},\ and\ \citenamefont {Banaszek}}]{ball2004exploiting}%
	\BibitemOpen
	\bibfield  {author} {\bibinfo {author} {\bibfnamefont {J.}~\bibnamefont
			{Ball}}, \bibinfo {author} {\bibfnamefont {A.}~\bibnamefont {Dragan}}, \ and\
		\bibinfo {author} {\bibfnamefont {K.}~\bibnamefont {Banaszek}},\ }\href@noop
	{} {\bibfield  {journal} {\bibinfo  {journal} {Physical Review A}\ }\textbf
		{\bibinfo {volume} {69}},\ \bibinfo {pages} {042324} (\bibinfo {year}
		{2004})}\BibitemShut {NoStop}%
	\bibitem [{\citenamefont {Banaszek}\ \emph {et~al.}(2004)\citenamefont
		{Banaszek}, \citenamefont {Dragan}, \citenamefont {Wasilewski},\ and\
		\citenamefont {Radzewicz}}]{banaszek2004experimental}%
	\BibitemOpen
	\bibfield  {author} {\bibinfo {author} {\bibfnamefont {K.}~\bibnamefont
			{Banaszek}}, \bibinfo {author} {\bibfnamefont {A.}~\bibnamefont {Dragan}},
		\bibinfo {author} {\bibfnamefont {W.}~\bibnamefont {Wasilewski}}, \ and\
		\bibinfo {author} {\bibfnamefont {C.}~\bibnamefont {Radzewicz}},\ }\href@noop
	{} {\bibfield  {journal} {\bibinfo  {journal} {Physical Review Letters}\
		}\textbf {\bibinfo {volume} {92}},\ \bibinfo {pages} {257901} (\bibinfo
		{year} {2004})}\BibitemShut {NoStop}%
	\bibitem [{\citenamefont {Bowen}\ and\ \citenamefont
		{Mancini}(2004)}]{bowen04correlations}%
	\BibitemOpen
	\bibfield  {author} {\bibinfo {author} {\bibfnamefont {G.}~\bibnamefont
			{Bowen}}\ and\ \bibinfo {author} {\bibfnamefont {S.}~\bibnamefont
			{Mancini}},\ }\href@noop {} {\bibfield  {journal} {\bibinfo  {journal}
			{Physical Review A}\ }\textbf {\bibinfo {volume} {69}},\ \bibinfo {pages}
		{012306} (\bibinfo {year} {2004})}\BibitemShut {NoStop}%
	\bibitem [{\citenamefont {Plenio}\ and\ \citenamefont
		{Virmani}(2007)}]{plenio2007spin}%
	\BibitemOpen
	\bibfield  {author} {\bibinfo {author} {\bibfnamefont {M.}~\bibnamefont
			{Plenio}}\ and\ \bibinfo {author} {\bibfnamefont {S.}~\bibnamefont
			{Virmani}},\ }\href@noop {} {\bibfield  {journal} {\bibinfo  {journal}
			{Physical Review Letters}\ }\textbf {\bibinfo {volume} {99}},\ \bibinfo
		{pages} {120504} (\bibinfo {year} {2007})}\BibitemShut {NoStop}%
	\bibitem [{\citenamefont {Bayat}\ \emph {et~al.}(2008)\citenamefont {Bayat},
		\citenamefont {Burgarth}, \citenamefont {Mancini},\ and\ \citenamefont
		{Bose}}]{bayat2008memory}%
	\BibitemOpen
	\bibfield  {author} {\bibinfo {author} {\bibfnamefont {A.}~\bibnamefont
			{Bayat}}, \bibinfo {author} {\bibfnamefont {D.}~\bibnamefont {Burgarth}},
		\bibinfo {author} {\bibfnamefont {S.}~\bibnamefont {Mancini}}, \ and\
		\bibinfo {author} {\bibfnamefont {S.}~\bibnamefont {Bose}},\ }\href@noop {}
	{\bibfield  {journal} {\bibinfo  {journal} {Physical Review A}\ }\textbf
		{\bibinfo {volume} {77}},\ \bibinfo {pages} {050306} (\bibinfo {year}
		{2008})}\BibitemShut {NoStop}%
	\bibitem [{\citenamefont {Karpov}\ \emph {et~al.}(2006)\citenamefont {Karpov},
		\citenamefont {Daems},\ and\ \citenamefont {Cerf}}]{karpov2006entanglement}%
	\BibitemOpen
	\bibfield  {author} {\bibinfo {author} {\bibfnamefont {E.}~\bibnamefont
			{Karpov}}, \bibinfo {author} {\bibfnamefont {D.}~\bibnamefont {Daems}}, \
		and\ \bibinfo {author} {\bibfnamefont {N.}~\bibnamefont {Cerf}},\ }\href@noop
	{} {\bibfield  {journal} {\bibinfo  {journal} {Physical Review A}\ }\textbf
		{\bibinfo {volume} {74}},\ \bibinfo {pages} {032320} (\bibinfo {year}
		{2006})}\BibitemShut {NoStop}%
	\bibitem [{\citenamefont {Memarzadeh}\ \emph {et~al.}(2011)\citenamefont
		{Memarzadeh}, \citenamefont {Macchiavello},\ and\ \citenamefont
		{Mancini}}]{memarzadeh2011recovering}%
	\BibitemOpen
	\bibfield  {author} {\bibinfo {author} {\bibfnamefont {L.}~\bibnamefont
			{Memarzadeh}}, \bibinfo {author} {\bibfnamefont {C.}~\bibnamefont
			{Macchiavello}}, \ and\ \bibinfo {author} {\bibfnamefont {S.}~\bibnamefont
			{Mancini}},\ }\href@noop {} {\bibfield  {journal} {\bibinfo  {journal} {New
				Journal of Physics}\ }\textbf {\bibinfo {volume} {13}},\ \bibinfo {pages}
		{103031} (\bibinfo {year} {2011})}\BibitemShut {NoStop}%
	\bibitem [{\citenamefont {Xiao}\ \emph {et~al.}(2016)\citenamefont {Xiao},
		\citenamefont {Yao}, \citenamefont {Xie}, \citenamefont {Wang},\ and\
		\citenamefont {Li}}]{xiao2016protecting}%
	\BibitemOpen
	\bibfield  {author} {\bibinfo {author} {\bibfnamefont {X.}~\bibnamefont
			{Xiao}}, \bibinfo {author} {\bibfnamefont {Y.}~\bibnamefont {Yao}}, \bibinfo
		{author} {\bibfnamefont {Y.-M.}\ \bibnamefont {Xie}}, \bibinfo {author}
		{\bibfnamefont {X.-H.}\ \bibnamefont {Wang}}, \ and\ \bibinfo {author}
		{\bibfnamefont {Y.-L.}\ \bibnamefont {Li}},\ }\href@noop {} {\bibfield
		{journal} {\bibinfo  {journal} {Quantum Information Processing}\ }\textbf
		{\bibinfo {volume} {15}},\ \bibinfo {pages} {3881} (\bibinfo {year}
		{2016})}\BibitemShut {NoStop}%
	\bibitem [{\citenamefont {D’Arrigo}\ \emph {et~al.}(2012)\citenamefont
		{D’Arrigo}, \citenamefont {Benenti},\ and\ \citenamefont
		{Falci}}]{darrigo2012transmission}%
	\BibitemOpen
	\bibfield  {author} {\bibinfo {author} {\bibfnamefont {A.}~\bibnamefont
			{D’Arrigo}}, \bibinfo {author} {\bibfnamefont {G.}~\bibnamefont {Benenti}},
		\ and\ \bibinfo {author} {\bibfnamefont {G.}~\bibnamefont {Falci}},\
	}\href@noop {} {\bibfield  {journal} {\bibinfo  {journal} {The European
				Physical Journal D}\ }\textbf {\bibinfo {volume} {66}},\ \bibinfo {pages}
		{147} (\bibinfo {year} {2012})}\BibitemShut {NoStop}%
	\bibitem [{\citenamefont {Aharonov}\ \emph {et~al.}(1990)\citenamefont
		{Aharonov}, \citenamefont {Anandan}, \citenamefont {Popescu},\ and\
		\citenamefont {Vaidman}}]{Aharonov1990}%
	\BibitemOpen
	\bibfield  {author} {\bibinfo {author} {\bibfnamefont {Y.}~\bibnamefont
			{Aharonov}}, \bibinfo {author} {\bibfnamefont {J.}~\bibnamefont {Anandan}},
		\bibinfo {author} {\bibfnamefont {S.}~\bibnamefont {Popescu}}, \ and\
		\bibinfo {author} {\bibfnamefont {L.}~\bibnamefont {Vaidman}},\ }\href@noop
	{} {\bibfield  {journal} {\bibinfo  {journal} {Physical Review Letters}\
		}\textbf {\bibinfo {volume} {64}},\ \bibinfo {pages} {2965} (\bibinfo {year}
		{1990})}\BibitemShut {NoStop}%
	\bibitem [{\citenamefont {Oi}(2003)}]{oi2003interference}%
	\BibitemOpen
	\bibfield  {author} {\bibinfo {author} {\bibfnamefont {D.~K.}\ \bibnamefont
			{Oi}},\ }\href@noop {} {\bibfield  {journal} {\bibinfo  {journal} {Physical
				Review Letters}\ }\textbf {\bibinfo {volume} {91}},\ \bibinfo {pages}
		{067902} (\bibinfo {year} {2003})}\BibitemShut {NoStop}%
	\bibitem [{\citenamefont {{\AA}berg}(2004{\natexlab{a}})}]{aaberg2004subspace}%
	\BibitemOpen
	\bibfield  {author} {\bibinfo {author} {\bibfnamefont {J.}~\bibnamefont
			{{\AA}berg}},\ }\href@noop {} {\bibfield  {journal} {\bibinfo  {journal}
			{Annals of Physics}\ }\textbf {\bibinfo {volume} {313}},\ \bibinfo {pages}
		{326} (\bibinfo {year} {2004}{\natexlab{a}})}\BibitemShut {NoStop}%
	\bibitem [{\citenamefont {Gisin}\ \emph {et~al.}(2005)\citenamefont {Gisin},
		\citenamefont {Linden}, \citenamefont {Massar},\ and\ \citenamefont
		{Popescu}}]{gisin2005error}%
	\BibitemOpen
	\bibfield  {author} {\bibinfo {author} {\bibfnamefont {N.}~\bibnamefont
			{Gisin}}, \bibinfo {author} {\bibfnamefont {N.}~\bibnamefont {Linden}},
		\bibinfo {author} {\bibfnamefont {S.}~\bibnamefont {Massar}}, \ and\ \bibinfo
		{author} {\bibfnamefont {S.}~\bibnamefont {Popescu}},\ }\href@noop {}
	{\bibfield  {journal} {\bibinfo  {journal} {Physical Review A}\ }\textbf
		{\bibinfo {volume} {72}},\ \bibinfo {pages} {012338} (\bibinfo {year}
		{2005})}\BibitemShut {NoStop}%
	\bibitem [{\citenamefont {Abbott}\ \emph {et~al.}(2020)\citenamefont {Abbott},
		\citenamefont {Wechs}, \citenamefont {Horsman}, \citenamefont {Mhalla},\ and\
		\citenamefont {Branciard}}]{abbott2018communication}%
	\BibitemOpen
	\bibfield  {author} {\bibinfo {author} {\bibfnamefont {A.~A.}\ \bibnamefont
			{Abbott}}, \bibinfo {author} {\bibfnamefont {J.}~\bibnamefont {Wechs}},
		\bibinfo {author} {\bibfnamefont {D.}~\bibnamefont {Horsman}}, \bibinfo
		{author} {\bibfnamefont {M.}~\bibnamefont {Mhalla}}, \ and\ \bibinfo {author}
		{\bibfnamefont {C.}~\bibnamefont {Branciard}},\ }\href@noop {} {\bibfield
		{journal} {\bibinfo  {journal} {Quantum}\ }\textbf {\bibinfo {volume} {4}},\
		\bibinfo {pages} {333} (\bibinfo {year} {2020})}\BibitemShut {NoStop}%
	\bibitem [{\citenamefont {Chiribella}\ and\ \citenamefont
		{Kristj\'ansson}(2019)}]{chiribella2019shannon2q}%
	\BibitemOpen
	\bibfield  {author} {\bibinfo {author} {\bibfnamefont {G.}~\bibnamefont
			{Chiribella}}\ and\ \bibinfo {author} {\bibfnamefont {H.}~\bibnamefont
			{Kristj\'ansson}},\ }\href@noop {} {\bibfield  {journal} {\bibinfo  {journal}
			{Proc.\ R.\ Soc.\ A}\ }\textbf {\bibinfo {volume} {475}},\ \bibinfo {pages}
		{20180903} (\bibinfo {year} {2019})}\BibitemShut {NoStop}%
	\bibitem [{\citenamefont {Dong}\ \emph {et~al.}(2019)\citenamefont {Dong},
		\citenamefont {Nakayama}, \citenamefont {Soeda},\ and\ \citenamefont
		{Murao}}]{dong2019controlled}%
	\BibitemOpen
	\bibfield  {author} {\bibinfo {author} {\bibfnamefont {Q.}~\bibnamefont
			{Dong}}, \bibinfo {author} {\bibfnamefont {S.}~\bibnamefont {Nakayama}},
		\bibinfo {author} {\bibfnamefont {A.}~\bibnamefont {Soeda}}, \ and\ \bibinfo
		{author} {\bibfnamefont {M.}~\bibnamefont {Murao}},\ }\href@noop {}
	{\bibfield  {journal} {\bibinfo  {journal} {arXiv preprint arXiv:1911.01645}\
		} (\bibinfo {year} {2019})}\BibitemShut {NoStop}%
	\bibitem [{\citenamefont {Vanrietvelde}\ and\ \citenamefont
		{Chiribella}(2021)}]{vanrietvelde2021universal}%
	\BibitemOpen
	\bibfield  {author} {\bibinfo {author} {\bibfnamefont {A.}~\bibnamefont
			{Vanrietvelde}}\ and\ \bibinfo {author} {\bibfnamefont {G.}~\bibnamefont
			{Chiribella}},\ }\href@noop {} {\bibfield  {journal} {\bibinfo  {journal}
			{arXiv preprint arXiv:2106.12463}\ } (\bibinfo {year} {2021})}\BibitemShut
	{NoStop}%
	\bibitem [{\citenamefont {Oreshkov}(2019)}]{oreshkov2019time}%
	\BibitemOpen
	\bibfield  {author} {\bibinfo {author} {\bibfnamefont {O.}~\bibnamefont
			{Oreshkov}},\ }\href@noop {} {\bibfield  {journal} {\bibinfo  {journal}
			{Quantum}\ }\textbf {\bibinfo {volume} {3}},\ \bibinfo {pages} {206}
		(\bibinfo {year} {2019})}\BibitemShut {NoStop}%
	\bibitem [{\citenamefont {Chiribella}\ \emph
		{et~al.}(2009{\natexlab{a}})\citenamefont {Chiribella}, \citenamefont
		{D’Ariano}, \citenamefont {Perinotti},\ and\ \citenamefont
		{Valiron}}]{chiribella2009beyond}%
	\BibitemOpen
	\bibfield  {author} {\bibinfo {author} {\bibfnamefont {G.}~\bibnamefont
			{Chiribella}}, \bibinfo {author} {\bibfnamefont {G.}~\bibnamefont
			{D’Ariano}}, \bibinfo {author} {\bibfnamefont {P.}~\bibnamefont
			{Perinotti}}, \ and\ \bibinfo {author} {\bibfnamefont {B.}~\bibnamefont
			{Valiron}},\ }\href@noop {} {\bibfield  {journal} {\bibinfo  {journal} {arXiv
				preprint arXiv:0912.0195}\ } (\bibinfo {year}
		{2009}{\natexlab{a}})}\BibitemShut {NoStop}%
	\bibitem [{\citenamefont {Chiribella}\ \emph {et~al.}(2013)\citenamefont
		{Chiribella}, \citenamefont {D’Ariano}, \citenamefont {Perinotti},\ and\
		\citenamefont {Valiron}}]{chiribella2013quantum}%
	\BibitemOpen
	\bibfield  {author} {\bibinfo {author} {\bibfnamefont {G.}~\bibnamefont
			{Chiribella}}, \bibinfo {author} {\bibfnamefont {G.~M.}\ \bibnamefont
			{D’Ariano}}, \bibinfo {author} {\bibfnamefont {P.}~\bibnamefont
			{Perinotti}}, \ and\ \bibinfo {author} {\bibfnamefont {B.}~\bibnamefont
			{Valiron}},\ }\href@noop {} {\bibfield  {journal} {\bibinfo  {journal}
			{Physical Review A}\ }\textbf {\bibinfo {volume} {88}},\ \bibinfo {pages}
		{022318} (\bibinfo {year} {2013})}\BibitemShut {NoStop}%
	\bibitem [{\citenamefont {Procopio}\ \emph {et~al.}(2015)\citenamefont
		{Procopio}, \citenamefont {Moqanaki}, \citenamefont {Ara{\'u}jo},
		\citenamefont {Costa}, \citenamefont {Calafell}, \citenamefont {Dowd},
		\citenamefont {Hamel}, \citenamefont {Rozema}, \citenamefont {Brukner},\ and\
		\citenamefont {Walther}}]{procopio2015experimental}%
	\BibitemOpen
	\bibfield  {author} {\bibinfo {author} {\bibfnamefont {L.~M.}\ \bibnamefont
			{Procopio}}, \bibinfo {author} {\bibfnamefont {A.}~\bibnamefont {Moqanaki}},
		\bibinfo {author} {\bibfnamefont {M.}~\bibnamefont {Ara{\'u}jo}}, \bibinfo
		{author} {\bibfnamefont {F.}~\bibnamefont {Costa}}, \bibinfo {author}
		{\bibfnamefont {I.~A.}\ \bibnamefont {Calafell}}, \bibinfo {author}
		{\bibfnamefont {E.~G.}\ \bibnamefont {Dowd}}, \bibinfo {author}
		{\bibfnamefont {D.~R.}\ \bibnamefont {Hamel}}, \bibinfo {author}
		{\bibfnamefont {L.~A.}\ \bibnamefont {Rozema}}, \bibinfo {author}
		{\bibfnamefont {{\v{C}}.}~\bibnamefont {Brukner}}, \ and\ \bibinfo {author}
		{\bibfnamefont {P.}~\bibnamefont {Walther}},\ }\href@noop {} {\bibfield
		{journal} {\bibinfo  {journal} {Nature Communications}\ }\textbf {\bibinfo
			{volume} {6}},\ \bibinfo {pages} {7913} (\bibinfo {year} {2015})}\BibitemShut
	{NoStop}%
	\bibitem [{\citenamefont {Rubino}\ \emph {et~al.}(2017)\citenamefont {Rubino},
		\citenamefont {Rozema}, \citenamefont {Feix}, \citenamefont {Ara{\'u}jo},
		\citenamefont {Zeuner}, \citenamefont {Procopio}, \citenamefont {Brukner},\
		and\ \citenamefont {Walther}}]{rubino2017experimental}%
	\BibitemOpen
	\bibfield  {author} {\bibinfo {author} {\bibfnamefont {G.}~\bibnamefont
			{Rubino}}, \bibinfo {author} {\bibfnamefont {L.~A.}\ \bibnamefont {Rozema}},
		\bibinfo {author} {\bibfnamefont {A.}~\bibnamefont {Feix}}, \bibinfo {author}
		{\bibfnamefont {M.}~\bibnamefont {Ara{\'u}jo}}, \bibinfo {author}
		{\bibfnamefont {J.~M.}\ \bibnamefont {Zeuner}}, \bibinfo {author}
		{\bibfnamefont {L.~M.}\ \bibnamefont {Procopio}}, \bibinfo {author}
		{\bibfnamefont {{\v{C}}.}~\bibnamefont {Brukner}}, \ and\ \bibinfo {author}
		{\bibfnamefont {P.}~\bibnamefont {Walther}},\ }\href@noop {} {\bibfield
		{journal} {\bibinfo  {journal} {Science Advances}\ }\textbf {\bibinfo
			{volume} {3}},\ \bibinfo {pages} {e1602589} (\bibinfo {year}
		{2017})}\BibitemShut {NoStop}%
	\bibitem [{\citenamefont {Goswami}\ \emph {et~al.}(2018)\citenamefont
		{Goswami}, \citenamefont {Giarmatzi}, \citenamefont {Kewming}, \citenamefont
		{Costa}, \citenamefont {Branciard}, \citenamefont {Romero},\ and\
		\citenamefont {White}}]{goswami2018indefinite}%
	\BibitemOpen
	\bibfield  {author} {\bibinfo {author} {\bibfnamefont {K.}~\bibnamefont
			{Goswami}}, \bibinfo {author} {\bibfnamefont {C.}~\bibnamefont {Giarmatzi}},
		\bibinfo {author} {\bibfnamefont {M.}~\bibnamefont {Kewming}}, \bibinfo
		{author} {\bibfnamefont {F.}~\bibnamefont {Costa}}, \bibinfo {author}
		{\bibfnamefont {C.}~\bibnamefont {Branciard}}, \bibinfo {author}
		{\bibfnamefont {J.}~\bibnamefont {Romero}}, \ and\ \bibinfo {author}
		{\bibfnamefont {A.}~\bibnamefont {White}},\ }\href@noop {} {\bibfield
		{journal} {\bibinfo  {journal} {Physical Review Letters}\ }\textbf {\bibinfo
			{volume} {121}},\ \bibinfo {pages} {090503} (\bibinfo {year}
		{2018})}\BibitemShut {NoStop}%
	\bibitem [{\citenamefont {Guo}\ \emph {et~al.}(2020)\citenamefont {Guo},
		\citenamefont {Hu}, \citenamefont {Hou}, \citenamefont {Cao}, \citenamefont
		{Cui}, \citenamefont {Liu}, \citenamefont {Huang}, \citenamefont {Li},
		\citenamefont {Guo},\ and\ \citenamefont {Chiribella}}]{guo2020experimental}%
	\BibitemOpen
	\bibfield  {author} {\bibinfo {author} {\bibfnamefont {Y.}~\bibnamefont
			{Guo}}, \bibinfo {author} {\bibfnamefont {X.-M.}\ \bibnamefont {Hu}},
		\bibinfo {author} {\bibfnamefont {Z.-B.}\ \bibnamefont {Hou}}, \bibinfo
		{author} {\bibfnamefont {H.}~\bibnamefont {Cao}}, \bibinfo {author}
		{\bibfnamefont {J.-M.}\ \bibnamefont {Cui}}, \bibinfo {author} {\bibfnamefont
			{B.-H.}\ \bibnamefont {Liu}}, \bibinfo {author} {\bibfnamefont {Y.-F.}\
			\bibnamefont {Huang}}, \bibinfo {author} {\bibfnamefont {C.-F.}\ \bibnamefont
			{Li}}, \bibinfo {author} {\bibfnamefont {G.-C.}\ \bibnamefont {Guo}}, \ and\
		\bibinfo {author} {\bibfnamefont {G.}~\bibnamefont {Chiribella}},\
	}\href@noop {} {\bibfield  {journal} {\bibinfo  {journal} {Physical Review
				Letters}\ }\textbf {\bibinfo {volume} {124}},\ \bibinfo {pages} {030502}
		(\bibinfo {year} {2020})}\BibitemShut {NoStop}%
	\bibitem [{\citenamefont {Goswami}\ \emph {et~al.}(2020)\citenamefont
		{Goswami}, \citenamefont {Cao}, \citenamefont {Paz-Silva}, \citenamefont
		{Romero},\ and\ \citenamefont {White}}]{goswami2018communicating}%
	\BibitemOpen
	\bibfield  {author} {\bibinfo {author} {\bibfnamefont {K.}~\bibnamefont
			{Goswami}}, \bibinfo {author} {\bibfnamefont {Y.}~\bibnamefont {Cao}},
		\bibinfo {author} {\bibfnamefont {G.~A.}\ \bibnamefont {Paz-Silva}}, \bibinfo
		{author} {\bibfnamefont {J.}~\bibnamefont {Romero}}, \ and\ \bibinfo {author}
		{\bibfnamefont {A.~G.}\ \bibnamefont {White}},\ }\href {\doibase
		10.1103/PhysRevResearch.2.033292} {\bibfield  {journal} {\bibinfo  {journal}
			{Physical Review Research}\ }\textbf {\bibinfo {volume} {2}},\ \bibinfo
		{pages} {033292} (\bibinfo {year} {2020})}\BibitemShut {NoStop}%
	\bibitem [{\citenamefont {Goswami}\ and\ \citenamefont
		{Romero}(2020)}]{goswami2020experiments}%
	\BibitemOpen
	\bibfield  {author} {\bibinfo {author} {\bibfnamefont {K.}~\bibnamefont
			{Goswami}}\ and\ \bibinfo {author} {\bibfnamefont {J.}~\bibnamefont
			{Romero}},\ }\href@noop {} {\bibfield  {journal} {\bibinfo  {journal} {AVS
				Quantum Science}\ }\textbf {\bibinfo {volume} {2}},\ \bibinfo {pages}
		{037101} (\bibinfo {year} {2020})}\BibitemShut {NoStop}%
	\bibitem [{\citenamefont {Rubino}\ \emph {et~al.}(2021)\citenamefont {Rubino},
		\citenamefont {Rozema}, \citenamefont {Ebler}, \citenamefont
		{Kristj{\'a}nsson}, \citenamefont {Salek}, \citenamefont {Gu{\'e}rin},
		\citenamefont {Abbott}, \citenamefont {Branciard}, \citenamefont {Brukner},
		\citenamefont {Chiribella} \emph {et~al.}}]{rubino2021experimental}%
	\BibitemOpen
	\bibfield  {author} {\bibinfo {author} {\bibfnamefont {G.}~\bibnamefont
			{Rubino}}, \bibinfo {author} {\bibfnamefont {L.~A.}\ \bibnamefont {Rozema}},
		\bibinfo {author} {\bibfnamefont {D.}~\bibnamefont {Ebler}}, \bibinfo
		{author} {\bibfnamefont {H.}~\bibnamefont {Kristj{\'a}nsson}}, \bibinfo
		{author} {\bibfnamefont {S.}~\bibnamefont {Salek}}, \bibinfo {author}
		{\bibfnamefont {P.~A.}\ \bibnamefont {Gu{\'e}rin}}, \bibinfo {author}
		{\bibfnamefont {A.~A.}\ \bibnamefont {Abbott}}, \bibinfo {author}
		{\bibfnamefont {C.}~\bibnamefont {Branciard}}, \bibinfo {author}
		{\bibfnamefont {{\v{C}}.}~\bibnamefont {Brukner}}, \bibinfo {author}
		{\bibfnamefont {G.}~\bibnamefont {Chiribella}},  \emph {et~al.},\ }\href@noop
	{} {\bibfield  {journal} {\bibinfo  {journal} {Physical Review Research}\
		}\textbf {\bibinfo {volume} {3}},\ \bibinfo {pages} {013093} (\bibinfo {year}
		{2021})}\BibitemShut {NoStop}%
	\bibitem [{\citenamefont {Ebler}\ \emph {et~al.}(2018)\citenamefont {Ebler},
		\citenamefont {Salek},\ and\ \citenamefont {Chiribella}}]{ebler2018enhanced}%
	\BibitemOpen
	\bibfield  {author} {\bibinfo {author} {\bibfnamefont {D.}~\bibnamefont
			{Ebler}}, \bibinfo {author} {\bibfnamefont {S.}~\bibnamefont {Salek}}, \ and\
		\bibinfo {author} {\bibfnamefont {G.}~\bibnamefont {Chiribella}},\
	}\href@noop {} {\bibfield  {journal} {\bibinfo  {journal} {Physical Review
				Letters}\ }\textbf {\bibinfo {volume} {120}},\ \bibinfo {pages} {120502}
		(\bibinfo {year} {2018})}\BibitemShut {NoStop}%
	\bibitem [{\citenamefont {Salek}\ \emph {et~al.}(2018)\citenamefont {Salek},
		\citenamefont {Ebler},\ and\ \citenamefont {Chiribella}}]{salek2018quantum}%
	\BibitemOpen
	\bibfield  {author} {\bibinfo {author} {\bibfnamefont {S.}~\bibnamefont
			{Salek}}, \bibinfo {author} {\bibfnamefont {D.}~\bibnamefont {Ebler}}, \ and\
		\bibinfo {author} {\bibfnamefont {G.}~\bibnamefont {Chiribella}},\
	}\href@noop {} {\bibfield  {journal} {\bibinfo  {journal} {arXiv preprint
				arXiv:1809.06655}\ } (\bibinfo {year} {2018})}\BibitemShut {NoStop}%
	\bibitem [{\citenamefont {Chiribella}\ \emph {et~al.}(2021)\citenamefont
		{Chiribella}, \citenamefont {Banik}, \citenamefont {Bhattacharya},
		\citenamefont {Guha}, \citenamefont {Alimuddin}, \citenamefont {Roy},
		\citenamefont {Saha}, \citenamefont {Agrawal},\ and\ \citenamefont
		{Kar}}]{chiribella2018indefinite}%
	\BibitemOpen
	\bibfield  {author} {\bibinfo {author} {\bibfnamefont {G.}~\bibnamefont
			{Chiribella}}, \bibinfo {author} {\bibfnamefont {M.}~\bibnamefont {Banik}},
		\bibinfo {author} {\bibfnamefont {S.~S.}\ \bibnamefont {Bhattacharya}},
		\bibinfo {author} {\bibfnamefont {T.}~\bibnamefont {Guha}}, \bibinfo {author}
		{\bibfnamefont {M.}~\bibnamefont {Alimuddin}}, \bibinfo {author}
		{\bibfnamefont {A.}~\bibnamefont {Roy}}, \bibinfo {author} {\bibfnamefont
			{S.}~\bibnamefont {Saha}}, \bibinfo {author} {\bibfnamefont {S.}~\bibnamefont
			{Agrawal}}, \ and\ \bibinfo {author} {\bibfnamefont {G.}~\bibnamefont
			{Kar}},\ }\href@noop {} {\bibfield  {journal} {\bibinfo  {journal} {New
				Journal of Physics}\ }\textbf {\bibinfo {volume} {23}},\ \bibinfo {pages}
		{033039} (\bibinfo {year} {2021})}\BibitemShut {NoStop}%
	\bibitem [{\citenamefont {Procopio}\ \emph {et~al.}(2019)\citenamefont
		{Procopio}, \citenamefont {Delgado}, \citenamefont {Enr{\'\i}quez},
		\citenamefont {Belabas},\ and\ \citenamefont
		{Levenson}}]{procopio2019communication}%
	\BibitemOpen
	\bibfield  {author} {\bibinfo {author} {\bibfnamefont {L.~M.}\ \bibnamefont
			{Procopio}}, \bibinfo {author} {\bibfnamefont {F.}~\bibnamefont {Delgado}},
		\bibinfo {author} {\bibfnamefont {M.}~\bibnamefont {Enr{\'\i}quez}}, \bibinfo
		{author} {\bibfnamefont {N.}~\bibnamefont {Belabas}}, \ and\ \bibinfo
		{author} {\bibfnamefont {J.~A.}\ \bibnamefont {Levenson}},\ }\href@noop {}
	{\bibfield  {journal} {\bibinfo  {journal} {Entropy}\ }\textbf {\bibinfo
			{volume} {21}},\ \bibinfo {pages} {1012} (\bibinfo {year}
		{2019})}\BibitemShut {NoStop}%
	\bibitem [{\citenamefont {Procopio}\ \emph {et~al.}(2020)\citenamefont
		{Procopio}, \citenamefont {Delgado}, \citenamefont {Enr\'{\i}quez},
		\citenamefont {Belabas},\ and\ \citenamefont
		{Levenson}}]{procopio2020sending}%
	\BibitemOpen
	\bibfield  {author} {\bibinfo {author} {\bibfnamefont {L.~M.}\ \bibnamefont
			{Procopio}}, \bibinfo {author} {\bibfnamefont {F.}~\bibnamefont {Delgado}},
		\bibinfo {author} {\bibfnamefont {M.}~\bibnamefont {Enr\'{\i}quez}}, \bibinfo
		{author} {\bibfnamefont {N.}~\bibnamefont {Belabas}}, \ and\ \bibinfo
		{author} {\bibfnamefont {J.~A.}\ \bibnamefont {Levenson}},\ }\href {\doibase
		10.1103/PhysRevA.101.012346} {\bibfield  {journal} {\bibinfo  {journal}
			{Physical Review A}\ }\textbf {\bibinfo {volume} {101}},\ \bibinfo {pages}
		{012346} (\bibinfo {year} {2020})}\BibitemShut {NoStop}%
	\bibitem [{\citenamefont {Chiribella}\ \emph {et~al.}(2020)\citenamefont
		{Chiribella}, \citenamefont {Wilson},\ and\ \citenamefont
		{Chau}}]{chiribella2020quantum}%
	\BibitemOpen
	\bibfield  {author} {\bibinfo {author} {\bibfnamefont {G.}~\bibnamefont
			{Chiribella}}, \bibinfo {author} {\bibfnamefont {M.}~\bibnamefont {Wilson}},
		\ and\ \bibinfo {author} {\bibfnamefont {H.-F.}\ \bibnamefont {Chau}},\
	}\href@noop {} {\bibfield  {journal} {\bibinfo  {journal} {arXiv:2005.00618}\
		} (\bibinfo {year} {2020})}\BibitemShut {NoStop}%
	\bibitem [{\citenamefont {Chiribella}\ \emph
		{et~al.}(2008{\natexlab{a}})\citenamefont {Chiribella}, \citenamefont
		{D'Ariano},\ and\ \citenamefont {Perinotti}}]{chiribella2008quantum}%
	\BibitemOpen
	\bibfield  {author} {\bibinfo {author} {\bibfnamefont {G.}~\bibnamefont
			{Chiribella}}, \bibinfo {author} {\bibfnamefont {G.~M.}\ \bibnamefont
			{D'Ariano}}, \ and\ \bibinfo {author} {\bibfnamefont {P.}~\bibnamefont
			{Perinotti}},\ }\href@noop {} {\bibfield  {journal} {\bibinfo  {journal}
			{Physical Review Letters}\ }\textbf {\bibinfo {volume} {101}},\ \bibinfo
		{pages} {060401} (\bibinfo {year} {2008}{\natexlab{a}})}\BibitemShut
	{NoStop}%
	\bibitem [{\citenamefont {Chiribella}\ \emph
		{et~al.}(2009{\natexlab{b}})\citenamefont {Chiribella}, \citenamefont
		{D'Ariano},\ and\ \citenamefont {Perinotti}}]{chiribella2009theoretical}%
	\BibitemOpen
	\bibfield  {author} {\bibinfo {author} {\bibfnamefont {G.}~\bibnamefont
			{Chiribella}}, \bibinfo {author} {\bibfnamefont {G.~M.}\ \bibnamefont
			{D'Ariano}}, \ and\ \bibinfo {author} {\bibfnamefont {P.}~\bibnamefont
			{Perinotti}},\ }\href@noop {} {\bibfield  {journal} {\bibinfo  {journal}
			{Physical Review A}\ }\textbf {\bibinfo {volume} {80}},\ \bibinfo {pages}
		{022339} (\bibinfo {year} {2009}{\natexlab{b}})}\BibitemShut {NoStop}%
	\bibitem [{\citenamefont {Breuer}\ \emph {et~al.}(2016)\citenamefont {Breuer},
		\citenamefont {Laine}, \citenamefont {Piilo},\ and\ \citenamefont
		{Vacchini}}]{breuer2016nonmarkov}%
	\BibitemOpen
	\bibfield  {author} {\bibinfo {author} {\bibfnamefont {H.-P.}\ \bibnamefont
			{Breuer}}, \bibinfo {author} {\bibfnamefont {E.-M.}\ \bibnamefont {Laine}},
		\bibinfo {author} {\bibfnamefont {J.}~\bibnamefont {Piilo}}, \ and\ \bibinfo
		{author} {\bibfnamefont {B.}~\bibnamefont {Vacchini}},\ }\href@noop {}
	{\bibfield  {journal} {\bibinfo  {journal} {Reviews of Modern Physics}\
		}\textbf {\bibinfo {volume} {88}},\ \bibinfo {pages} {021002} (\bibinfo
		{year} {2016})}\BibitemShut {NoStop}%
	\bibitem [{\citenamefont {Humphreys}\ \emph {et~al.}(2013)\citenamefont
		{Humphreys}, \citenamefont {Metcalf}, \citenamefont {Spring}, \citenamefont
		{Moore}, \citenamefont {Jin}, \citenamefont {Barbieri}, \citenamefont
		{Kolthammer},\ and\ \citenamefont {Walmsley}}]{humphreys2013linear}%
	\BibitemOpen
	\bibfield  {author} {\bibinfo {author} {\bibfnamefont {P.~C.}\ \bibnamefont
			{Humphreys}}, \bibinfo {author} {\bibfnamefont {B.~J.}\ \bibnamefont
			{Metcalf}}, \bibinfo {author} {\bibfnamefont {J.~B.}\ \bibnamefont {Spring}},
		\bibinfo {author} {\bibfnamefont {M.}~\bibnamefont {Moore}}, \bibinfo
		{author} {\bibfnamefont {X.-M.}\ \bibnamefont {Jin}}, \bibinfo {author}
		{\bibfnamefont {M.}~\bibnamefont {Barbieri}}, \bibinfo {author}
		{\bibfnamefont {W.~S.}\ \bibnamefont {Kolthammer}}, \ and\ \bibinfo {author}
		{\bibfnamefont {I.~A.}\ \bibnamefont {Walmsley}},\ }\href@noop {} {\bibfield
		{journal} {\bibinfo  {journal} {Physical Review Letters}\ }\textbf {\bibinfo
			{volume} {111}},\ \bibinfo {pages} {150501} (\bibinfo {year}
		{2013})}\BibitemShut {NoStop}%
	\bibitem [{\citenamefont {Donohue}\ \emph {et~al.}(2013)\citenamefont
		{Donohue}, \citenamefont {Agnew}, \citenamefont {Lavoie},\ and\ \citenamefont
		{Resch}}]{donohue2013coherent}%
	\BibitemOpen
	\bibfield  {author} {\bibinfo {author} {\bibfnamefont {J.~M.}\ \bibnamefont
			{Donohue}}, \bibinfo {author} {\bibfnamefont {M.}~\bibnamefont {Agnew}},
		\bibinfo {author} {\bibfnamefont {J.}~\bibnamefont {Lavoie}}, \ and\ \bibinfo
		{author} {\bibfnamefont {K.~J.}\ \bibnamefont {Resch}},\ }\href@noop {}
	{\bibfield  {journal} {\bibinfo  {journal} {Physical Review Letters}\
		}\textbf {\bibinfo {volume} {111}},\ \bibinfo {pages} {153602} (\bibinfo
		{year} {2013})}\BibitemShut {NoStop}%
	\bibitem [{\citenamefont {Li}\ and\ \citenamefont
		{Ghose}(2015)}]{li2015hyperentanglement}%
	\BibitemOpen
	\bibfield  {author} {\bibinfo {author} {\bibfnamefont {X.-H.}\ \bibnamefont
			{Li}}\ and\ \bibinfo {author} {\bibfnamefont {S.}~\bibnamefont {Ghose}},\
	}\href@noop {} {\bibfield  {journal} {\bibinfo  {journal} {Physical Review
				A}\ }\textbf {\bibinfo {volume} {91}},\ \bibinfo {pages} {062302} (\bibinfo
		{year} {2015})}\BibitemShut {NoStop}%
	\bibitem [{\citenamefont {Donohue}\ \emph {et~al.}(2014)\citenamefont
		{Donohue}, \citenamefont {Lavoie},\ and\ \citenamefont
		{Resch}}]{donohue2014ultrafast}%
	\BibitemOpen
	\bibfield  {author} {\bibinfo {author} {\bibfnamefont {J.~M.}\ \bibnamefont
			{Donohue}}, \bibinfo {author} {\bibfnamefont {J.}~\bibnamefont {Lavoie}}, \
		and\ \bibinfo {author} {\bibfnamefont {K.~J.}\ \bibnamefont {Resch}},\
	}\href@noop {} {\bibfield  {journal} {\bibinfo  {journal} {Physical Review
				Letters}\ }\textbf {\bibinfo {volume} {113}},\ \bibinfo {pages} {163602}
		(\bibinfo {year} {2014})}\BibitemShut {NoStop}%
	\bibitem [{\citenamefont {Ghafari}\ \emph {et~al.}(2019)\citenamefont
		{Ghafari}, \citenamefont {Tischler}, \citenamefont {Di~Franco}, \citenamefont
		{Thompson}, \citenamefont {Gu},\ and\ \citenamefont
		{Pryde}}]{ghafari2019interfering}%
	\BibitemOpen
	\bibfield  {author} {\bibinfo {author} {\bibfnamefont {F.}~\bibnamefont
			{Ghafari}}, \bibinfo {author} {\bibfnamefont {N.}~\bibnamefont {Tischler}},
		\bibinfo {author} {\bibfnamefont {C.}~\bibnamefont {Di~Franco}}, \bibinfo
		{author} {\bibfnamefont {J.}~\bibnamefont {Thompson}}, \bibinfo {author}
		{\bibfnamefont {M.}~\bibnamefont {Gu}}, \ and\ \bibinfo {author}
		{\bibfnamefont {G.~J.}\ \bibnamefont {Pryde}},\ }\href@noop {} {\bibfield
		{journal} {\bibinfo  {journal} {Nature Communications}\ }\textbf {\bibinfo
			{volume} {10}},\ \bibinfo {pages} {1} (\bibinfo {year} {2019})}\BibitemShut
	{NoStop}%
	\bibitem [{\citenamefont {Marcikic}\ \emph {et~al.}(2002)\citenamefont
		{Marcikic}, \citenamefont {de~Riedmatten}, \citenamefont {Tittel},
		\citenamefont {Scarani}, \citenamefont {Zbinden},\ and\ \citenamefont
		{Gisin}}]{marcikic2002time}%
	\BibitemOpen
	\bibfield  {author} {\bibinfo {author} {\bibfnamefont {I.}~\bibnamefont
			{Marcikic}}, \bibinfo {author} {\bibfnamefont {H.}~\bibnamefont
			{de~Riedmatten}}, \bibinfo {author} {\bibfnamefont {W.}~\bibnamefont
			{Tittel}}, \bibinfo {author} {\bibfnamefont {V.}~\bibnamefont {Scarani}},
		\bibinfo {author} {\bibfnamefont {H.}~\bibnamefont {Zbinden}}, \ and\
		\bibinfo {author} {\bibfnamefont {N.}~\bibnamefont {Gisin}},\ }\href@noop {}
	{\bibfield  {journal} {\bibinfo  {journal} {Physical Review A}\ }\textbf
		{\bibinfo {volume} {66}},\ \bibinfo {pages} {062308} (\bibinfo {year}
		{2002})}\BibitemShut {NoStop}%
	\bibitem [{\citenamefont {Birkhoff}(1946)}]{birkhoff1946three}%
	\BibitemOpen
	\bibfield  {author} {\bibinfo {author} {\bibfnamefont {G.}~\bibnamefont
			{Birkhoff}},\ }\href@noop {} {\bibfield  {journal} {\bibinfo  {journal}
			{Univ. Nac. Tacuman, Rev. Ser. A}\ }\textbf {\bibinfo {volume} {5}},\
		\bibinfo {pages} {147} (\bibinfo {year} {1946})}\BibitemShut {NoStop}%
	\bibitem [{\citenamefont {Horodecki}\ \emph {et~al.}(2003)\citenamefont
		{Horodecki}, \citenamefont {Shor},\ and\ \citenamefont
		{Ruskai}}]{horodecki2003entanglement}%
	\BibitemOpen
	\bibfield  {author} {\bibinfo {author} {\bibfnamefont {M.}~\bibnamefont
			{Horodecki}}, \bibinfo {author} {\bibfnamefont {P.~W.}\ \bibnamefont {Shor}},
		\ and\ \bibinfo {author} {\bibfnamefont {M.~B.}\ \bibnamefont {Ruskai}},\
	}\href@noop {} {\bibfield  {journal} {\bibinfo  {journal} {Reviews in
				Mathematical Physics}\ }\textbf {\bibinfo {volume} {15}},\ \bibinfo {pages}
		{629} (\bibinfo {year} {2003})}\BibitemShut {NoStop}%
	\bibitem [{\citenamefont {Shor}(2002)}]{shor2002additivity}%
	\BibitemOpen
	\bibfield  {author} {\bibinfo {author} {\bibfnamefont {P.~W.}\ \bibnamefont
			{Shor}},\ }\href@noop {} {\bibfield  {journal} {\bibinfo  {journal} {Journal
				of Mathematical Physics}\ }\textbf {\bibinfo {volume} {43}},\ \bibinfo
		{pages} {4334} (\bibinfo {year} {2002})}\BibitemShut {NoStop}%
	\bibitem [{\citenamefont {Davies}(1978)}]{davies1978information}%
	\BibitemOpen
	\bibfield  {author} {\bibinfo {author} {\bibfnamefont {E.}~\bibnamefont
			{Davies}},\ }\href@noop {} {\bibfield  {journal} {\bibinfo  {journal} {IEEE
				Transactions on Information Theory}\ }\textbf {\bibinfo {volume} {24}},\
		\bibinfo {pages} {596} (\bibinfo {year} {1978})}\BibitemShut {NoStop}%
	\bibitem [{\citenamefont {Fuchs}(1997)}]{fuchs1997nonorthogonal}%
	\BibitemOpen
	\bibfield  {author} {\bibinfo {author} {\bibfnamefont {C.~A.}\ \bibnamefont
			{Fuchs}},\ }\href@noop {} {\bibfield  {journal} {\bibinfo  {journal}
			{Physical Review Letters}\ }\textbf {\bibinfo {volume} {79}},\ \bibinfo
		{pages} {1162} (\bibinfo {year} {1997})}\BibitemShut {NoStop}%
	\bibitem [{\citenamefont {King}\ \emph {et~al.}(2002)\citenamefont {King},
		\citenamefont {Nathanson},\ and\ \citenamefont {Ruskai}}]{king2002qubit}%
	\BibitemOpen
	\bibfield  {author} {\bibinfo {author} {\bibfnamefont {C.}~\bibnamefont
			{King}}, \bibinfo {author} {\bibfnamefont {M.}~\bibnamefont {Nathanson}}, \
		and\ \bibinfo {author} {\bibfnamefont {M.~B.}\ \bibnamefont {Ruskai}},\
	}\href@noop {} {\bibfield  {journal} {\bibinfo  {journal} {Physical Review
				Letters}\ }\textbf {\bibinfo {volume} {88}},\ \bibinfo {pages} {057901}
		(\bibinfo {year} {2002})}\BibitemShut {NoStop}%
	\bibitem [{\citenamefont {Hayashi}\ \emph {et~al.}(2004)\citenamefont
		{Hayashi}, \citenamefont {Imai}, \citenamefont {Matsumoto}, \citenamefont
		{Ruskai},\ and\ \citenamefont {Shimono}}]{hayashi2004qubit}%
	\BibitemOpen
	\bibfield  {author} {\bibinfo {author} {\bibfnamefont {M.}~\bibnamefont
			{Hayashi}}, \bibinfo {author} {\bibfnamefont {H.}~\bibnamefont {Imai}},
		\bibinfo {author} {\bibfnamefont {K.}~\bibnamefont {Matsumoto}}, \bibinfo
		{author} {\bibfnamefont {M.~B.}\ \bibnamefont {Ruskai}}, \ and\ \bibinfo
		{author} {\bibfnamefont {T.}~\bibnamefont {Shimono}},\ }\href@noop {}
	{\bibfield  {journal} {\bibinfo  {journal} {arXiv preprint quant-ph/0403176}\
		} (\bibinfo {year} {2004})}\BibitemShut {NoStop}%
	\bibitem [{\citenamefont {Kristj{\'a}nsson}\ \emph {et~al.}()\citenamefont
		{Kristj{\'a}nsson}, \citenamefont {Zhong}, \citenamefont {Munson},\ and\
		\citenamefont {Chiribella}}]{kristjansson2021network}%
	\BibitemOpen
	\bibfield  {author} {\bibinfo {author} {\bibfnamefont {H.}~\bibnamefont
			{Kristj{\'a}nsson}}, \bibinfo {author} {\bibfnamefont {Y.}~\bibnamefont
			{Zhong}}, \bibinfo {author} {\bibfnamefont {A.}~\bibnamefont {Munson}}, \
		and\ \bibinfo {author} {\bibfnamefont {G.}~\bibnamefont {Chiribella}},\
	}\href@noop {} {\bibinfo  {journal} {In preparation}\ }\BibitemShut {NoStop}%
	\bibitem [{\citenamefont {Del~Santo}\ and\ \citenamefont
		{Daki{\'c}}(2018)}]{del2018two}%
	\BibitemOpen
	\bibfield  {journal} {  }\bibfield  {author} {\bibinfo {author} {\bibfnamefont
			{F.}~\bibnamefont {Del~Santo}}\ and\ \bibinfo {author} {\bibfnamefont
			{B.}~\bibnamefont {Daki{\'c}}},\ }\href@noop {} {\bibfield  {journal}
		{\bibinfo  {journal} {Physical Review Letters}\ }\textbf {\bibinfo {volume}
			{120}},\ \bibinfo {pages} {060503} (\bibinfo {year} {2018})}\BibitemShut
	{NoStop}%
	\bibitem [{\citenamefont
		{{\AA}berg}(2004{\natexlab{b}})}]{aaberg2004operations}%
	\BibitemOpen
	\bibfield  {author} {\bibinfo {author} {\bibfnamefont {J.}~\bibnamefont
			{{\AA}berg}},\ }\href@noop {} {\bibfield  {journal} {\bibinfo  {journal}
			{Physical Review A}\ }\textbf {\bibinfo {volume} {70}},\ \bibinfo {pages}
		{012103} (\bibinfo {year} {2004}{\natexlab{b}})}\BibitemShut {NoStop}%
	\bibitem [{\citenamefont {Zhou}\ \emph {et~al.}(2011)\citenamefont {Zhou},
		\citenamefont {Ralph}, \citenamefont {Kalasuwan}, \citenamefont {Zhang},
		\citenamefont {Peruzzo}, \citenamefont {Lanyon},\ and\ \citenamefont
		{O'Brien}}]{zhou2011adding}%
	\BibitemOpen
	\bibfield  {author} {\bibinfo {author} {\bibfnamefont {X.-Q.}\ \bibnamefont
			{Zhou}}, \bibinfo {author} {\bibfnamefont {T.~C.}\ \bibnamefont {Ralph}},
		\bibinfo {author} {\bibfnamefont {P.}~\bibnamefont {Kalasuwan}}, \bibinfo
		{author} {\bibfnamefont {M.}~\bibnamefont {Zhang}}, \bibinfo {author}
		{\bibfnamefont {A.}~\bibnamefont {Peruzzo}}, \bibinfo {author} {\bibfnamefont
			{B.~P.}\ \bibnamefont {Lanyon}}, \ and\ \bibinfo {author} {\bibfnamefont
			{J.~L.}\ \bibnamefont {O'Brien}},\ }\href@noop {} {\bibfield  {journal}
		{\bibinfo  {journal} {Nature Communications}\ }\textbf {\bibinfo {volume}
			{2}},\ \bibinfo {pages} {1} (\bibinfo {year} {2011})}\BibitemShut {NoStop}%
	\bibitem [{\citenamefont {Chiribella}\ \emph
		{et~al.}(2008{\natexlab{b}})\citenamefont {Chiribella}, \citenamefont
		{D'Ariano},\ and\ \citenamefont {Perinotti}}]{chiribella2008transforming}%
	\BibitemOpen
	\bibfield  {author} {\bibinfo {author} {\bibfnamefont {G.}~\bibnamefont
			{Chiribella}}, \bibinfo {author} {\bibfnamefont {G.~M.}\ \bibnamefont
			{D'Ariano}}, \ and\ \bibinfo {author} {\bibfnamefont {P.}~\bibnamefont
			{Perinotti}},\ }\href@noop {} {\bibfield  {journal} {\bibinfo  {journal} {EPL
				(Europhys. Lett.)}\ }\textbf {\bibinfo {volume} {83}},\ \bibinfo {pages}
		{30004} (\bibinfo {year} {2008}{\natexlab{b}})}\BibitemShut {NoStop}%
	\bibitem [{\citenamefont {Peres}(1996)}]{peres1996separability}%
	\BibitemOpen
	\bibfield  {author} {\bibinfo {author} {\bibfnamefont {A.}~\bibnamefont
			{Peres}},\ }\href@noop {} {\bibfield  {journal} {\bibinfo  {journal}
			{Physical Review Letters}\ }\textbf {\bibinfo {volume} {77}},\ \bibinfo
		{pages} {1413} (\bibinfo {year} {1996})}\BibitemShut {NoStop}%
	\bibitem [{\citenamefont {Horodecki}\ \emph {et~al.}(2001)\citenamefont
		{Horodecki}, \citenamefont {Horodecki},\ and\ \citenamefont
		{Horodecki}}]{horodecki2001separability}%
	\BibitemOpen
	\bibfield  {author} {\bibinfo {author} {\bibfnamefont {M.}~\bibnamefont
			{Horodecki}}, \bibinfo {author} {\bibfnamefont {P.}~\bibnamefont
			{Horodecki}}, \ and\ \bibinfo {author} {\bibfnamefont {R.}~\bibnamefont
			{Horodecki}},\ }\href@noop {} {\bibfield  {journal} {\bibinfo  {journal}
			{Physics Letters A}\ }\textbf {\bibinfo {volume} {283}},\ \bibinfo {pages}
		{1} (\bibinfo {year} {2001})}\BibitemShut {NoStop}%
	\bibitem [{\citenamefont {Holevo}(2002)}]{holevo2002remarks}%
	\BibitemOpen
	\bibfield  {author} {\bibinfo {author} {\bibfnamefont {A.~S.}\ \bibnamefont
			{Holevo}},\ }\href@noop {} {\bibfield  {journal} {\bibinfo  {journal} {arXiv
				preprint quant-ph/0212025}\ } (\bibinfo {year} {2002})}\BibitemShut {NoStop}%
	\bibitem [{\citenamefont {Chiribella}\ and\ \citenamefont
		{Mauro~D’Ariano}(2006)}]{chiribella2006extremal}%
	\BibitemOpen
	\bibfield  {author} {\bibinfo {author} {\bibfnamefont {G.}~\bibnamefont
			{Chiribella}}\ and\ \bibinfo {author} {\bibfnamefont {G.}~\bibnamefont
			{Mauro~D’Ariano}},\ }\href@noop {} {\bibfield  {journal} {\bibinfo
			{journal} {Journal of Mathematical Physics}\ }\textbf {\bibinfo {volume}
			{47}},\ \bibinfo {pages} {092107} (\bibinfo {year} {2006})}\BibitemShut
	{NoStop}%
\end{thebibliography}
